\documentclass[aps,amssymb,amsmath,notitlepage,superscriptaddress,prx,twocolumn,nofootinbib]{revtex4-1}

\usepackage{amsfonts}
\usepackage{graphicx,graphics,epsfig,times,bm,bbm,amssymb,amsmath,amsfonts,mathrsfs}
\usepackage[normalem]{ulem}
\usepackage{wrapfig}
\usepackage{setspace}
\usepackage{subfigure}
\usepackage{dsfont}
\usepackage{braket}
\usepackage[pdftex]{color}
\usepackage{upgreek }
\usepackage[x11names]{xcolor}
\usepackage{tikz}
\usepackage{natbib}
\usepackage{chngcntr}

\newcommand{\beq}{\begin{equation}}
\newcommand{\eeq}{\end{equation}}
\newcommand{\beqnn}{\begin{equation*}}
\newcommand{\eeqnn}{\end{equation*}}
\newcommand{\beann}{\begin{eqnarray*}}
\newcommand{\eeann}{\end{eqnarray*}}

\newcommand{\mc}{\mathcal}

\newcommand{\bes} {\begin{subequations}}
\newcommand{\ees} {\end{subequations}}
\newcommand{\bea} {\begin{eqnarray}}
\newcommand{\eea} {\end{eqnarray}}

\newcommand{\ignore}[1]{}

\begin{document}

\title{Quantum Annealing Correction with Minor Embedding}

\author{Walter Vinci}
\affiliation{Department of Electrical Engineering, University of Southern California, Los Angeles, California 90089, USA}
\affiliation{Department of Physics and Astronomy, University of Southern California, Los Angeles, California 90089, USA}
\affiliation{Center for Quantum Information Science \& Technology, University of Southern California, Los Angeles, California 90089, USA}
\author{Tameem Albash}
\affiliation{Department of Physics and Astronomy, University of Southern California, Los Angeles, California 90089, USA}
\affiliation{Center for Quantum Information Science \& Technology, University of Southern California, Los Angeles, California 90089, USA}
\affiliation{Information Sciences Institute, University of Southern California, Marina del Rey, California 90292, USA}
\author{Gerardo Paz-Silva}
\affiliation{Centre for Quantum Computation and Communication Technology (Australian Research Council), Griffith University, Brisbane, Queensland 4111, Australia}
\affiliation{Centre for Quantum Dynamics, Griffith University, Brisbane, Queensland 4111, Australia}

\author{Itay Hen}
\affiliation{Center for Quantum Information Science \& Technology, University of Southern California, Los Angeles, California 90089, USA}
\affiliation{Information Sciences Institute, University of Southern California, Marina del Rey, California 90292, USA}

\author{Daniel A. Lidar}
\affiliation{Department of Electrical Engineering, University of Southern California, Los Angeles, California 90089, USA}
\affiliation{Department of Physics and Astronomy, University of Southern California, Los Angeles, California 90089, USA}
\affiliation{Center for Quantum Information Science \& Technology, University of Southern California, Los Angeles, California 90089, USA}
\affiliation{Department of Chemistry, University of Southern California, Los Angeles, California 90089, USA}

\begin{abstract}

Quantum annealing provides a promising route for the development of quantum optimization devices, but the usefulness of such devices will be limited in part by the range of implementable problems as dictated by hardware constraints. To overcome constraints imposed by restricted connectivity between qubits, a larger set of interactions can be approximated using minor embedding techniques whereby several physical qubits are used to represent a single logical qubit. However, minor embedding introduces new types of errors due to its approximate nature. We introduce and study quantum annealing correction schemes designed to improve the performance of quantum annealers in conjunction with minor embedding, thus leading to a hybrid scheme defined over an encoded graph. We argue that this scheme can be efficiently decoded using an energy minimization technique provided the density of errors does not exceed the per-site percolation threshold of the encoded graph.
We test the hybrid scheme using a D-Wave Two processor on problems for which the encoded graph is a $2$-level grid and the Ising model is known to be NP-hard. The problems we consider are frustrated Ising model problem instances with ``planted" (a priori known) solutions. Applied in conjunction with optimized energy penalties and decoding techniques, we find that this approach enables the quantum annealer to solve minor embedded instances with significantly higher success probability than it would without error correction. 
Our work demonstrates that quantum annealing correction can and should be used to improve the robustness of quantum annealing not only for natively embeddable problems, but also when minor embedding is used to extend the connectivity of physical devices.  
\end{abstract}
\maketitle
%
%%%%%%%%%%%%%%%%%%%%%%%%%%%%%%%%%%
\section{Introduction}
%%%%%%%%%%%%%%%%%%%%%%%%%%%%%%%%%%

Quantum computers are in principle able to solve computational problems that are intractable for their classical counterparts \cite{Bacon:review,Harrow:2014mz}. 
Remarkably, rudimentary quantum annealers---a special type of adiabatic quantum computer \cite{farhi_quantum_2000} specifically engineered to implement quantum annealing \cite{finnila_quantum_1994,kadowaki_quantum_1998,Brooke1999,brooke_tunable_2001,2002quant.ph.11152K,Dwave} to solve hard optimization problems \cite{farhi_quantum_2001}---are already commercially available \cite{Johnson:2010ys,Berkley:2010zr,Harris:2010kx},
and their availability to the scientific community has recently stimulated a productive debate \cite{EPJ-ST:2015,Aaronson-blog,q-sig,q108,Smolin,comment-SS,SSSV,Albash:2014if,q-sig2,SSSV-comment,Crowley:2014qp}. The question at stake is whether current experimental quantum annealing can really be considered a form of quantum computation, even when the decoherence time of the component qubits is much smaller than the overall computational time. While theory suggests that the answer to this question can be positive \cite{Albash:2015nx,Kechedzhi:2015ul}, as it concerns quantum annealing experiments the answer is still open, in particular due to the ``black-box" nature of quantum annealing, which makes it difficult to unambiguously determine the nature of the physical processes taking place between the initialization and readout steps. However, convincing evidence is mounting that quantum effects, including entanglement \cite{DWave-entanglement} and multi-qubit coherent tunneling \cite{Boixo:2014yu}, play a functional role in explaining the behavior of available annealers. Moreover, open system quantum dynamical models \cite{ABLZ:12-SI,Boixo:2014yu} have been  successful in reproducing experimental data from quantum annealing experiments \cite{q-sig,DWave-16q,q-sig2,Boixo:2014yu,Albash:2014if}. A closely related question is whether quantum annealers are able to provide quantum speedup, a theoretical possibility \cite{Somma:2012kx} despite skepticism based, e.g., on the appearance of exponentially small gaps \cite{Jorg:2010qa}. While compelling evidence of a quantum speedup is still lacking with current physical annealers, it is also generally recognized that  technological improvements (in particular reduced control errors and shorter annealing times) in the next generation of quantum annealers are necessary in order to unambiguously address the issue \cite{speedup,Hen:2015rt,King:2014uq,King:2015zr,Amin:2015qf}. A careful choice of the benchmark problems is another important consideration \cite{2014Katzgraber,Katzgraber:2015gf}, in particular in light of the fragility of spin glasses to small perturbations (temperature chaos) \cite{Martin-Mayor:2015dq,Zhu:2015pd}.

Ultimately, the usefulness of quantum annealers relies on the promise of achieving quantum information processing on a large scale. This goal can only be met through the use of some form of quantum error correction (QEC), which is necessary to protect the performance of any quantum computation against decoherence \cite{Gottesman:10,Gaitan:book,Lidar-Brun:book}. The scalability of quantum computing via QEC has been established in the gate model in the form of the celebrated threshold theorem (e.g, Refs.~\cite{Knill:98,Aharonov:05}).  Adiabatic quantum computation and quantum annealing also require some form of error correction in order to preserve their advantages over classical computation. This is true even though adiabatic dynamics is intrinsically robust to some forms of decoherence \cite{childs_robustness_2001,PhysRevLett.95.250503,TAQC}. Several error suppression techniques suitable for adiabatic quantum computing (AQC) have been  proposed \cite{jordan2006error,PhysRevLett.100.160506,PhysRevA.86.042333,Ganti:13,Bookatz:2014uq}. However, despite valiant theoretical attempts  \cite{Mizel:2014sp} much less is known about how to achieve fault tolerant AQC, and some negative results have been established \cite{Young:13,Sarovar:2013kx,Marvian:2014nr}. Still, it has recently been shown experimentally that it is possible to improve the performance of physical quantum annealers with the help of tailored quantum annealing correction \cite{PAL:13,PAL:14,Young:2013fk}, a method we shall employ and generalize in this work.

Typically,  realistic implementations of quantum annealing are subject to restricted (local) interactions between qubits specified by the physical hardware.  Minor embedding (ME) techniques overcome this limitation by using several physical qubits to represent a single logical qubit with a larger set of interactions \cite{Choi1,Choi2,klymko_adiabatic_2012,Cai:2014nx}. A minor embedding replaces a given optimization problem with an approximate yet implementable one. As a consequence of this approximation the performance of a quantum annealer is typically degraded. 
ME also adds two extra steps to the computations performed with a quantum annealer. One is an encoding procedure that determines the mapping between the logical problem and its physically embeddable representation, which in turn requires a choice of energy penalties to align the physical qubits of each logical qubit. It turns out that performance is sensitive to the mapping chosen and it is thus important to optimize it. The other is a decoding procedure that recovers the logical answer from the physical output. Decoding is nontrivial whenever the values of the physical qubits corresponding to the same logical qubit disagree, and will be discussed in detail in this work.

Quantum annealing correction (QAC) was proposed in Ref.~\cite{PAL:13}, where it was demonstrated on anti-ferromagnetic chains and further discussed for random Ising problems in Ref.~\cite{PAL:14}. These works showed that QAC can improve the performance of quantum annealers. Implementing several independent copies of the same problem enables errors to be corrected using, e.g., majority vote as in classical repetition codes. Thus, importantly, the requirement of a purely adiabatic evolution can be relaxed since excited states can be tolerated as long as they can be correctly decoded. 
Quantum error suppression is obtained with the use of ferromagnetic energy penalties connecting the various copies, that are turned on gradually along with the final Hamiltonian. Since these can also be understood as the stabilizers of the quantum repetition code, that detect and penalize bit-flip errors during the course of the evolution, agreement is enhanced within each copy by energetically penalizing errors that would result in disagreement.
%\red{TA: This statement is misleading because it suggests we have shown this to be the case for the full quantum spectrum.}.
The combined effect of encoding into logical qubits and quantum error suppression can be seen as an increase of the effective energy scale with which the logical problem is physically implemented. This increases the relevant minimum energy gaps that protect quantum annealing from thermal and dynamical excitations. Moreover, it improves the response of quantum annealing devices to intrinsic control errors (due to the  limited accuracy with which the physical fields %$J$ and $h$ fields 
 can be tuned) by using several physical couplings to specify a single logical interaction \cite{PAL:13,PAL:14,Young:2013fk}.

The main goal of this work is to develop and study QAC techniques for improving the performance of a quantum annealer following the expected degradation due to the implementation of ME techniques. In doing so, we explore the efficacy and applicability of several encoding and decoding strategies. Our tests are performed using a D-Wave Two (DW2) ``Vesuvius" quantum annealer with $504$ functional qubits.

Our paper is organized as follows. In section~\ref{sec:MEQA}, after a brief review of quantum annealing, we give an overview of ME and QAC. We explain how ME can be viewed as a map from a given logical problem to the layout dictated by the physical implementation, while QAC reverses this direction and maps from the physical level to an encoding that provides protection. Thus, while different in scope, the two approaches can be seamlessly concatenated in what we call quantum annealing correction with minor embedding (QAC-ME). We also discuss the different role of frustration effects on ME and QAC. 
In Sec.~\ref{sec:ES} we discuss different strategies for choosing the energy penalties in both ME and QAC, and with an eye to optimality, introduce a nonuniform strategy.
In Sec.~\ref{sec:DS} we discuss decoding strategies. These are required in order to assign a value to a logical or encoded qubit when the corresponding physical qubits have different values. A majority vote on the physical qubits is a simple choice but that may not always justified. We thus propose energy minimization as a more effective decoding strategy and provide a general argument based on percolation theory to determine whether it can be efficiently implemented. Section~\ref{sec:SC} introduces the ``square code,'' a QAC-ME scheme defined on a $2$-level grid that can be implemented with the physical connectivity (``Chimera" graph)  of the DW2 processor. The Ising problem on the $2$-level grid is of particular interest since it was the first example of an NP-hard Ising model problem \cite{Barahona1982}. The square code is of independent interest since it is amenable to concatenation. 
Section~\ref{sec:ER} presents our experimental results, obtained using the DW2 processor. The problem instances we study are constructed using the planted ground states method for frustrated loops introduced in Ref.~\cite{Hen:2015rt}. This method allows for a tunable hardness parameter in the form of the effective clause density. The experimental results demonstrate the significant performance boost that can be achieved using the QAC-ME strategy, by exhibiting higher ground state success probabilities, with an improvement that is more pronounced for the harder problem instances.
Section~\ref{sec:dec-sc} reports a detailed comparative experimental analysis of the different decoding strategies discussed in Sec.~\ref{sec:DS}, specialized to the square code. 
In Sec.~\ref{sec:NS} we present the results of a simulated quantum annealing study that numerically investigates the role of temperature and control noise on the performance of the ME and QAC-ME strategies, as these are parameters that we cannot control experimentally.  
We conclude and discuss possible developments of our work in Sec.~\ref{sec:CD}. Additional supporting information is presented in the Appendix.

\begin{figure*}[ht]
\begin{center}
\subfigure[\ USC-ISI DW2 Chimera graph]{ \includegraphics[width=0.45\textwidth]{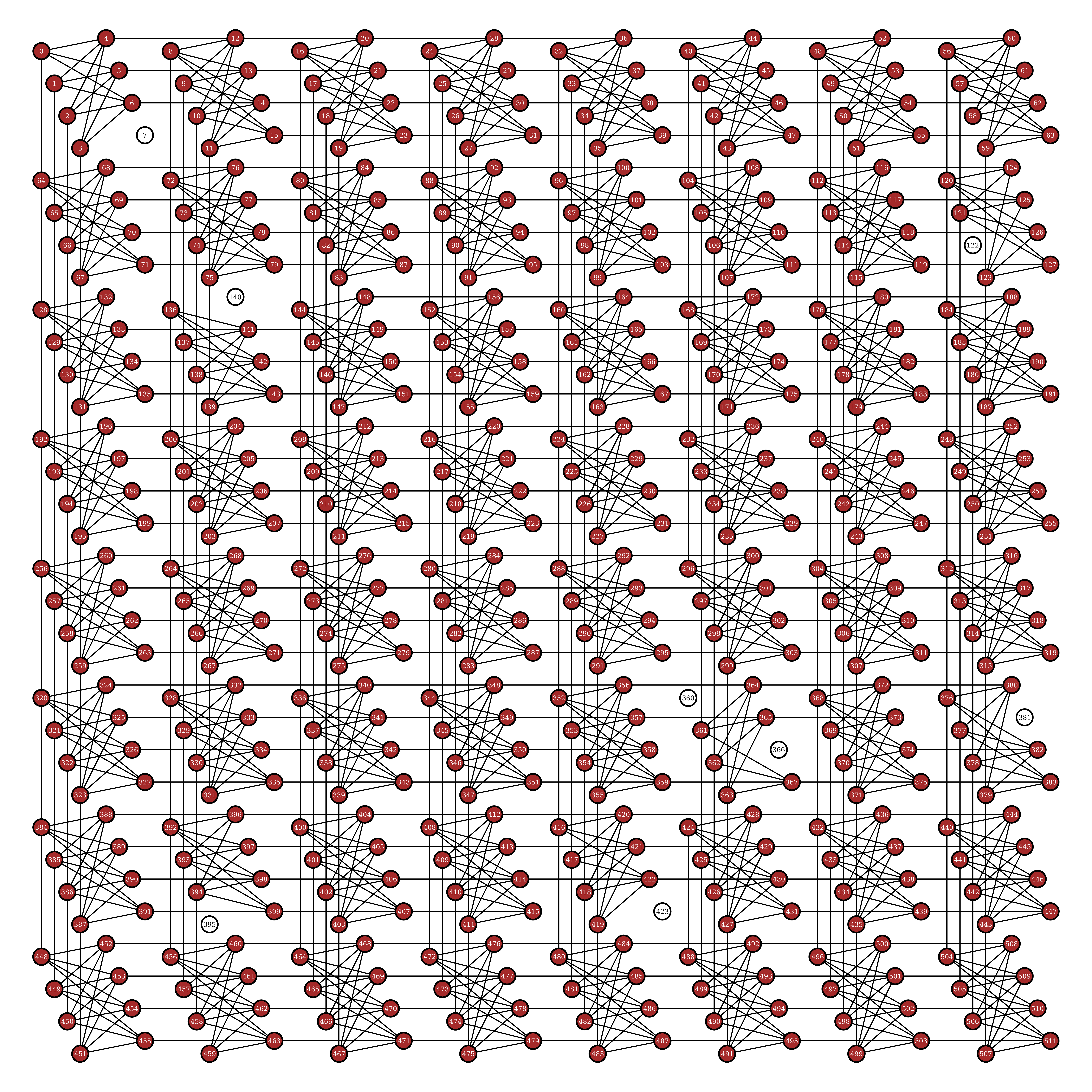}\label{fig:Vesuvius}}
\subfigure[\ Annealing schedule]{ \hspace{-2cm} \includegraphics[width=0.6\textwidth]{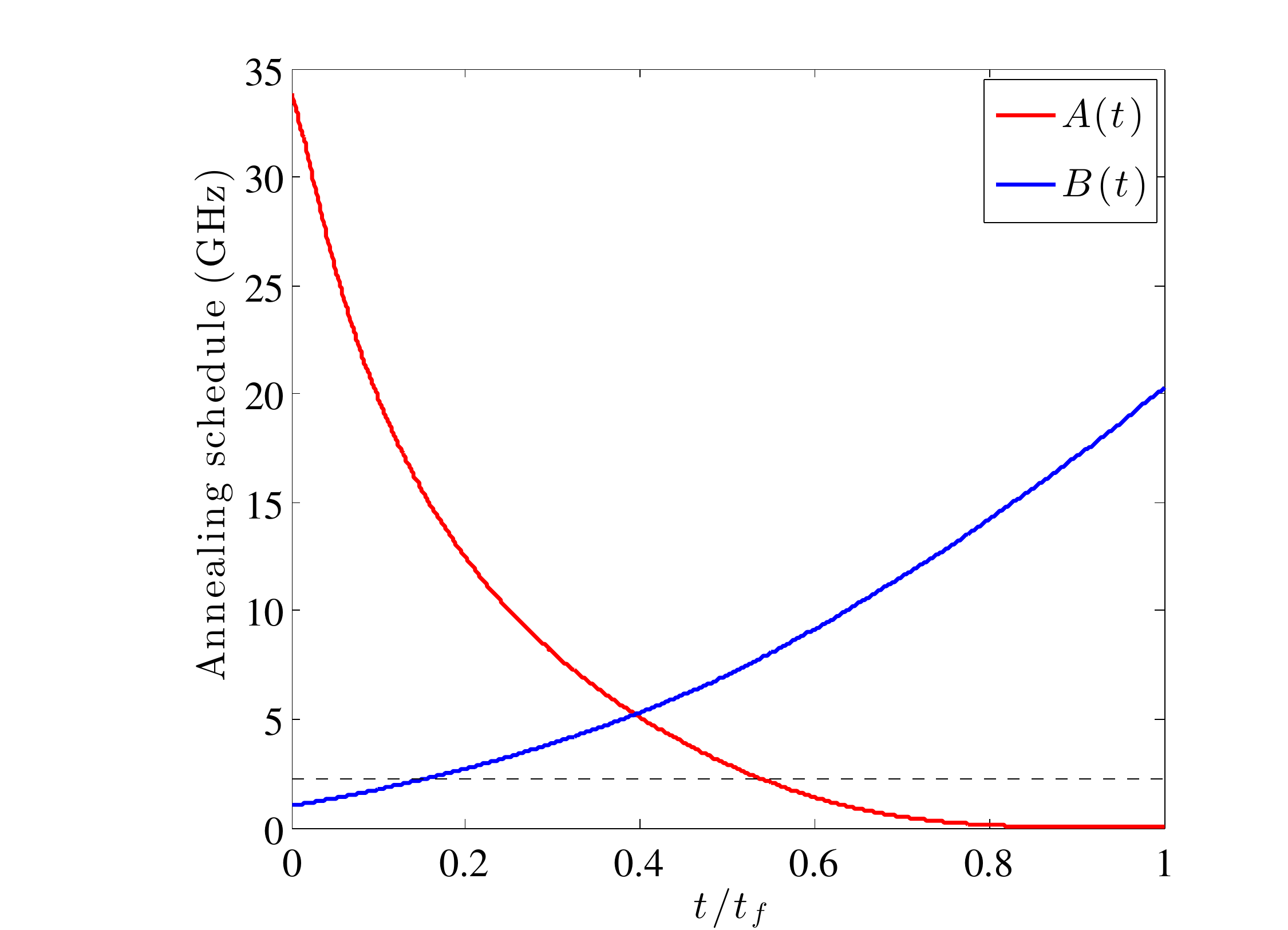}\label{fig:Schedule}}
\caption{(a) Physical connectivity of the ``Chimera" graph of the $512$-qubit DW2 processor. White circles correspond to the $8$ unusable qubits in the USC-ISI DW2 processor. (b) Annealing schedule of the DW2 processor. The dashed horizontal line is the physical temperature of $17$mK at which the device is kept.} 
\label{fig:schedule}
\end{center}
\end{figure*}

%%%%%%%%%%%%%%%%%%%%%%%%%%%%%%%%%%%%%%%%%
\section{Quantum Annealing, Minor Embedding, and Quantum Annealing Correction}
\label{sec:MEQA}
%%%%%%%%%%%%%%%%%%%%%%%%%%%%%%%%%%%%%%%%%
%
\subsection{Quantum annealing}
The term quantum annealing was first coined (as far as we know) in 1994 by Finnila \textit{et al}. \cite{finnila_quantum_1994}, though the underlying idea was 
proposed at least as early as 1984 \cite{Ebeling:1984tg} as a numerical algorithm inspired by simulated annealing \cite{kirkpatrick_optimization_1983}, where the role of thermal fluctuations was conjectured to be advantageously replaced by quantum tunneling. This was soon followed by numerous other studies based on a similar idea (see Refs.~\cite{Rujan:1988hc,Somorjai:1991oq,Olszewski:1992kl,PhysRevB.39.11828,kadowaki_quantum_1998} for some of the earliest works and Ref.~\cite{RevModPhys.80.1061} for a recent review). 
Numerical and experimental results have suggested that quantum tunneling can be more effective than thermal fluctuations for reaching the ground state \cite{Brooke1999,brooke_tunable_2001,2002quant.ph.11152K,Dwave}. This conclusion  has recently been revisited \cite{Heim:2014jf}, and remains the subject of active scrutiny (e.g., Ref.~\cite{Muthukrishnan:2015ff}). %Adiabatic quantum computation provides a universal framework for quantum computation that relies on the adiabatic theorem and includes quantum annealing \cite{farhi_quantum_2000,aharonov_adiabatic_2007,PhysRevLett.99.070502,Gosset:2014rp}.  

Quantum annealing for spin systems is typically achieved by the following time-dependent, transverse-field Ising Hamiltonian:
\beq
H(t) = A(t) H_X + B(t)  H_{\mathrm{P}}\ , \qquad t\in[0,t_f] \ .
\label{eq:adiabatic}
\eeq
The term $H_X = -\sum_i \sigma_i^x$ (with $\sigma_i^x$ being the Pauli matrix acting on qubit $i$) is a transverse field whose amplitude controls the tunneling rate. The solution to an optimization problem of interest is encoded in the ground state of the problem Hamiltonian $H_{\mathrm{P}}$, given by
\beq 
\label{eq:HP}
H_{\mathrm{P}} = \sum_{i \in \mc{V}_h} h_i \sigma^z_i + \sum_{(i,j) \in \mc{E}_J} J_{ij}\sigma^z_i\sigma^z_j\ ,
\eeq
where the sums run over the vertices $\mc{V}$ and edges $\mc{E}$ of the weighted, undirected graph $G = (\mc{V}_h,\mc{E}_J)$, where the local fields $h_i$ and couplings $J_{ij}$ are the weights. 
In a closed system, the adiabatic theorem ensures that if the system is initialized in the ground state of $H(0) = A(0) H_X$ (assuming $B(0) =0)$, a sufficiently slow (adiabatic) evolution results at the end of the computation in the ground state of the final Hamiltonian $H(t_f) =B(t_f) H_{\mathrm{P}}$ (assuming $A(t_f) = 0$). In an open system annealing at a non-zero temperature one expects the system to evolve to a mixed final state, e.g., the Gibbs state of $H(t_f)$, if equilibration is possible throughout the annealing process (see, e.g., Ref.~\cite{Albash:2015nx}).

The D-Wave devices use an array of superconducting niobium-based flux qubits in order to physically realize quantum annealing \cite{Johnson:2010ys,Berkley:2010zr,Harris:2010kx}. Each vertex of the D-Wave ``Chimera'' graph is a physical flux qubit and each edge is a physical inductive coupling between two qubits. Fig.~\ref{fig:Vesuvius} shows a pictorial representation of such a graph. In the DW2 processor used in this work the temperature is set to $17{\rm mK}$ and the annealing schedule is given by the  functions $A(t)$ and $B(t)$ shown in Fig.~\ref{fig:Schedule}. %The function $B(t)$ also determines the maximum strength of the allowed couplings: $|h_i^{\max}| = 2 B(t_f)$ and  $|J_{ij}^{\max}| = B(t_f)$. In the following, we specify the values of the couplings $h_i$ and $J_{ij}$ in dimensionless units of $B(t_f)$.
In the following, we specify the values of the couplings $h_i$ and $J_{ij}$ in dimensionless units, with the maximum strength of the allowed couplings given by $|h_i^{\max}| = 2$ and  $|J_{ij}^{\max}| = 1$.

\subsection{Mapping to and from the logical problem: ME and QAC}

The problem Hamiltonian defined in Eq.~\eqref{eq:HP} may or may not be specified in terms that directly correspond to physical variables representing the actual degrees of freedom of a physical device. When such a specification is possible we have a \emph{direct} (D) embedding. An example is the class of random Ising model problems studied in Refs.~\cite{q108,speedup}, which simply require a specification of the $h_i$ and $J_{ij}$ on the Chimera graph of the D-Wave device. 

However, often one is interested in solving problems that do not have such a direct embedding. E.g., consider a problem defined on a complete graph, which requires a minor embedding on a graph that is not complete. In addition one might want an encoding of the Hamiltonian in order to perform error correction. To deal with this more general scenario we shall proceed to define two maps: one for ME, another for QAC. We shall also consider the concatenated map, of ME followed by QAC, which we refer to as QAC-ME. 

\subsubsection{ME: from the logical Hamiltonian to the physical Hamiltonian}

When the problem Hamiltonian is defined in terms of variables that do not have a direct representation and requires a minor embedding we shall refer to it as a ``logical Hamiltonian'' and write
\beq
\tilde{H}_{\mathrm{P}} = \sum_{\tilde i \in \mc{\tilde V}_{\tilde h}}\tilde h_{\tilde i} \tilde{\sigma}^z_{\tilde i} + \sum_{(\tilde i,\tilde j) \in \mc{\tilde E}_{\tilde J}} \tilde{J}_{\tilde i\tilde j}\tilde{\sigma}^z_{\tilde i} \tilde{\sigma}^z_{\tilde j}\ .
\eeq
Quantities decorated with a tilde represent the abstract logical local fields, couplings, Pauli operators, etc. 
%The $\tilde{N} = |\tilde{\mc{V}}_{\tilde h}|$ logical qubits are binary $\pm 1$ variables.\footnote{It would be more accurate to refer to the classical binary logical variables $\tilde{\sigma}^z_{\tilde i}$ in $\tilde{H}_{\mathrm{P}}$ as spins rather than qubits, but since ultimately we are interested in the total Hamiltonian, which includes the transverse field $H_X$, we prefer to use the qubit terminology throughout.} 
The logical Hamiltonian is defined on the vertices and edges of the logical graph $\tilde{G} = (\tilde{\mc{V}}_{\tilde h},\tilde{\mc{E}}_{\tilde J})$. Given the logical Hamiltonian which specifies the optimization problem of interest in terms of abstract variables, the task is to map it to a physical problem Hamiltonian that has a direct embedding. This is the minor embedding map
\beq
\tilde{H}_{\mathrm{P}} \overset{\mathcal{M}_{\mathrm{ME}}}{\longmapsto} {H}_{\mathrm{P}}\ .
\eeq
This map exists when $\tilde{G}$ is a \emph{minor} of $G$, i.e., it has a minor embedding in $G$.\footnote{A minor graph is obtained from $G$ by collapsing groups of connected (physical) vertices of $G$ into single (logical) vertices of $\tilde G$ and possibly removing edges from $G$ \cite{Wilson:book}. Note that $\mathcal{M}_{\mathrm{ME}}$ is a multivalued map (one-to-many).}

%%%%%%%
\begin{figure}[]
\begin{center}
\includegraphics[width=0.45\textwidth]{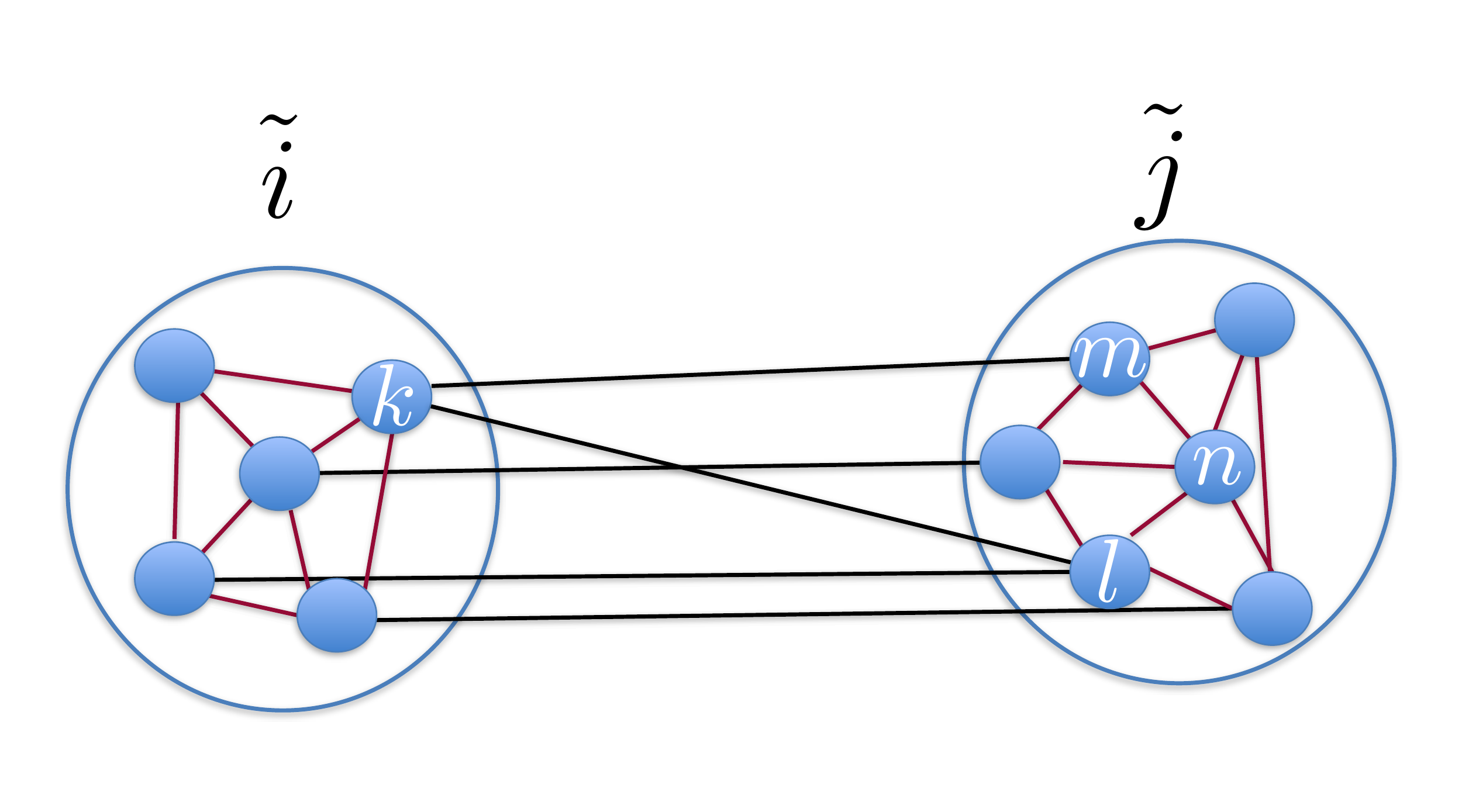} 
\caption{ME: Schematic of two logical groups $\tilde{i},\tilde{j}$ (large circles) representing two logical qubits $\tilde{\sigma}^z_{\tilde i}, \tilde{\sigma}^z_{\tilde j}$. Physical qubits are denoted by blue circles; physical problem couplings are denoted by black lines, penalty couplings by red lines. In the example shown the penalty coupling $J_{\tilde{j}_m \tilde{j}_n}$ and the physical problem couplings $J_{\tilde{i}_k \tilde{j}_l},J_{\tilde{i}_k \tilde{j}_m}$ are active, among others.}
\label{fig:lqc}
\end{center}
\end{figure}
%%%%%%%

%%%%%%%%
\begin{figure}[t]
\begin{center}
\includegraphics[width=0.45\textwidth]{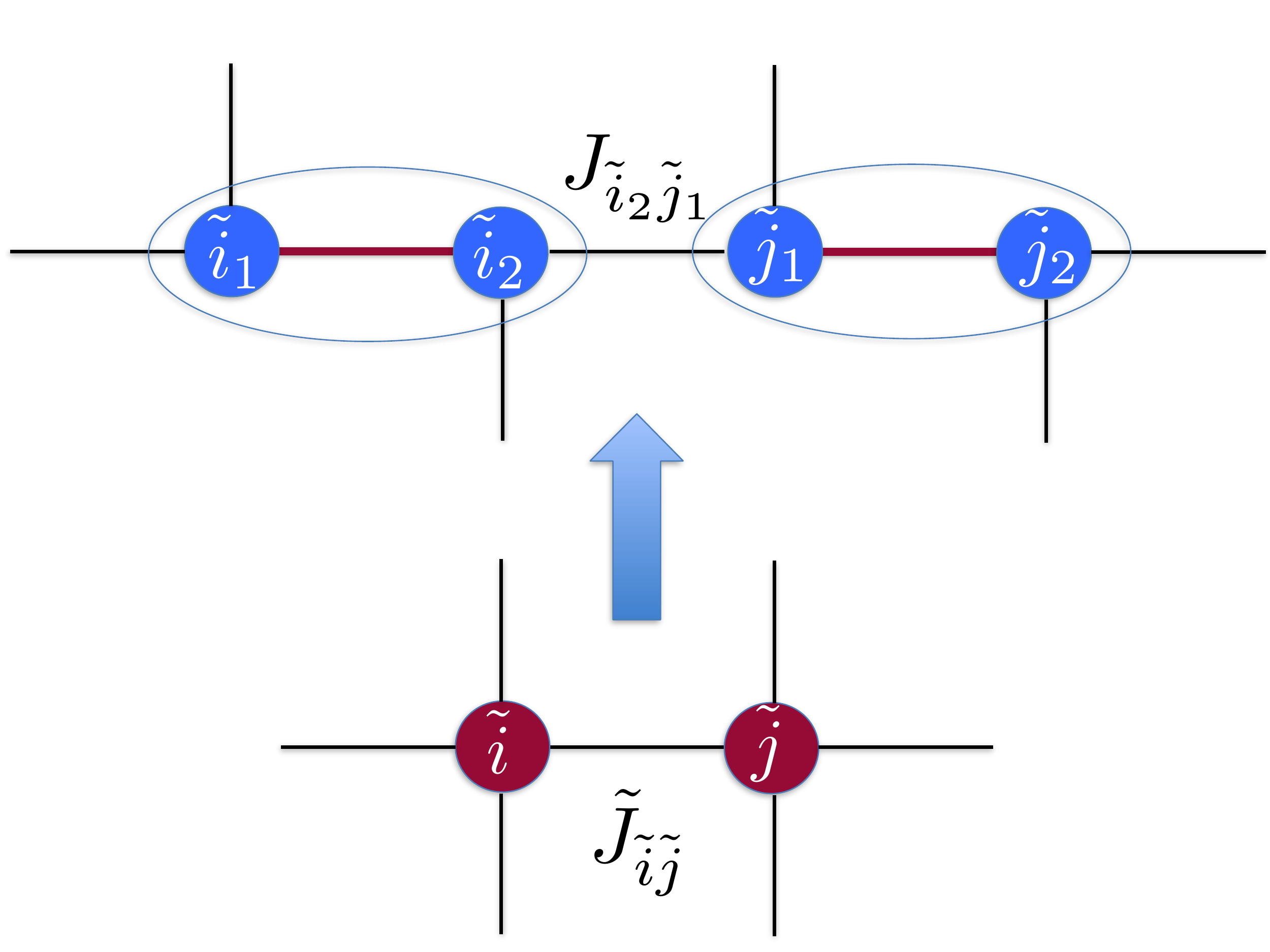} 
\caption{Schematic example of the map $\mathcal{M}_{\mathrm{ME}}$. Two logical qubits $\tilde{i}$ and $\tilde{j}$ (red dots) with four logical couplings each (black lines) are expanded into two physical qubits each (circled blue dots) with three physical couplings each. The red lines 
%\red{TA: These are hard to distinguish.  Perhaps we can make the line more red rather than dark red.}
connecting the physical qubits are the penalty couplings $J_{\tilde{i}_1 \tilde{i}_2}$ and $J_{\tilde{j}_1 \tilde{j}_2}$. In this example the logical coupling $\tilde{J}_{\tilde{i}\tilde{j}}$ is replaced by a single physical coupling $J_{\tilde{i}_2 \tilde{j}_1}$, a special case of Eq.~\eqref{eq:MEcoups}.}
\label{fig:schematics}
\end{center}
\end{figure}
%%%%%%%%
Each of the $\tilde{N} = |\tilde{\mc{V}}_{\tilde h}|$ vertices of $\tilde{G}$ is occupied by a classical binary ($\pm 1$) variable, which we call a logical qubit.\footnote{Technically this is a misnomer since the variable is classical. However, we have in mind the entire time-dependent Hamiltonian of quantum annealing, and moreover, we can always measure the logical qubit in the logical $\sigma^z$ basis.} The map $\mathcal{M}_{\mathrm{ME}}$ replaces these by $N = |\mc{V}_{h}|$  physical qubits.
A minor embedding defines a partitioning of all $N$ physical qubits into $\tilde{N}$ disjoint logical groups \{$\tilde{i}$\}, each comprising $N_{\tilde{i}}$ physical qubits, whose union covers all physical qubits, i.e, $N = \sum_{\tilde{i}=1}^{\tilde{N}} N_{\tilde{i}}$. Note that we distinguish between a \emph{logical group} and a \emph{logical qubit}. The latter was defined above; the former is a set of physical qubits representing the logical qubit.
What identifies a logical group is a set of ferromagnetic ``penalty" couplings which tie the physical qubits representing a given logical qubit together, as illustrated in Fig.~\ref{fig:lqc}. Penalty couplings are always ``intra", i.e., couple physical qubits of a given logical group $\tilde{i}$, and can thus always be written as 
\beq
J_{\mathrm{pen}} \in \{J_{\tilde{i}_k \tilde{i}_l}\}\ .
\eeq
The physical qubits comprising two logical groups are coupled in a manner that depends on the problem Hamiltonian. The logical local fields and couplings are defined via
\beq
\tilde{h}_{\tilde{i}} = \sum_{k=1}^{N_{\tilde{i}}} h_{\tilde{i}_k}\ , 
\qquad 
\tilde{J}_{\tilde{i}\tilde{j}} = \sum_{k=1}^{N_{\tilde{i}}} \sum_{l=1}^{N_{\tilde{j}}} J_{\tilde{i}_k\tilde{j}_l}\quad (\tilde i\neq \tilde j)\ ,
\label{eq:MEcoups}
\eeq
where some of the $J_{\tilde{i}_k\tilde{j}_l}$ may vanish, as illustrated in Fig.~\ref{fig:lqc}. Note that Eq.~\eqref{eq:MEcoups} does not uniquely specify the physical local fields and couplings (which are always ``inter''), allowing for a large arbitrariness in fixing the mapping $\mathcal{M}_{\mathrm{ME}}$ between the logical and physical problems. This arbitrariness is in fact useful, since the efficacy of an ME implementation can be very sensitive to the specific choice of the physical couplings.  

Fig.~\ref{fig:schematics} shows schematically how a logical-graph vertex of degree four can be minor embedded into two physical-graph vertices of degree three, and how two logical qubits are correspondingly replaced by four physical qubits.

Ideally, a minor embedding should give a faithful representation of the original logical problem, in the sense that it should encode the solution to the logical problem in its ground state. Intuitively, this can be achieved by implementing large penalty couplings, though it is clear that a balance must be struck so as not to overwhelm the problem couplings. Consequently the optimal choice for the energy penalties is not obvious. Very large penalties make logical groups behave as independent ferromagnetic clusters. These penalties can overwhelm the relatively weaker interactions between different clusters that represent the logical problem. When the logical groups are small, theoretical and experimental analyses suggest that the optimal value of the energy penalties is of the same order of magnitude as that of the logical couplings \cite{PAL:13,PAL:14}:\footnote{When long chains are used as logical groups, the optimal value for the energy penalties  seems to scale linearly with the chain length   \cite{Venturelli:2014nx}.}
\beq 
\label{eq:penalty}
J_{\tilde{i}_k \tilde{i}_l} \sim \tilde{J}_{\tilde{i}\tilde{j}}\ .
\eeq
We discuss how this freedom can be exploited in order to improve the performance of ME implementations in Section~\ref{sec:ES}.

An important consequence of the finiteness of the energy penalties within each logical group is that, informally, logical groups can ``break", i.e., the values of the physical qubits (measured in the $\sigma^z$ eigenbasis at $t_f$) representing a logical qubit can disagree. This may result in an unfaithful representation of the logical qubit by its logical group. We call logical groups with internal disagreement ``broken qubits'' (BQs). More formally, the logical qubit belongs to the code subspace defined by the two ``all-agree" states of the logical group, and any ``disagreement" is a non-code state (more on this in our discussion of QAC below). Because of Eq.~\eqref{eq:penalty} BQs are ubiquitous in every ME implementation. Classical post-processing (decoding) strategies that assign appropriate logical values to BQs are thus a necessary aspect of minor embedded quantum annealing. We discuss several such strategies in Section~\ref{sec:DS}.

\subsubsection{QAC: from the physical Hamiltonian to the encoded Hamiltonian}
\label{sec:QAC}

%%%%%%%
\begin{figure}[]
\begin{center}
\includegraphics[width=0.45\textwidth]{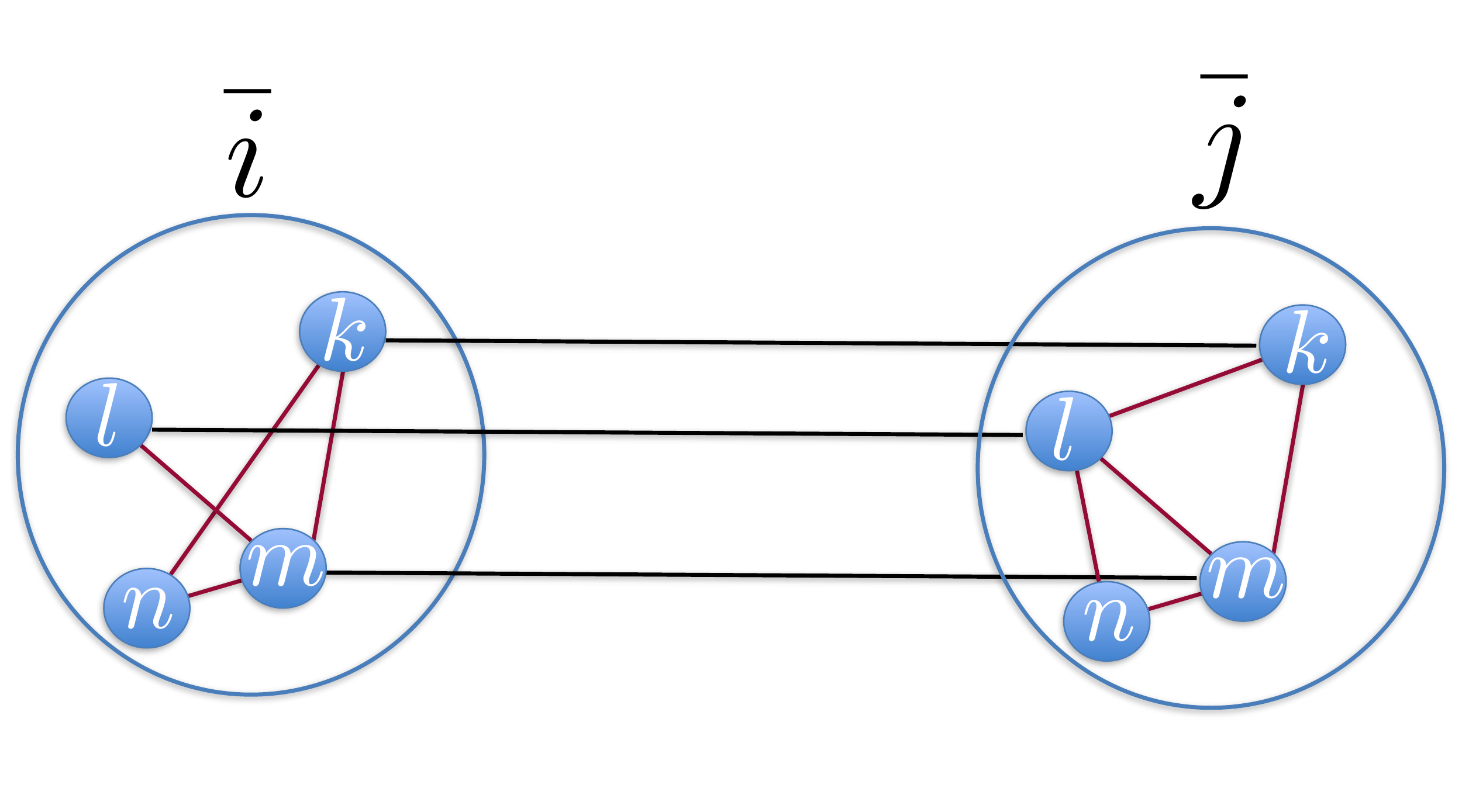} 
\caption{QAC: Schematic of two encoded groups ${\bar{i}},{\bar{j}}$ (large circles) representing two encoded qubits $\bar{\sigma}^z_{i}, \bar{\sigma}^z_{j}$. Physical qubits are denoted by blue circles; physical problem couplings are denoted by black lines, penalty couplings by red lines
%\red{TA:These are hard to distinguish.  Perhaps we can make the line more red rather than dark red.}. 
In the QAC case the problem couplings are an injection (one-to-one but not necessarily onto) between the physical qubits of each encoded group. As depicted, the penalty couplings may involve all or a subset of all physical qubits of each encoded group, and may vary from group to group.}
\label{fig:eqc}
\end{center}
\end{figure}
%%%%%%%

When the problem Hamiltonian $H_{\mathrm{P}}$ in Eq.~\eqref{eq:HP} needs to be encoded in order to provide protection against decoherence and noise we shall refer to the result as the ``encoded Hamiltonian" and write
\beq 
\label{eq:HP-QAC}
\bar{H}_{\mathrm{P}} = \sum_{\bar{i} \in \bar{\mc{V}}_h} h_{\bar{i}} \bar{\sigma}^z_{\bar{i}} + \sum_{({\bar{i}},{\bar{j}}) \in \bar{\mc{E}}_J} J_{{\bar{i}}{\bar{j}}}\bar{\sigma}^z_{\bar{i}}\bar{\sigma}^z_{\bar{j}}\ .
\eeq
Quantities decorated with a bar represent the encoded operators, acting on the $\bar{N}$ encoded qubits. The encoded Hamiltonian is defined on the vertices and edges of the encoded graph $\bar{G} = (\bar{\mc{V}}_{h},\bar{\mc{E}}_{J})$. Note that unlike the ME case, the same local fields and couplings are used in the encoded Hamiltonian as in the physical Hamiltonian (i.e., $h$ and $J$ do not have a bar), except that they are defined on the encoded graph. In particular, $h_{\bar{i}} = h_i$ and $J_{{\bar{i}}{\bar{j}}}=J_{ij}$. Given the physical Hamiltonian, the task is to map it to an encoded Hamiltonian that also has a direct embedding. This is the quantum annealing correction map
\beq
{H}_{\mathrm{P}} \overset{\mathcal{M}_{\mathrm{QAC}}}{\longmapsto} \bar{H}_{\mathrm{P}}\ .
\eeq
This map exists when $\bar{G}$ is a \emph{subgraph} of $G$. I.e., QAC requires several copies of the logical graph $\bar G$ to be directly embeddable into the physical graph $G$.

Each of the ${N} = |{\mc{V}}_{h}|$ vertices of ${G}$ is occupied by a classical binary ($\pm 1$) variable. The map $\mathcal{M}_{\mathrm{QAC}}$ replaces these by $\bar{N} = |\bar{\mc{V}}_{h}|$  encoded qubits.
Similarly to the ME case, we identify encoded groups as sets of physical qubits tied together by ferromagnetic penalty couplings, as depicted in Fig.~\ref{fig:eqc}. Unlike the ME case, we impose that the problem couplings between encoded groups are injective (one-to-one) and that all encoded groups are of identical size: $N_{\bar{i}} = K$ $\forall \bar{i} \in \{1,\dots,\bar{N}\}$. The penalty couplings may be non-uniform in order to allow for their optimization, e.g., as in Eq.~\eqref{eq:penalty}.

In the QAC case encoding acts to provide protection. To see how this works, we briefly digress to frame our discussion in a more general setting. Let us recall that one can define a quantum error correction code via a set of stabilizers, $\{ S_s\}$, (or parity checks in classical error correction language) \cite{Gottesman:1996fk,Lidar-Brun:book}. This set of commuting operators on $n$ qubits defines a collection of syndrome subspaces, each encoding a fixed number of encoded qubits $k$ with corresponding encoded operators $\{\bar{\sigma}^x_{\kappa},\bar{\sigma}^z_{\kappa}\}_{\kappa=1}^k$ (we temporarily drop the bar from the qubit subscripts; we will sometimes return to this notation if there is no danger of confusion). These subspaces are defined according to their eigenvalues with respect to the stabilizer operators. Since our focus is on encoding the final Hamiltonian $H_{\mathrm{P}}$ we only consider bit flip errors and thus restrict our attention to quantum repetition codes. 

An $[[n,k,d]]$ quantum repetition code (encoding $n$ physical qubits into $k$ encoded qubits and having distance\footnote{A code can detect any Pauli errors error $E$ having weight $\leq d-1$, i.e., $\bra{\psi_a}E\ket{\psi_b}= c\,\delta_{ab}$ for any pair of codewords $\ket{\psi_a},\ket{\psi_b}$ and constant $c$. Equivalently, the distance is the minimum weight of a Pauli operator $E$ such that $\bra{\psi_a}E\ket{\psi_b}\neq c\,\delta_{ab}$.} $d=n$ can be defined by the stabilizers $\sigma^z_i\sigma^z_j$ where $\{i,j\}$ can be chosen as all the nearest neighbor pairs. Such a code can correct $t=\lfloor d/2 \rfloor$ errors, and can be reliably decoded using majority voting if the number of errors does not exceed $t$. The decoding is undecided in the case of a tie, i.e., for $n/2$ errors ($n$ even) and fails if the number of errors exceeds $t$. 

For simplicity we consider the case of a single encoded qubit ($k=1$) per encoded group of $n$, and consider multiple encoded groups $\{\ell\}$. Such a code has the basis states 
\beq
\ket{\bar{0}} = \ket{0}^{\otimes n} \,\,;\,\, \ket{\bar{1}} = \ket{1}^{\otimes n}\ .
\eeq
The encoded operators are $\bar{\sigma}^x = ({\sigma}^x)^{\otimes n}$ and, e.g., $\bar{\sigma}^z=\sigma^z_1$.

The QAC protection idea is to exploit the various realizations of the encoded operators in order to provide an energy boost, and to include stabilizer terms in such a way that errors causing transitions outside of the code space are penalized. In particular, writing the unencoded final Hamiltonian [Eq.~\eqref{eq:HP}] in the form
\beq
H =\sum_\ell  h_\ell {\sigma}^z_\ell + \sum_{\ell,\ell'} J_{\ell\ell'}{\sigma}^z_\ell {\sigma}^z_{\ell'}\ ,
\eeq
the encoded final Hamiltonian becomes
\beq
\bar{H} = \sum_\ell h_\ell \sum_r [{\sigma}^z_{\ell}]_r + \sum_{\ell,\ell'} \sum_r J_{\ell\ell'} [{\sigma}^z_{\ell} {\sigma}^z_{\ell'}]_r \ ,
\eeq
where $[A]_r$ is the $r$th realization of the operator $A$. These multiple realizations exist since the logical Pauli operators are equivalent under multiplication by the code stabilizers, and for our repetition code any tensor product of an odd number of $\sigma^z$ operators is an equivalent realization of $\bar{\sigma}^z$. This multiplicity boosts the energy scale of $\bar{H}$ relative to $H$, and it increases the gap against thermal excitations. This is complemented by adding to $\bar{H}$ a  penalty term that is a sum over the available stabilizers $S_\alpha$ for each encoded qubit $\ell$: 
\beq
H_{\mathrm{pen}} = -J_{\mathrm{pen}}\sum_{\ell,s} [S_s]_\ell\ .
\eeq
Each bit-flip error anticommutes with at least one term in $H_{\mathrm{pen}}$ and hence incurs an energy cost of at least $2J_{\mathrm{pen}}$. Such errors are thus thermally suppressed \cite{jordan2006error,Young:13,Sarovar:2013kx,Ganti:13,Bookatz:2014uq,PAL:13,PAL:14,Young:2013fk}.

Returning now to our specific setting and previous notation, note that whereas in the ME case the $\tilde{\sigma}^z_{\tilde i}$ were abstract binary variables, in the QAC case we choose the encoded Pauli operators as the total Pauli operators of each encoded qubit, i.e., 
\beq
\bar{\sigma}^\alpha_{\bar{i}} = 
\sum_{k=1}^{K'} \sigma^\alpha_{\bar{i}_k}\ , \qquad K' \leq K ,
\label{eq:enc-pauli}
\eeq
where we allow for the possibility that not all $K$ physical qubits in each encoded group participate in defining the encoded Pauli operators (i.e., $K'<K$). In this manner the encoded problem Hamiltonian is expressed in terms of the encoded operators of the quantum repetition code. This means that $J_{ij} \sigma_i^z \sigma_j^z \mapsto J_{{\bar{i}}{\bar{j}}}\bar{\sigma}^z_{\bar{i}}\bar{\sigma}^z_{\bar{j}} = J_{{\bar{i}}{\bar{j}}} \sum_{k=1}^{K'} \sigma^z_{\bar{i}_k} \sigma^z_{\bar{j}_k}$, i.e., each coupling $J_{ij}$ is boosted by a factor of $K'$, since on the encoded graph each $J_{{\bar{i}}{\bar{j}}}$ is represented by $K'$ coupled pairs of physical qubits. 
We stress again that this boost is a useful feature since it increases the energy scale of the encoded problem Hamiltonian, a feature that works to suppress both thermal excitations and random control errors \cite{PAL:13,PAL:14,Young:2013fk}.

Broken encoded qubits appear in the QAC case just as BQs appear in the ME case, manifesting as encoded groups whose physical qubits yield different results when measured at $t_f$, resulting in a measured eigenvalue of  $\bar{\sigma}^\alpha_{\bar{i}}$ that is smaller than $K'$ in absolute value. Our QAC code can be viewed as a $[[K',1,K']]$ repetition code, i.e., a code with distance $K'$ encoding one qubit into $K'$. 
The penalty terms $J_{\bar{i}_k,\bar{i}_l}\sigma^z_{\bar{i}_k}\sigma^z_{\bar{i}_l}$ implicit in $\bar{H}_{\mathrm{P}}$ can be understood as elements of the stabilizers of the repetition code. Since these stabilizers detect bit-flip errors, as mentioned above these errors are energetically suppressed.

The transverse field is invariant under the transformation to the encoded $\sigma^x$ operator as in Eq.~\eqref{eq:enc-pauli}: 
\beq
H_X = \sum_{i=1}^N \sigma_i^x = \sum_{\bar{i}=1}^{\bar{N}} \sum_{k=1}^{K} \sigma_{\bar{i}_k}^x = \sum_{\bar{i}=1}^{\bar{N}} \bar{\sigma}^x_{\bar{i}} = \bar{H}_X \ .
\eeq
This observation is motivated by the fact that the D-Wave processors do not allow control over the transverse field.
However, we note that for the purposes of error correction this represents a significant limitation, since (as mentioned above) the encoded $\sigma^x$ operator of the repetition code is in fact $\bar{\sigma}^x = \otimes_k \sigma_k^x$, and the inability to implement this (many-body) operator compromises the efficacy of QAC \cite{PAL:13,PAL:14} as well as other strategies \cite{jordan2006error,Young:13}.

%Fig.~\ref{fig:schematics} (right) shows schematically how a physical problem qubit and its problem couplings can be represented by three encoded qubits, each of which is a copy of the problem qubit and its couplings. \mg{check}

\subsection{ME \textit{vs} QAC: frustration effects}

%%%%%%%%
\begin{figure}[t]
\begin{center}
{\includegraphics[width=0.45\textwidth]{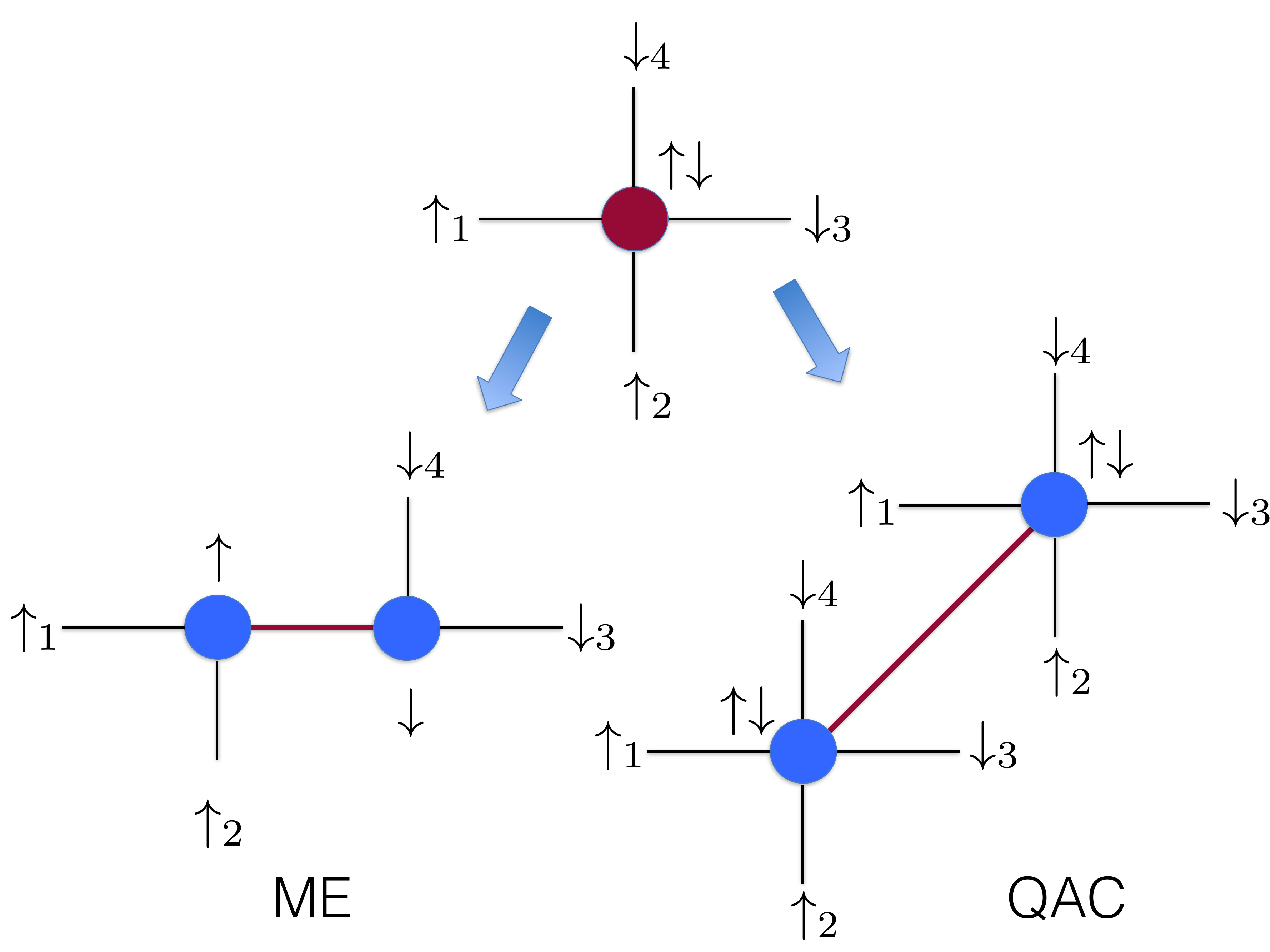}}
\caption{Contrasting ME and QAC via frustration effects. ME: The top configuration displays an example of a frustrated logical qubit (red) with ferromagnetic logical interactions (black). Frustration is a consequence of the opposite orientations of the outer logical qubits (represented as numbered arrows). Frustration of the logical interactions can be removed by the appearance of a broken qubit (lower left) after minor embedding. This configuration can be energetically favored if the penalty coupling (red line) is sufficiently weak. Thus, in the ME case, frustration may favor the appearance of broken qubits. QAC: Now the top configuration represents a frustrated physical qubit (red) with ferromagnetic couplings (black). In this example the QAC map replaces this configuration with two copies (bottom right), coupled with a ferromagnetic penalty (red). After the QAC map it is energetically favorable for the penalty-coupled qubits (blue) to be aligned for arbitrarily small values of the penalty. Thus, in the QAC case, frustration does not favor the appearance of broken qubits. Instead, configurations with broken qubits are excited states of the encoded problem for arbitrarily small values of the penalty terms.}
\label{fig:frustration}
\end{center}
\end{figure}
%%%%%%%%

QAC is known to be effective in improving the performance of a quantum annealer when compared to a direct (D) embedding of optimization problems \cite{PAL:13,PAL:14}. On the other hand, ME typically reduces performance relative to a direct implementation of a given optimization problem. This opposite response to ME and QAC has an intuitive explanation in how ME and QAC behave under frustration, which can be useful once the transverse field is off and the evolution is effectively classical [i.e., when $A(t)\ll kT < B(t)$, e.g., for $t/t_f \approx 0.8$ in Fig.~\ref{fig:Schedule}]. The problem Hamiltonian corresponding to a hard optimization problem typically represents a strongly frustrated spin system, i.e., wherein the ground state configuration cannot satisfy all logical couplings. An example of what happens when a frustrated logical qubit is minor embedded or QAC embedded is given in Fig.~\ref{fig:frustration}. In the ME case (where we assume that the logical problem cannot be directly embedded), if the penalty term is sufficiently weak, frustration results in a broken qubit, thus introducing new logical errors (to-be-decoded broken qubits). In the QAC case (where we assume that the problem can be directly embedded), in contrast, each copy of the problem bears the same frustration independently, and it is energetically favorable to satisfy the penalty coupling, no matter how weak it is. Therefore in QAC frustration is not the reason for the appearance of logical errors (whose source can instead be, e.g., thermal). Rather, thanks to their connection via energy penalties, the different copies of the problem help each other to relax more efficiently to the logical ground state. 
%\blue{[TA:We have to be careful here.  This is very much classical intuition.  It is not clear what is going on in the quantum evolution]}

\subsection{Quantum annealing correction on minor embeddings (QAC-ME) }
\label{sec:MEQA-frustr}

%%%%%%%%
\begin{figure}[t]
\begin{center}
\includegraphics[width=0.4\textwidth]{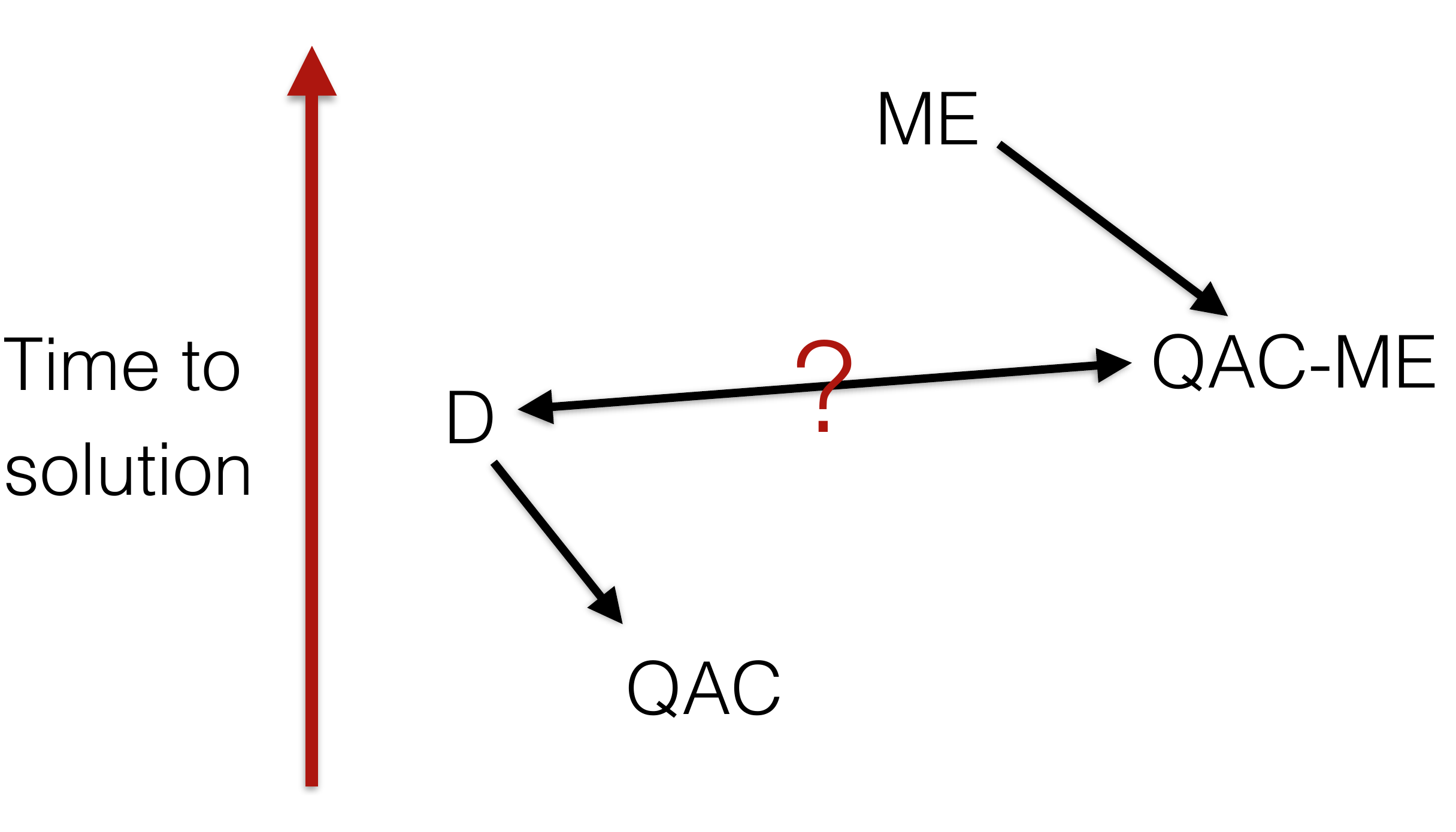}
%\vspace{-2cm}
\caption{Schematic comparing the performance of different embedding schemes. The time to solution (TTS) axis is not to be taken literally and represents wishful thinking. The goal of QAC is to reduce TTS (or increase the success probability) with respect to a D embedding, while QAC-ME is meant to improve over ME. Due to various tradeoffs it is not generally clear whether QAC-ME improves upon D. Of course, a D embedding may not be available, in which case QAC-ME is the only option.}
\label{fig:schematics-TTS}
\end{center}
\end{figure}
%%%%%%%%

It is natural to consider a concatenation of the ME and QAC maps:

\beq
\tilde{H}_{\mathrm{P}} \overset{\mathcal{M}_{\mathrm{ME}}}{\longmapsto} {H}_{\mathrm{P}} \overset{\mathcal{M}_{\mathrm{QAC}}}{\longmapsto} \bar{H}_{\mathrm{P}} \ .
\eeq
This is the QAC-ME map. It replaces a logical problem Hamiltonian by an encoded problem Hamiltonian, after a minor embedding of the former. 

The ME and QAC maps both expand the number of qubits. In the ME case a logical qubit is expanded into several physical qubits in order to enable an embedding of the logical problem Hamiltonian into a physical problem Hamiltonian that obeys the constraints of a given hardware graph. In the QAC case a physical qubit is expanded into several physical qubits in order to generate protection against decoherence and noise. 
%It is thus clear that QAC-ME uses more resources than strictly needed for the embedding of the logical problem. However, as suggested by Fig.~\ref{fig:schematics-TTS}, and as we shall demonstrate in detail using experimental data, this can be advantageous when a direct embedding of the logical problem is not possible.

Figure~\ref{fig:schematics-TTS} summarizes the various embeddings we have discussed so far. A problem formulated on a logical graph $\bar G$ can be directly embedded (D) on a physical device with given hardware graph $G$ if $\bar G$ is a subgraph of $G$. QAC aims to improve the success probability of solving this problem as compared to a D implementation. On the other hand, for problems formulated on a logical graph that can only be minor embedded, the goal of QAC-ME is to improve the success with respect to ME. It is typically not possible to directly compare D/QAC and ME/QAC-ME, simply because direct implementations of a given optimization problem are generically unavailable. It is however of great interest to consider cases where both D and ME/QAC-ME implementations are possible. This gives us an estimate of what performance decay is due to ME , and how much of it can be recovered by quantum annealing correction schemes such as QAC-ME. We perform such comparisons below.

%%%%%%%%%%%%%%%%%%%%%%%%%%%%%%%%%%%%%%%%%
\section{Choosing the penalties}
\label{sec:ES}
%%%%%%%%%%%%%%%%%%%%%%%%%%%%%%%%%%%%%%%%%
The penalty strengths should be optimized in order to maximize the performance of the embedding procedures. It is impractical to perform this optimization on an instance-by-instance basis, so it is important to find a general rule for determining the optimal choice of penalties. This is an open problem that we address below.
%%%%%%%%%%%%%%%%%%%%%%%%%%%%%%%%%%%%%%%%%
\subsection{Homogeneous penalties}
\label{sec:hom-pen}
%%%%%%%%%%%%%%%%%%%%%%%%%%%%%%%%%%%%%%%%%
Recent work has shown both experimentally and theoretically that even the simplest choice---a constant---for the penalties, 
\beq
\label{eq:pen-uni}
J_{\mathrm{pen}} = - \gamma\ ,
\eeq
allows for successful QAC \cite{PAL:13,PAL:14} or ME \cite{Venturelli:2014nx}. The optimal value of $\gamma$ is found to be roughly proportional to the value of the physical couplings that represent logical connections and typically grows with the size of the logical (or encoded) groups.
 
%%%%%%%%%%%%%%%%%%%%%%%%%%%%%%%%%%%%%%%%%
\subsection{Nonuniform penalties}
\label{sec:nonuni-pen}
%%%%%%%%%%%%%%%%%%%%%%%%%%%%%%%%%%%%%%%%%
The optimal choice of penalty strengths achieves a delicate balance of tying the cluster of physical qubits forming a logical group or encoded group together while not overwhelming the problem couplings. 
We can expect that a better balance can be achieved with a choice of penalties that depends on the local structure of the logical problem.  One possibility is to choose the energy penalties as follows:
\bes
\label{eq:pen-nonuni}
\begin{align}
&\textrm{ME: }\,\,\, \forall k,l\qquad J_{\tilde{i}_k,\tilde{i}_l} \equiv J_{\mathrm{pen},\tilde{i}} = -  \gamma\, \mathrm{mean}_{\tilde j }| \tilde J_{\tilde i,\tilde j}|\\
&\textrm{QAC: }\forall k,l\qquad J_{\bar{i}_k,\bar{i}_l} \equiv J_{\mathrm{pen},\bar{i}} = -  \gamma\, \mathrm{mean}_{\bar j }| J_{\bar i,\bar j}|\ ,
\end{align}
\ees
These expressions assign a value to the energy penalties associated with logical qubit $\tilde i$ (encoded qubit $\bar i$) that is proportional to the mean of the absolute strength of the logical couplings connected to that logical qubit (encoded qubit). The overall constant factor $\gamma$ is then optimized as in the uniform case for the best results. Typically, practical problems translate into Ising Hamiltonians with a certain amount of structure, both in the connectivity and in the value of the logical couplings. We thus expect this nonuniform choice to be generally more effective. In section~\ref{sec:ER} we present a study of the efficacy of the uniform and nonuniform encoding strategies described above using the DW2 device.

%%%%%%%%%%%%%%%%%%%%%%%%%%%%%%%%%%%%%%%%%
\section{Decoding Strategies}
\label{sec:DS}
%%%%%%%%%%%%%%%%%%%%%%%%%%%%%%%%%%%%%%%%%

%As well as an encoding strategy, minor embedded quantum annealing requires a 
A decoding strategy is a rule that assigns a value to each logical or encoded qubit given the values of the physical qubits retrieved from the annealer in the read-out process. As already mentioned in the general discussion about ME and QAC, in the decoding process we first distinguish between ``broken" and ``unbroken" logical or encoded qubits. The unbroken case corresponds to a logical or encoded group of physical qubits that all have the same measured final values. In this case, decoding is trivial and consists in assigning to the logical or encoded qubit the common value of the corresponding physical qubits. The broken case, on the other hand, corresponds to a logical or encoded group where there is at least one physical qubits whose value differs from the others. Decoding strictly applies to BQs, and it can be done in different ways. We will consider two general approaches (and a combination therefore). The first one is ``local", in the sense that the decoding assignment is performed on each logical or encoded group independently. It is the usual strategy for decoding classical repetition codes and relies on majority vote and coin tossing. The other can be considered a ``global" approach, in which the value of a logical qubit is decoded by a global energy-minimization procedure that simultaneously involves all BQs. 

%%%%%%%%%%%%%%%%%%%%%%%%%%%%%%%%%%%%%%%%%
\subsection{Coin tossing (CT)}
\label{sec:CT}
The simplest decoding strategy is random decoding (``coin tossing"), wherein we assign a random $\pm 1$ value to each logical or encoded qubit. This strategy merely serves as a baseline against which we compare all other strategies.
%%%%%%%%%%%%%%%%%%%%%%%%%%%%%%%%%%%%%%%%%

%%%%%%%%%%%%%%%%%%%%%%%%%%%%%%%%%%%%%%%%%
\subsection{Majority vote (MV)}
\label{sec:MV}
%%%%%%%%%%%%%%%%%%%%%%%%%%%%%%%%%%%%%%%%%
In the majority vote strategy the value of each encoded qubit is chosen as the most recurrent value within an encoded group:
\beq
\bar s_{\bar i} = {\rm sign} (\sum_{k} s_{\bar{i}_k})\,,
\eeq
where $\bar s$ and $s$ stand for the values of the encoded and physical qubits in the $\sigma^z$ basis at $t=t_f$, respectively.\footnote{From hereon, if there is no danger of confusion, we use the QAC terminology and notation to denote both QAC and ME.} 
MV decoding is most successful under the assumption of independence of the error probability $p_{\mathrm{err}}$ for each physical bit. A repetition code of length $n$ can be reliably decoded with up to $d=\lfloor n/2 \rfloor$ errors, an event that occurs with probability $p_{\mathrm{err}}^{d}$, which is exponentially decreasing in the code distance. An important caveat of applying MV to ME and QAC is that physical excitations above the ground state occurring during the evolution cannot be considered as generating independent errors on the physical qubits. MV itself thus does not ensure exponentially improving resiliency of the ME implementation as a function of the size of the encoded group. 
%\blue{[TA: While I totally agree with this, this is the first time we've really mentioned the difference between the classical problem and the quantum evolution.]}

In QAC schemes, where the encoded qubits are implemented as several equivalent physical copies, MV will be successful if errors in the different copies can be considered as independent. A quantum contribution in the QAC scheme arises from the energy penalties, which provide further suppression of errors. MV has indeed proven to be an effective approach to decoding QAC \cite{PAL:13,PAL:14}. 

\begin{figure}[t]
\begin{center}
\includegraphics[width=0.45\textwidth]{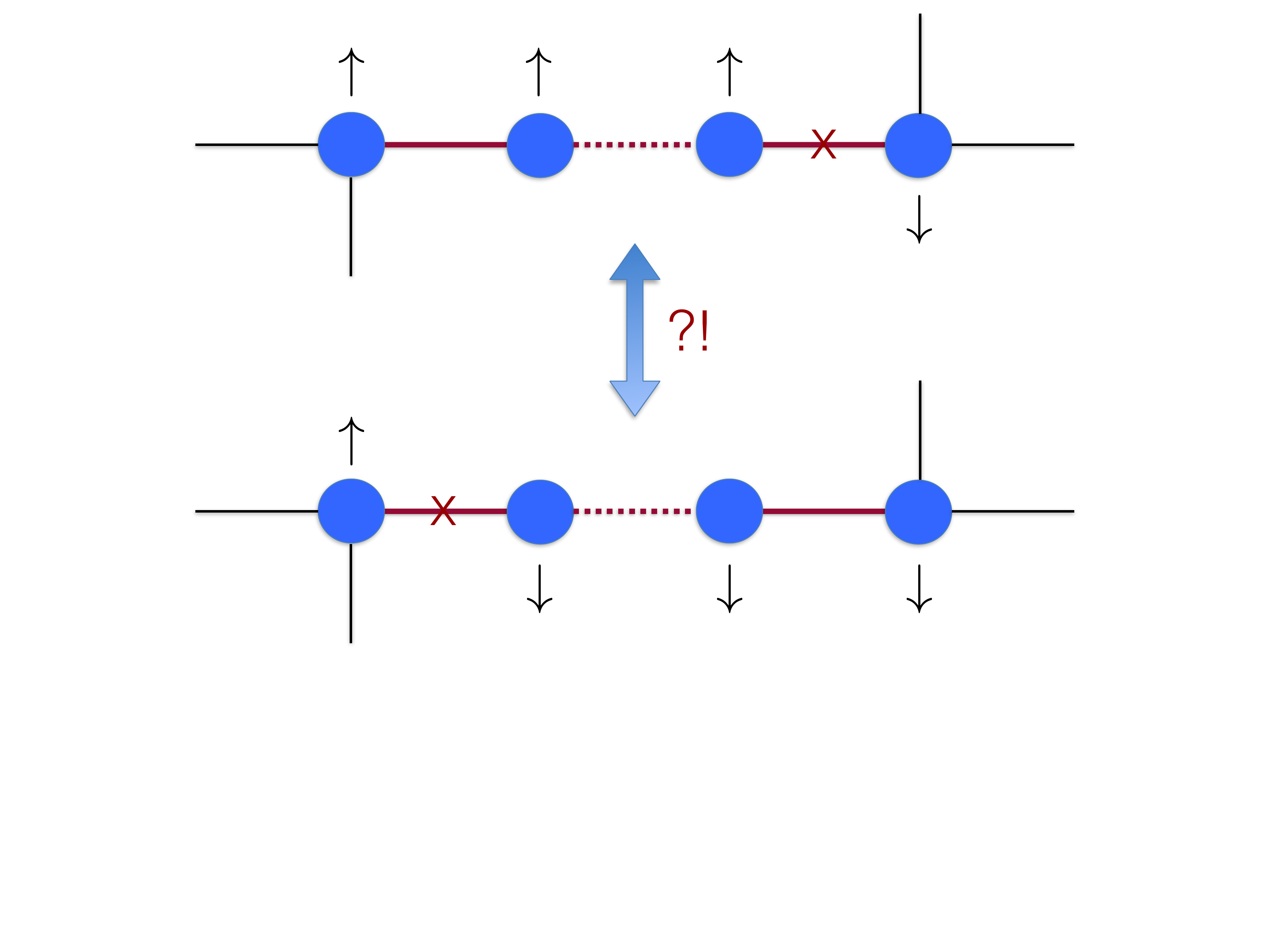} 
\vspace{-2cm}
\caption{Failure modes for majority vote in ME. A logical qubit is represented here by a physical chain. MV decoding on the bottom chain gives a spin-down result, on the top it gives spin-up. The chain can break iso-energetically in different places and decoding by majority vote, in this case, is likely to give no better results than coin tossing.}
\label{fig:broken_chain}
\end{center}
\end{figure}

However, MV is problematic for most ME schemes. In this case
%, the physical qubits used to represent a logical qubit are all independent, each contributing differently to the logical connectivity, and 
the absence of BQs is crucial for the correct representation of the logical problem as a minor embedded problem, since a logical qubit can break in several different ways, giving rise to different MV-decoded logical values. These different ``breaking modes" may have similar energy cost, making them similarly probable. A typical example of this is shown in Fig.~\ref{fig:broken_chain}, where a logical qubit is represented by a physical chain. The chain can break in several places, all of them costing the same energy, giving different results through MV. We should not expect, in this simple case, MV to be more effective than decoding by coin tossing. 

%%%%%%%%
%%%%%%%%

Another issue that arises for MV is how to decode a logical qubit comprising an even number $n$ of physical qubits. A tie ($n/2$ errors) cannot be decoded by MV, or by any ``local" decision rule other than coin tossing.

%%%%%%%%%%%%%%%%%%%%%%%%%%%%%%%%%%%%%%%%%
\subsection{Energy minimization (EM)}
\label{sec:EM}
%%%%%%%%%%%%%%%%%%%%%%%%%%%%%%%%%%%%%%%%%
Since we are interested in optimization problems, a natural alternative to MV is a ``global" decoding strategy wherein all BQs are decoded together in such a way as to minimize the energy of the encoded qubits. Decoding a particular state with this energy minimization strategy corresponds to solving an optimization problem on the subgraph of the encoded graph induced by the set of BQs, i.e., the qubits that require decoding. This problem is equivalent to minimizing the following ``decoding'' Hamiltonian
\beq
H_{\mathrm{D}} = \sum_{\bar i \in \mathrm{\mathrm{BQ}} }\left(h_{\bar i}  + \sum_{\bar j  \notin \mathrm{\mathrm{BQ}}} J_{\bar i\bar j}  \bar s_{\bar j} \right)  \bar \sigma^z_{\bar i} + \sum_{\bar i,\bar j \in \mathrm{\mathrm{BQ}}} J_{\bar i\bar j}\bar \sigma^z_{ \bar i} \bar \sigma^z_{ \bar j}\,.
\label{eq:DecHam}
\eeq
This energy minimization strategy is rather general and can be used to decode ME, QAC or QAC-MEC schemes. Moreover, it is applicable to cases where logical qubits comprise an even number of physical qubits, since it can also resolve ties. 

A disadvantage of EM is that, since it requires solving a global optimization problem, it can be computationally expensive. If the number of BQs is small (in a sense we quantify below), this optimization can be done exactly; otherwise, heuristic methods are necessary. We discuss EM decoding via simulated annealing, along with experimental results, in Appendix~\ref{app:EMD}. In the next section we argue that for a sufficiently small number of BQs, EM decoding can be implemented efficiently. 

%%%%%%%%%%%%%%%%%%%%%%%%%%%%%%%%%%%%%%%%%
\subsection{Efficient optimal decoding and the percolation threshold}
\label{sec:EOD}
%%%%%%%%%%%%%%%%%%%%%%%%%%%%%%%%%%%%%%%%%
We now give a general criterion to determine whether energy minimization decoding can be done efficiently. We use well-known results of percolation theory \cite{1957PCPS...53..629B} and rely on the simplifying assumption that an encoded qubit is broken with a qubit-independent probability $p_{\mathrm{\mathrm{BQ}}}$. 
This probability depends on the strength of the energy penalties used for the encoding. Eventually, the relevant value of $p_\mathrm{BQ}$ is obtained when it is evaluated at the optimal penalty value.

Let us start from two extreme cases. First, if $p_{\mathrm{\mathrm{BQ}}}\ll 1$ errors are sparse, and it is unlikely that two BQs are coupled. In this case the decoding effort grows only linearly with the size of the problem because each BQ can be optimized independently. In such a regime, the EM strategy can be implemented efficiently, since the decoding time scales linearly in the number of encoded qubits
\begin{equation}
T_{\mathrm{D}}(\bar N)  \sim  \bar N  \ .
\end{equation}
Second, if $p_{\mathrm{\mathrm{BQ}}}$ is close to $1$, most BQs are likely to be coupled. In this regime the EM decoding strategy requires one to solve an optimization problem that is as hard as the original problem:
\begin{equation}
T_{\mathrm{D}}(\bar N)  \sim   T_{\mathrm{P}}  ( \bar N)\,,
\end{equation}
where $T_{\mathrm{P}}$ is the time to solve the original problem.  In this regime, EM decoding cannot be implemented efficiently.

In general,  percolation theory can be used to establish a ``decoding threshold" $p_{\mathrm{D}}$, such that EM decoding can be implemented efficiently (in the large $\bar N$ limit) if and only if $p_{\mathrm{BQ}}< p_{\mathrm{D}}$ [this assumes that $p_{\mathrm{BQ}}$ is roughly constant across the graph, an assumption we test and confirm experimentally below (see Fig.~\ref{fig:BQ_density})]. It turns out that $p_{\mathrm{D}}$ is equal to the ``per-site" percolation threshold $p_{\mathrm{th}}$ of the encoded graph. To see this, we note that the decoding hamiltonian $H_{\mathrm{D}} $ is defined on a ``per-site" percolation subgraph, (i.e., the subgraph induced by vertices selected with some probability $p$).  A fundamental result of percolation theory is that (for $\bar N\gg 1$) there exists a threshold probability $p_{\mathrm{th}}$, determined by the local structure of the graph, above which an infinite (i.e., spanning) cluster is present with probability $1$. This result implies that EM decoding cannot be done efficiently if $p_{\mathrm{\mathrm{BQ}}} > p_{\mathrm{th}}$, since it requires minimizing $H_{\mathrm{D}} $ for a system of BQs that is as large as the original problem. 

%%%%%%
\begin{figure*}[ht]
\begin{center}
\subfigure[\ 2LG logical graph connectivity]{\includegraphics[width=0.32\textwidth]{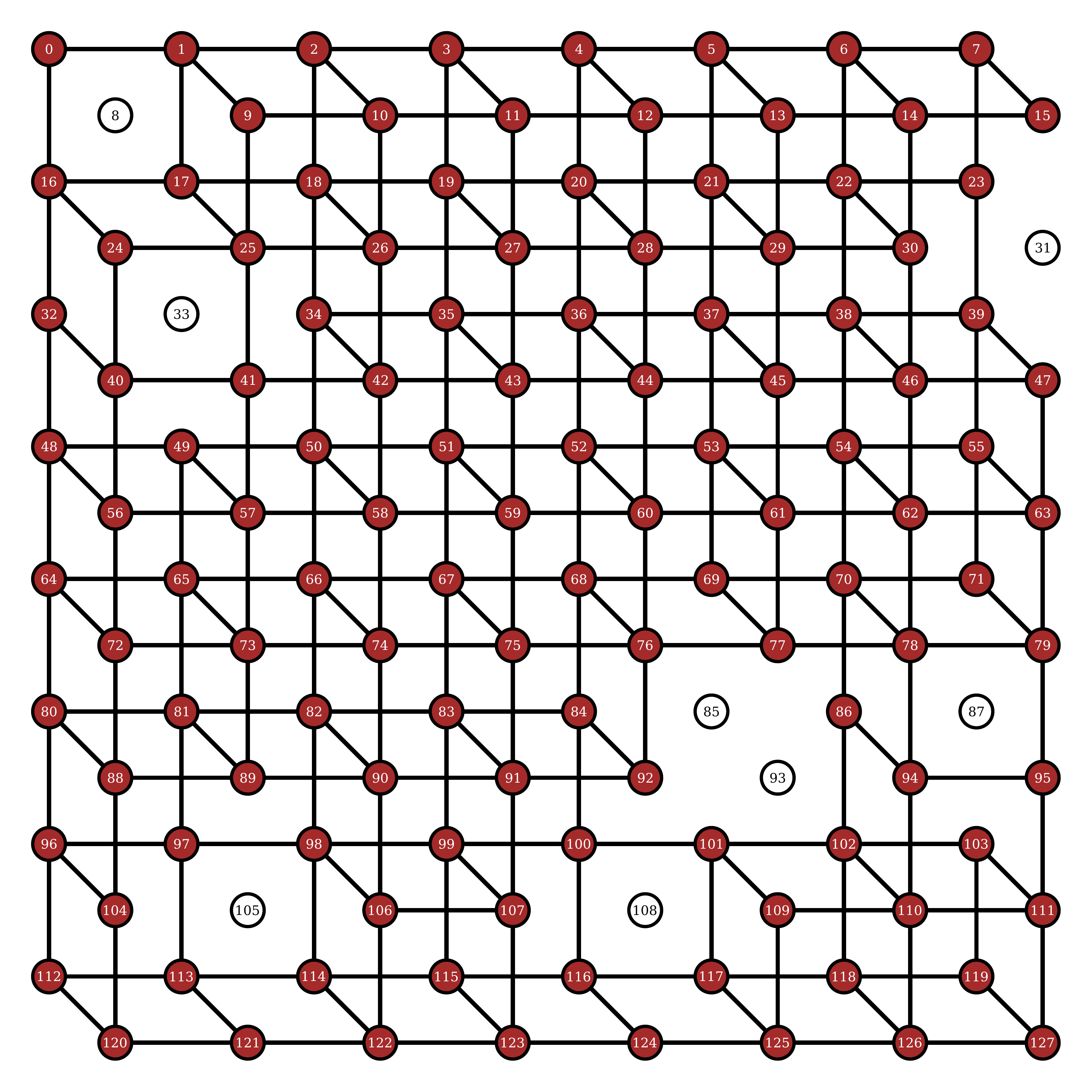}\label{fig:square-code-a}}
\subfigure[\ Minor embedding into Chimera graph]{\includegraphics[width=0.32\textwidth]{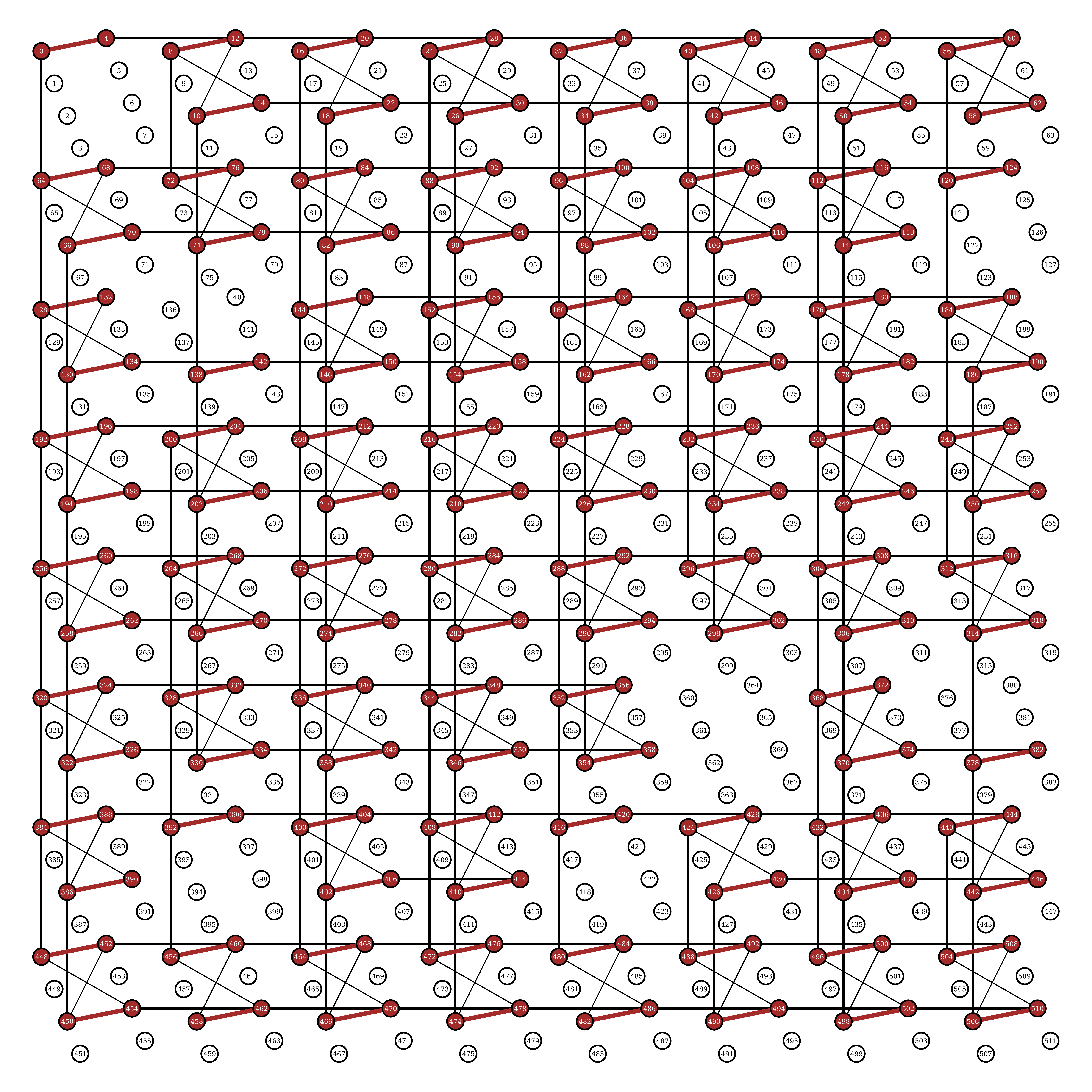}\label{fig:square-code-b}}
\subfigure[\ QAC-ME]{\includegraphics[width=0.32\textwidth]{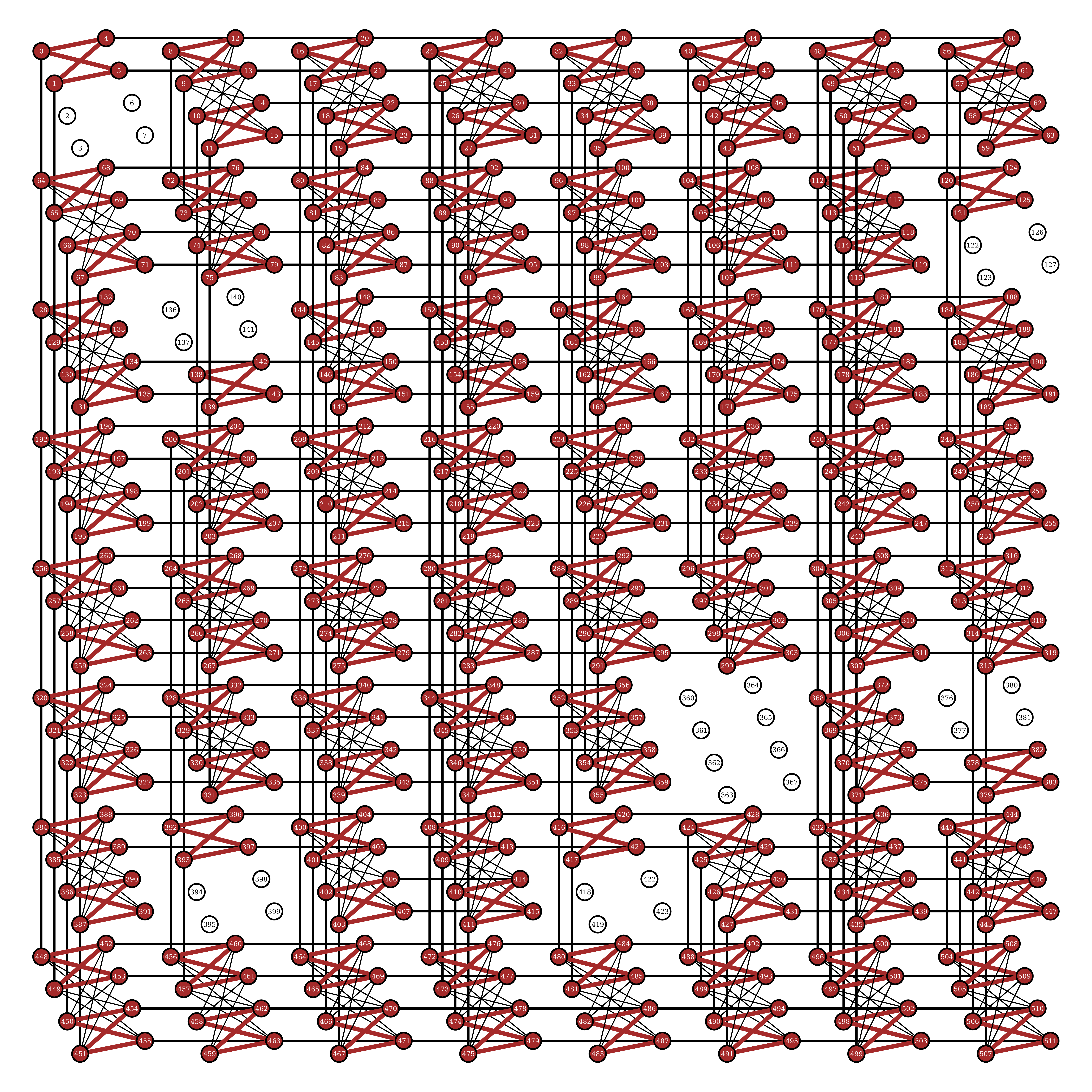}\label{fig:square-code-c}}
\caption{From a logical problem to a QAC-ME embedding. (a) Barahona's ``two-level-grid'' (2LG) graph, for which the Ising spin glass problem with couplings in $\{-1,0,1\}$ is NP-hard \cite{Barahona1982}. Disconnected vertices indicate spins with couplings set to zero, as well as  unused logical qubits in the logical DW2 Chimera graph. (b) Minor embedding of the 2LG graph into the physical DW2 Chimera graph. White circles correspond to unused or unusable qubits. (c) QAC-ME embedding of the 2LG problem using the ``square'' code.
In (b) and (c) penalties are represented by red (thick) couplings between groups of two (ME) and four (QAC-ME) physical qubits. Black (thin) links implement logical couplings. See also Fig.~\ref{fig:squareCodeLattice} for a more detailed description of the QAC-ME encoding.} 
\label{fig:square-code}
\end{center}
\end{figure*}
%%%%%%

On the other hand, below the threshold $p_{\mathrm{\mathrm{BQ}}} < p_{\mathrm{th}}$, the probability to have an infinitely large cluster vanishes. More precisely it can be shown that the probability that a given BQ is  connected to at least $\bar r$ other BQs decays exponentially with the size $\bar r$ of the connected domain \cite{Menshikov}. Assuming that the $\bar N$ encoded qubits grow their domains independently, we can thus express the total probability to have no domains of more than $\bar r$ connected encoded qubits as
\begin{equation}
P(\bar r)  =  (1- \alpha \, e^{-\gamma \bar r})^{\bar N} \approx 1-\alpha \, {\bar N}  e^{-\gamma \bar r}
\ , \quad (p_{\mathrm{\mathrm{BQ}}} < p_{\mathrm{th}})
\label{eq:P(bar-r)}
\end{equation}
for some constants $\alpha$ and $\gamma$ and for a sufficiently small exponential factor.  
Let us now fix the largest size of a decodable domain to be $\bar r$. If the encoded state contains a domain of larger size, we deem it ``undecodable," and we reject it. An undecodable event will happen with probability $1-P(\bar r)$. Let us again denote by $T_{\mathrm{P}}(\bar r)$ the time needed to optimize over a domain of size $\bar r$ (assuming that the decoding optimization is at most as hard as the original optimization problem). The total time required for decoding $T_{\mathrm{D}}(\bar N)$ is proportional to the typical number $\frac{\bar N}{\bar r}$ of independent domains of size $\bar{r}$ in the system, multiplied by the optimization time $T_{\mathrm{P}}(\bar r)$ of a region of size $\bar r$. Also, we note that the higher the probability of having  no domains of more than $r$ connected encoded qubits, the shorter is the total decoding time. Thus:
\begin{equation}
T_{\mathrm{D}}(\bar N)  \sim  \frac{\bar N}{\bar r} \, \frac{T_{\mathrm{P}}(\bar r)}{P(\bar r)} \ .
\end{equation}
Equation~\eqref{eq:P(bar-r)} shows that if we take $\bar r \sim \log (\bar N)$, $P(\bar  r)$ can be kept constant and close to 1 as $\bar N \rightarrow \infty$. Thus we have
\begin{equation}
T_{\mathrm{D}}(\bar N)  \sim   \frac{\bar  N}{\log(\bar N)} \, T_{\mathrm{P}}(\log(\bar N)) \ ,
\end{equation}
which is the main result of this section.  Below the percolation threshold, as long as we do not attempt to decode domains whose size scales faster than $\log(\bar{N})$, the time required for optimal decoding through energy minimization scales  sub-exponentially as compared to the time to find a solution to the original problem. In the large size limit, therefore, the decoding effort is negligible. 

We note that, in principle, the constant prefactor needed to have a small $P(\bar{r})$ can be very large and it diverges when we approach the percolation threshold.
% (equivalently, the size of the largest decodable domain $\bar r$ diverges). 
However, because the percolation transition is usually very sharp, we can expect the prefactor to be small if we are well below the threshold. 

We revisit the percolation threshold experimentally below, and show (see Fig.~\ref{fig:exptie}) that for the problem instances we considered in this work we are well below the threshold, so that efficient decoding is possible.

%%%%%%%%
\begin{figure*}[t]
\begin{center}
 \subfigure[\ ]{\includegraphics[width=0.45\textwidth]{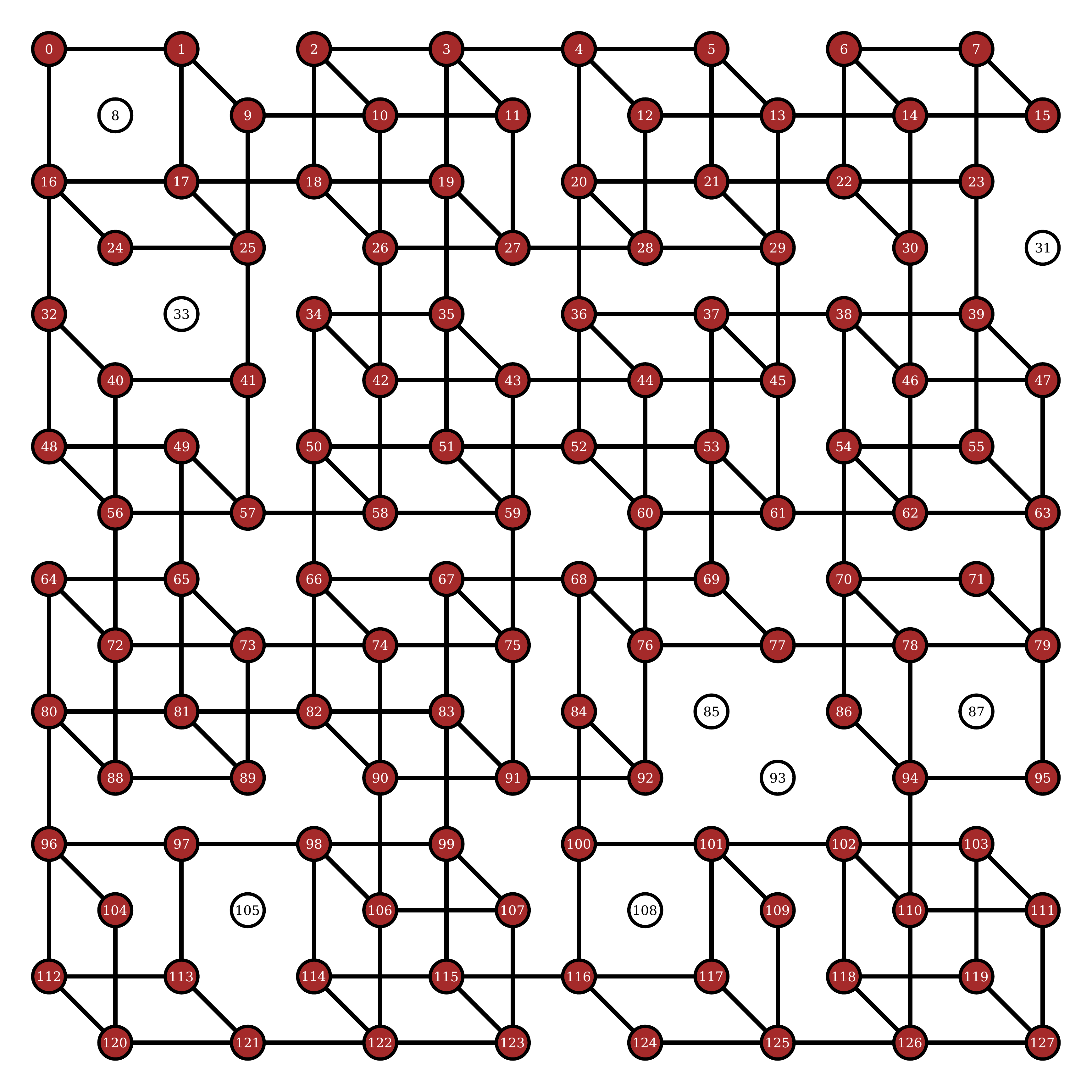}\label{fig:2LG-sub}} \quad\quad
 \subfigure[\ ]{\includegraphics[width=0.45\textwidth]{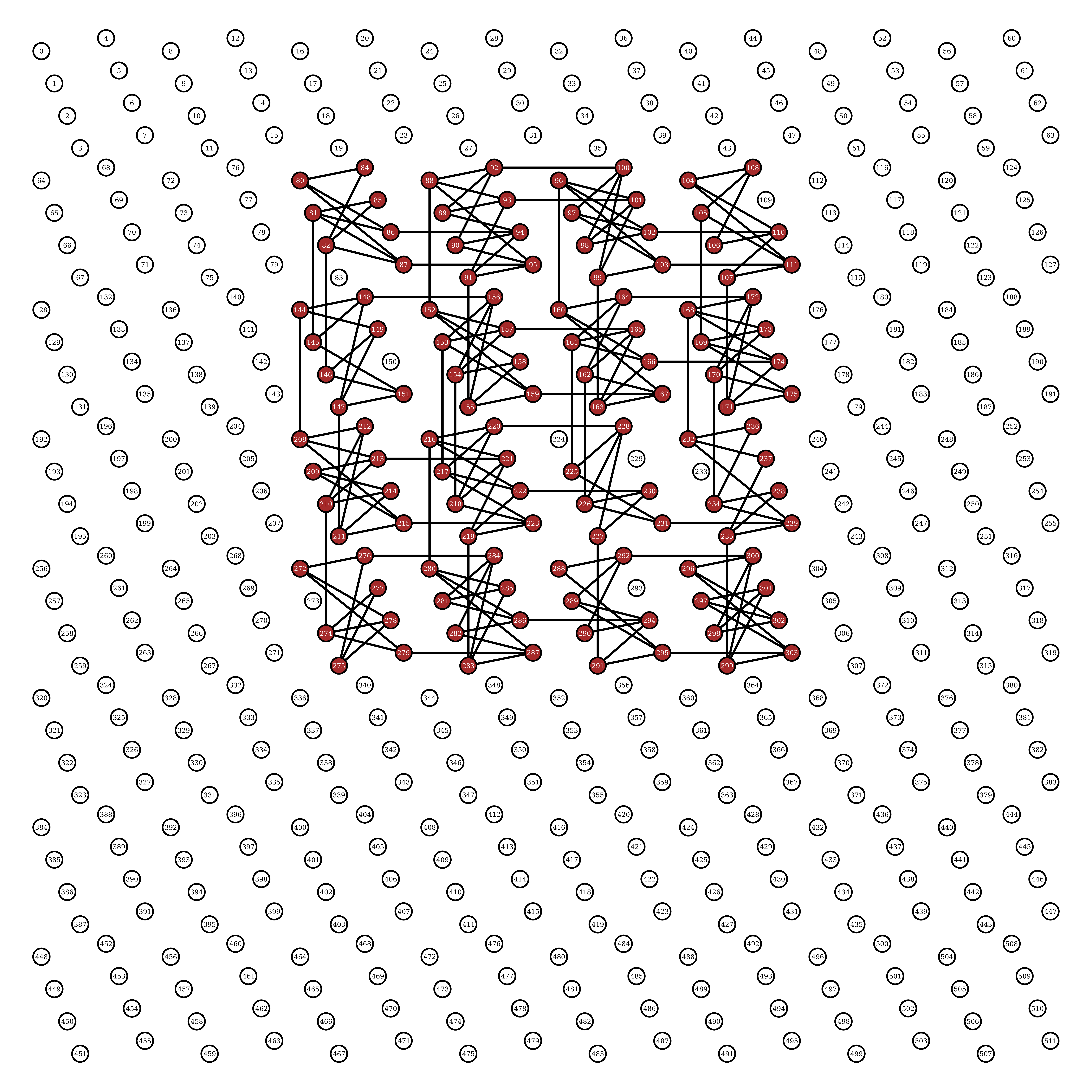}\label{fig:2LG-sub-direct}}
\caption{(a) The largest subgraph of the 2LG (encoded graph of the square code) that can be directly embedded into the USC-ISI DW2 Chimera graph. To see that this is a subgraph, note that some of the edges depicted in Fig.~\ref{fig:square-code-a} are missing, e.g., the edge between vertices $1$ and $2$. (b) Direct embedding of (a) into the Chimera graph. Each physical qubit in (b) corresponds to a logical qubit in (a).} 
\label{fig:subgraph}
\end{center}
\end{figure*}
%%%%%%%%

%%%%%%%%%%%%%%%%%%%%%%%%%%%%%%%%%%%%%%%%%
\section{QAC-ME of an NP-hard problem}
\label{sec:SC}
%%%%%%%%%%%%%%%%%%%%%%%%%%%%%%%%%%%%%%

In this section we describe an explicit QAC-ME scheme that combines a well-known NP-hard problem with a new QAC code.

\subsection{The two-level-grid (2LG) graph and its embedding}
Consider a logical graph $\tilde{G}$ consisting of two connected square lattices, i.e., a two-level-grid (2LG) as depicted in Fig.~\ref{fig:square-code-a}. The Ising spin glass problem on this graph is of interest, among other reasons, since it was perhaps the first shown to be NP-hard, for couplings in $\{-1,0,1\}$ and no local fields (Barahona's problem P3 \cite{Barahona1982}). This problem does not have a direct embedding in the Chimera graph, but it can be minor-embedded, as illustrated in Fig.~\ref{fig:square-code-b}.

The ME employs logical groups with pairs of physical qubits connected by a ferromagnetic coupling. Logical couplings are represented by one or two physical couplings per physical qubit. Note that fewer than half of the physical couplings are used in this ME. The remaining physical couplings can be used to implement a QAC-ME scheme as shown in Fig.~\ref{fig:square-code-c}, using a new QAC ``square'' code we describe below. 
%This code can also be understood as two MEs that are implemented in parallel and are further connected by two additional energy penalties. Each logical or encoded qubit is now represented by a group of four physical qubits, while the logical couplings are represented by two or eight physical couplings. 

%Square code construction
\begin{figure}[t]
\begin{center}
 \subfigure[\ Arrangement of physical qubits in the Chimera graph]{\includegraphics[width=0.22\textwidth]{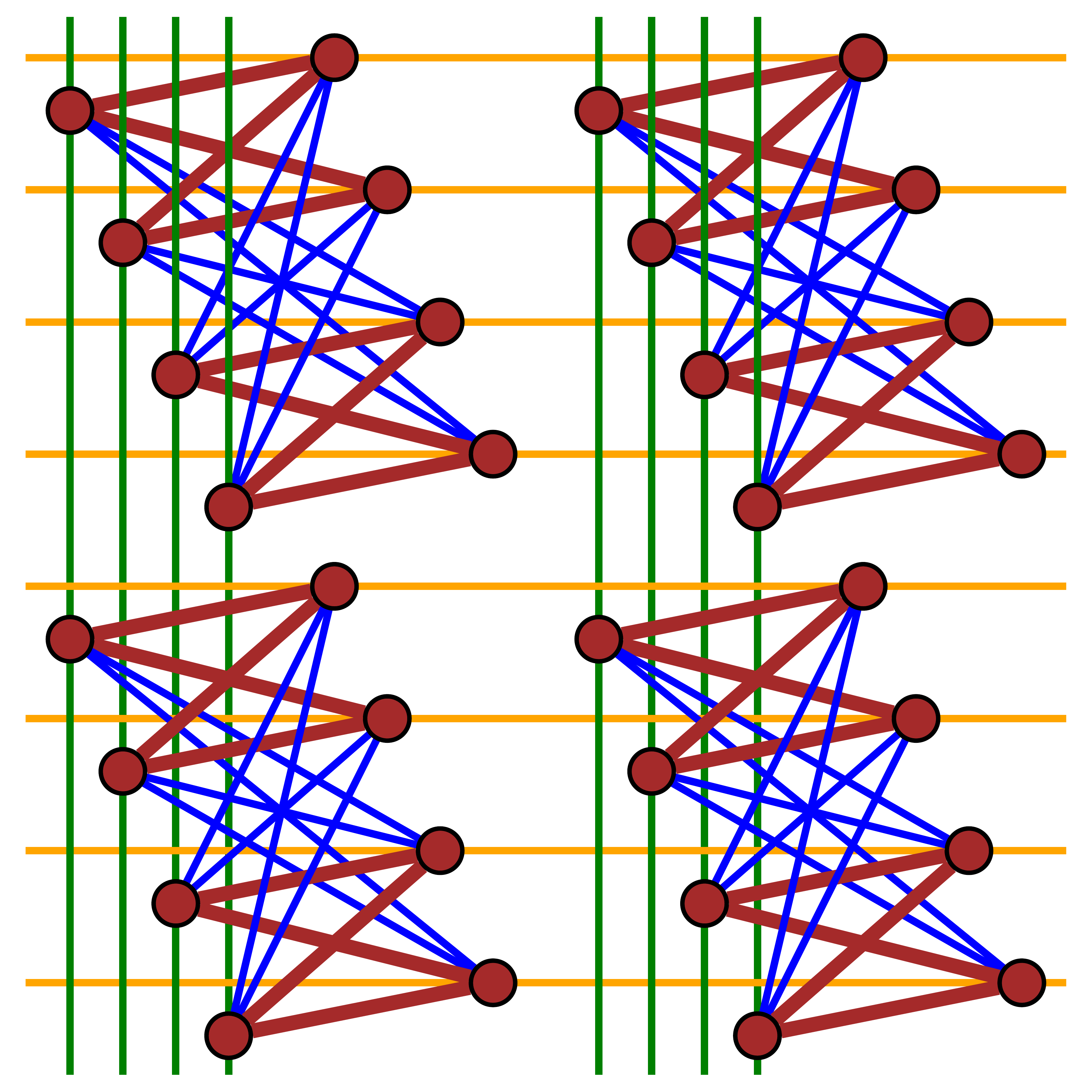}\label{l1encoding}}
 \quad
 \subfigure[\ Section of the corresponding encoded graph]{\includegraphics[width=0.17\textwidth]{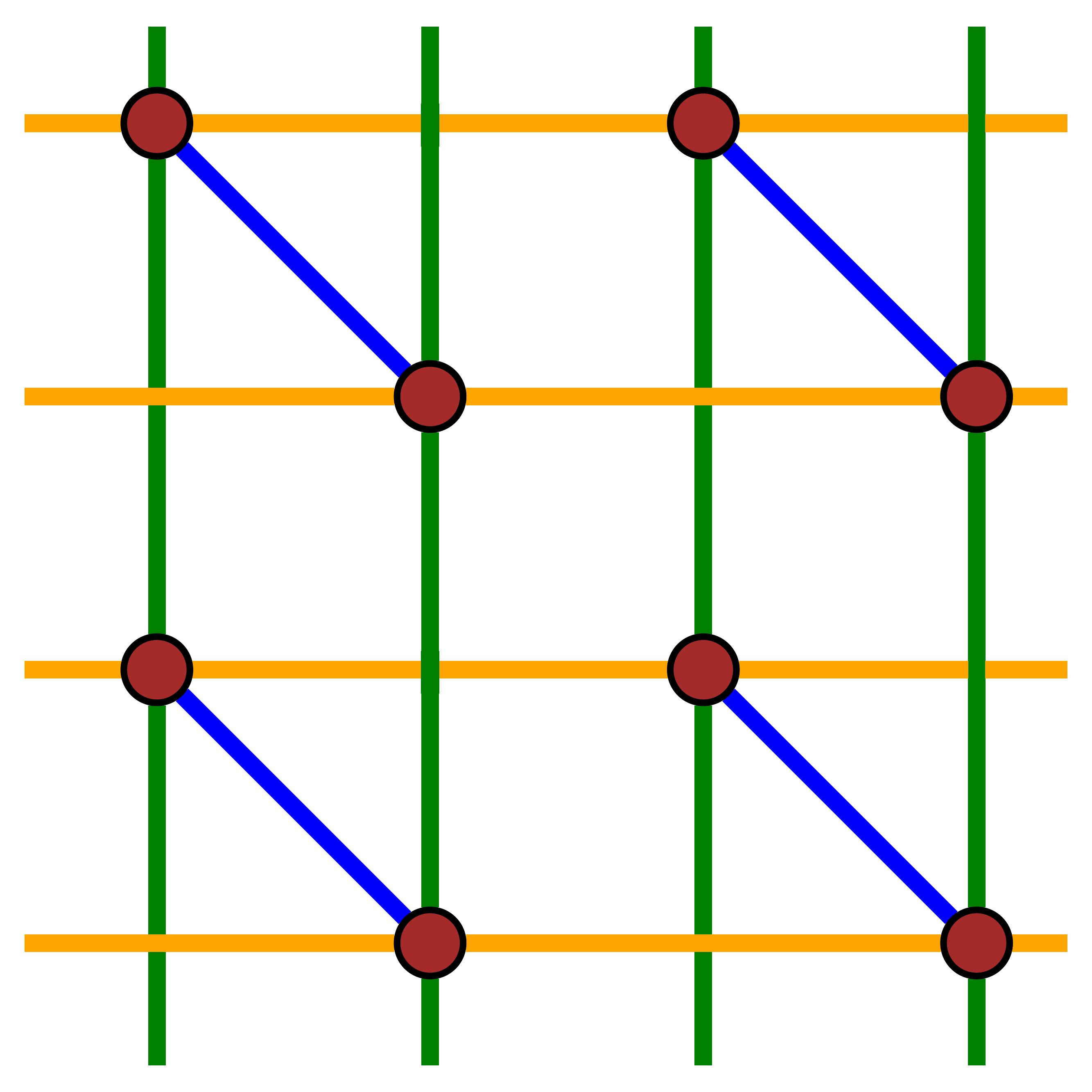}\label{fig:sqCode-latticeSection}}
  \caption{{ Construction of the ``square'' ${[[4,1,4]]}_{0}$ code.} Using the four physical qubits in the upper and the lower half of the Chimera unit cell, we construct two encoded qubits. In (a), the red (thick oblique) links represent the penalty couplings (stabilizers), while orange (horizontal), green (vertical) and blue (thin oblique) links form the logical Hamiltonian couplings. (b) The section of the encoded graph formed by (a).  This is also the connectivity induced at level $1$, where each link represents the collection of level $0$ representations in (a).}
  \label{fig:squareCodeLattice}
\end{center}
\end{figure}

\subsection{Direct embedding of the 2LG} 
Since we anticipate that minor embedding will perform poorly when compared  to a direct embedding, we are also interested in studying a direct embedding of the 2LG. An interesting set of such problems includes all instances that can be defined on the subgraph of the encoded graph of the square code. An example is shown in Fig.~\ref{fig:2LG-sub}, which depicts the largest 2LG subgraph directly embeddable into the USC-ISI DW2 Chimera graph. An example of a direct embedding of this subgraph is shown in Fig.~\ref{fig:2LG-sub-direct}.

\subsection{The square code and concatenation}
\label{sec:square-code}
We now describe the square code in detail. Using the notation ${[[n,k,d]]}_{p}$ to denote a distance $d$ code that uses $n$ physical data qubits and $p$ additional penalty qubits to encode $k$ logical qubits, this code can be understood as a ${[[4,1,4]]}_{0}$ code.\footnote{The code introduced and used in Refs.~\cite{PAL:13,PAL:14} is a ${[[3,1,3]]}_{1}$ code in this notation.}

The square code has a direct embedding into the Chimera graph, as shown in Fig.~\ref{l1encoding}. This yields a 2LG encoded graph, as shown in Fig.~\ref{fig:sqCode-latticeSection}, that can be extended as shown in Fig.~\ref{fig:square-code-a}, with extent bounded only by the size of the physical Chimera graph. 

\begin{figure}[t]
\centering
\includegraphics[width=0.4\textwidth]{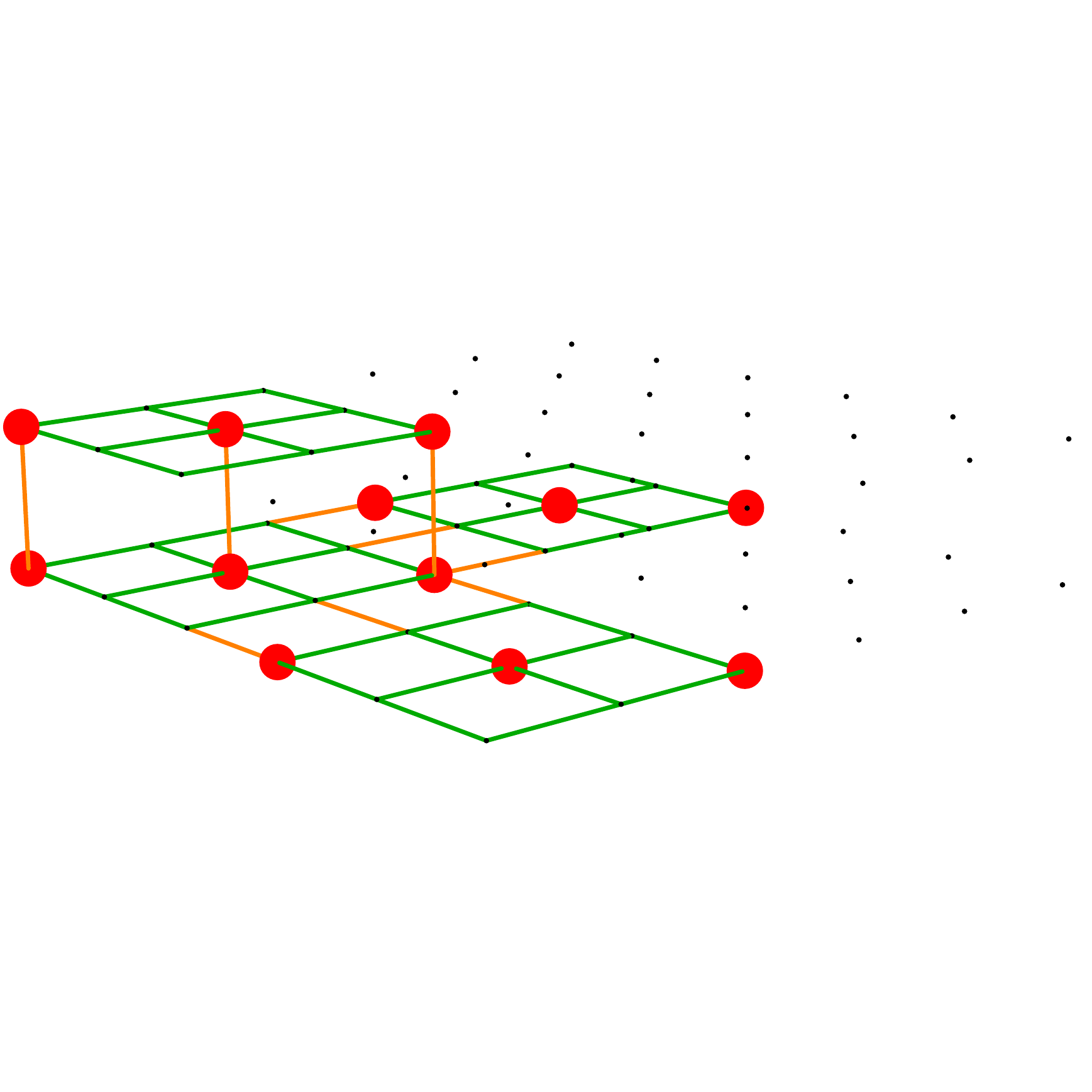}
\vspace{-2cm}
\caption{Level-2 encoding. Green lines represent level-$1$ couplings promoted to penalty couplings (stabilizers), while orange lines represent level-$1$ encoded $\sigma^z\otimes\sigma^z$ couplings ($2$ physical couplings). Each vertex represents a level-$1$ encoded qubit ($4$ physical qubits). Red dots represent realizations of level-$1$ $\sigma^z$ encoded operators. The encoding uses $n \times n$ level-$1$ encoded qubits, with the $n=3$ case depicted here.}
\label{SampleEncodedL1}
\end{figure}

As indicated by the pairs of {orange and green} lines connecting pairs of encoded qubits in Fig.~\ref{l1encoding}, each encoded qubit can couple to its neighbor via a $\sigma^z\otimes\sigma^z$ coupling in two distinct ways, resulting in a boost by a factor of $2$ at most. On the other hand, there are $8$ ways of realizing a {blue} logical $\sigma^z\otimes\sigma^z$ coupling, but, since we wish to maintain the relative strength between qubit-qubit coupling terms, we use only $2$ of these realizations, or alternatively we can use all $8$ but with $1/4$ the value of $J$ per physical coupling.  Similarly, there are $4$ ways of realizing a logical $\sigma^z$ operator by addressing each of the physical qubits comprising each encoded qubit, but we use only $2$ of these realizations, or alternatively we can use all $4$ but with half the value of $h$ per qubit. 
Thus
\beq
\bar{h}^{(1)}_\ell =  2 h_\ell\ , \qquad \bar{J}^{(1)}_{\ell,\ell'} =  2 J_{\ell\ell'}\ ,
\eeq
where we have interpreted the encoding as a level-$1$ code. This code induces an encoded 2LG graph of level-$1$ which we can exploit for concatenation. As a second layer of encoding, we use $n\times n$ encoded level-$1$ qubits to form an encoded level-$2$ qubit on each plane of the 2LG, as depicted in Fig.~\ref{SampleEncodedL1}. Note that with this encoding the 2LG graph structure of level-$1$ is maintained, and that an encoded level-$2$ qubit comprises $4n^2$ physical qubits.

The level-$2$ encoding can correct $t=2n^2-1$ errors, i.e., has distance $4 n^2$ as expected from the code concatenation, and will be undecided if there are $2n^2$ errors. To see why this is so, consider the level-$2$ code as a whole, and interpret it as a distance $d=4n^2$ qubit repetition code. A majority vote on the level-$2$ encoded 2LG graph can give the wrong result if more than half of the qubits are flipped, i.e., it can recover from $2 n^2-1$ flips, it is undecided for $2n^2$ errors, and will give the wrong answer for more than $2n^2$ errors. The level-$2$ code is thus a $[[4n^2,1,4n^2]]_0$ code. 

We note that an additional benefit of the encoding is the ability to fine-tune different couplings or local fields in the problem Hamiltonian. For example, suppose $J_{ij}$ and $h_i$ can only take values $\{-1,0,1\}$. How can we generate general problems where some couplings have fractional values? To do this we use the fact that at the level-$2$ encoding each coupling or local field has $2n$ different realizations. It is now straightforward to rescale all couplings with respect to the largest coupling. E.g., suppose a given coupling $J$ is $9/10$ of the largest coupling $J_{\max}$, then with a sufficiently large $n$, one can turn off $1/10$ of the realizations of $J$ and thus achieve the desired $J/J_{\max}$ ratio. The larger $n$ is, the greater the accuracy of this ratio. If the quantum annealing device has other possible values of $J_{ij}$ and $h_i$, then this reduces the necessary value of $n$ for a desired ratio.

It is clear that the square code can be concatenated further. The resulting concatenated code can be made to have arbitrarily large distance, i.e., is capable of dealing with increasingly more errors, at the cost of more physical qubits per encoded qubit. It also exhibits a potentially beneficial growth in the encoded local fields and couplings. We describe this in Appendix~\ref{app:concat}.

Decoding of the square code relies on the techniques we discussed in Sec.~\ref{sec:DS}. In Appendix~\ref{app:recursive} we propose an alternative, recursive decoding scheme.

%%%%%%%%%%%%%%%%%%%%%%%%%%%%%%%%%%%%%%%%%
\section{Experimental Results}
\label{sec:ER}
%%%%%%%%%%%%%%%%%%%%%%%%%%%%%%%%%%%%%%%%%

\subsection{Frustrated-loop Hamiltonians with planted solutions}
We now restrict our attention to a specific set of random Ising problem Hamiltonians with ``planted" ground states \cite{PhysRevLett.88.188701,PhysRevLett.102.238701}, which were recently introduced in the context of quantum annealing and studied in detail using the DW2 device \cite{Hen:2015rt}, albeit without ME and QAC considerations. We briefly review how these problem Hamiltonians are constructed; for a detailed explanation see Ref.~\cite{Hen:2015rt}. 

By ``planting" we mean that the planted state is, by construction, a ground state of the problem Hamiltonian. The planted state need not be the unique ground state (in fact it is typically degenerate), but knowing it means that we also know the ground state energy. This advance knowledge circumvents the need to resort to computationally expensive exact solvers that would otherwise be required to determine such properties of the ground state. 
Without loss of generality we can always choose to plant the all-zero state:
\beq
\ket{\Psi_{0}} =  \ket{0_1,\dots,0_{\bar N} }\ .
\eeq 

We generate random loops (cycles over the Chimera graph) with all ferromagnetic couplings, except one randomly chosen anti-ferromagnetic coupling. In this manner each loop is a frustrated Ising problem whose ground state is the all-zero state of the qubits covered by the loop. A problem instance is then built as a sum of such randomly generated frustrated loops:\footnote{In somewhat confusing terminology, this Hamiltonian is ``frustration-free'' in the sense that the ground state of the total Hamiltonian is also the ground state of each (frustrated) term in the sum \cite{Bravyi:2009sp}.} 
\beq
H_{\mathrm{P}} = \sum_{i}^L H_i^{\mathrm{loop}}\ ,
\label{eq:planting}
\eeq 

Similarly to the case of constrained satisfiability problems (SAT), the hardness of a problem can be tuned by varying the number of terms in the sum  of Eq.~\eqref{eq:planting} (i.e., the number of clauses in the terminology of SAT). Equivalently, hardness depends on a clause density $\alpha$ defined as the ratio between the number of loops (clauses) $L$ and the total number of qubits:
\beq
\alpha = \frac{L}{\bar N}\,.
\eeq
This turns out to generate problems with an easy-hard-easy structure as a function of the  clause density. We loosely refer to the hardness peak as occurring at the critical clause density $\alpha_c$.\footnote{We use the term ``critical clause density'' loosely here, in the sense that we do \emph{not} mean to imply the presence of a phase transition, but merely to signal the empirically observed hardness peak.} Typically, frustration and hardness peak both at the critical clause density; see Appendix~\ref{sec:rang&frust} for an estimate of frustration. 

While the length of each loop is arbitrary, here we only consider loops of length $4$ and $6$. Note that such short loops were not considered in Ref.~\cite{Hen:2015rt} as they generated instances that were too hard (and thus resulted in insufficient solutions to be statistically meaningful) with a direct embedding on the full (then $503$-qubit) Chimera graph. Instead Ref.~\cite{Hen:2015rt} used loops of length $\geq 8$. Note further that recent work~\cite{Katzgraber:2015gf} has generated significantly harder instances for the Chimera graph, and we use the current method primarily due to its aforementioned convenient feature of advance knowledge of the ground state energy. Moreover, while the construction we have described obviously generates only a subset of the NP-hard spin glass problem instances that can be defined over the 2LG, it features the important advantage that the clause density is a tunable parameter that allows us to probe the quantum annealer on instances of varying and controllable hardness.

\begin{figure*}[t]
\begin{center}
\subfigure[\ Planted (full 2LG graph)]{\includegraphics[width=0.45\textwidth]{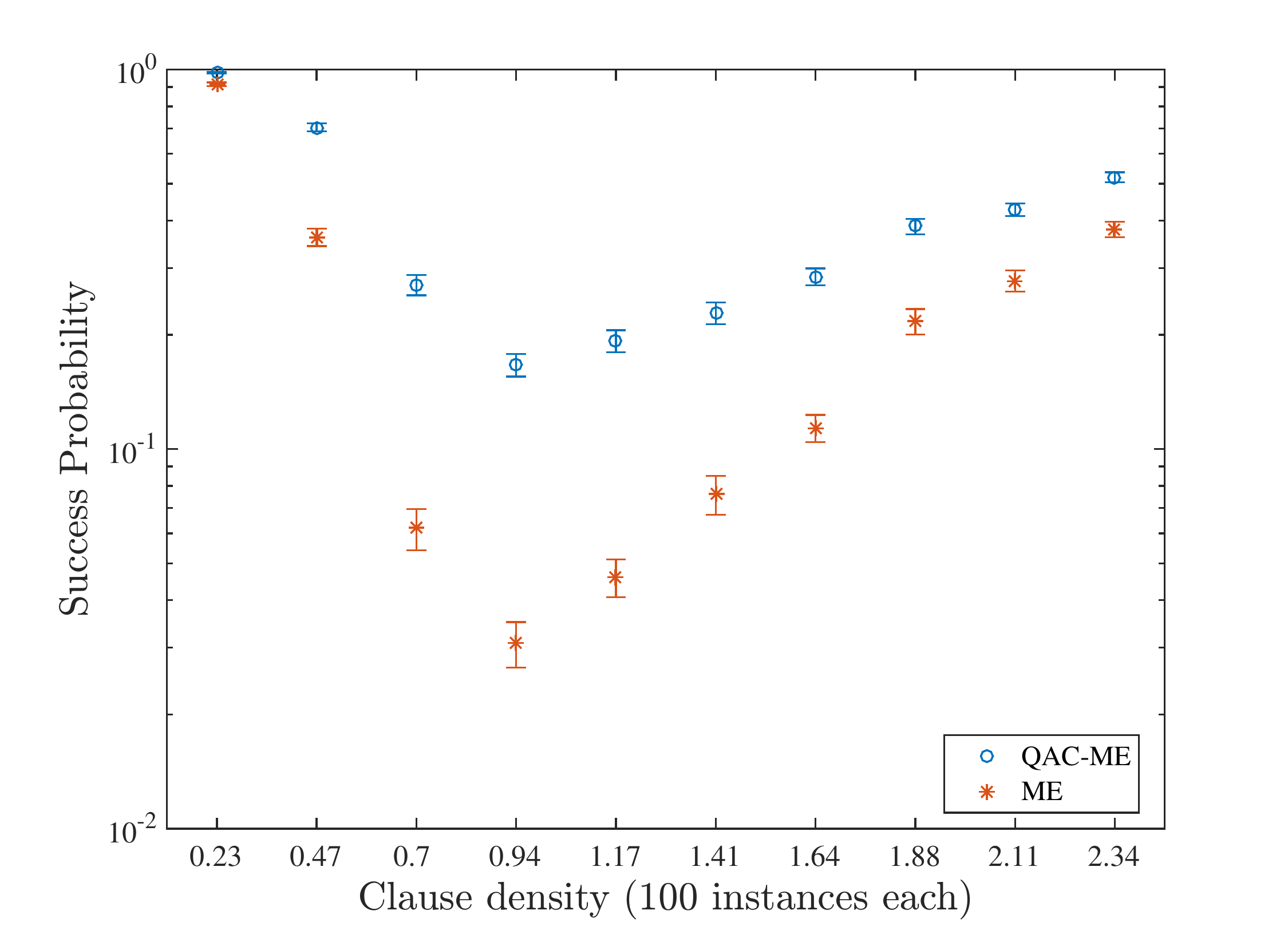}\label{fig:QAC-ME_best-Planted}}
\subfigure[\ Embeddable planted (embeddable 2LG subgraph)]{\includegraphics[width=0.45\textwidth]{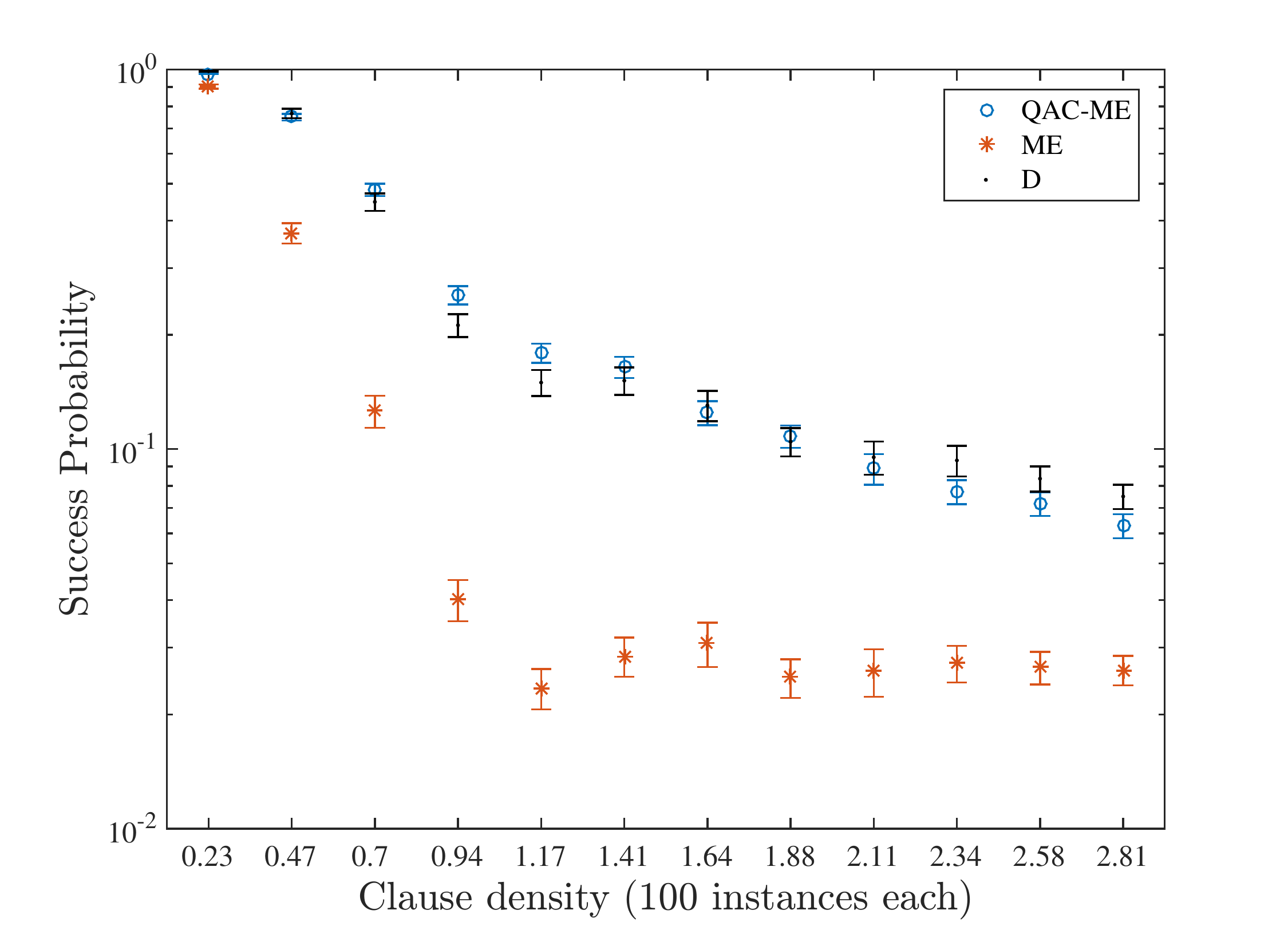}\label{fig:QAC-ME_best-EmbeddablePlanted}}
\caption{Mean success probabilities [after the renormalization of Eq.~\eqref{eq:Pcorrection}] of instances generated by planting solutions on (a) the square code encoded graph, (b) on the embeddable subgraph. Decoding used the strategy of energy minimization. In (a) the comparison shown is between the ME of Fig.~\ref{fig:square-code-b} and the QAC-ME of Fig.~\ref{fig:square-code-c}. QAC-ME is seen to improve the ME performance by almost an order of magnitude, with the largest improvement obtained at the critical clause density. In (b) the comparison shown is between D of Fig.~\ref{fig:2LG-sub-direct}, ME obtained by restricting Fig.~\ref{fig:square-code-b} to the directly embeddable subgraph, and QAC-ME obtained by similarly restricting Fig.~\ref{fig:square-code-c}. It is seen that the performance loss of ME with respect to D is completely recovered by QAC-ME. Errors bars denote $1\sigma$ confidence intervals here and in all subsequent plots.} 
\label{fig:planted}
\end{center}
\end{figure*}

\subsection{Experimental methods}
\label{sec:ER-method}

Our experiments were performed on the DW2 quantum annealing processor installed at the University of Southern California Information Sciences Institute (USC-ISI), which has been described in detail before (see, e.g., Ref.~\cite{speedup}). The D-Wave devices have been designed in such a way that complete logical graphs $K_n$ of $n$ vertices can be minor embedded into the Chimera graph representing the physical connectivity [see Fig.~\ref{fig:Vesuvius}] \cite{Choi1,Choi2}. The largest complete graph that can be minor embedded into the DW2 hardware graph is $K_{32}$. 

Apart from specifying the problem Hamiltonian [Eq.~\eqref{eq:HP}], the user interface enables us to choose the annealing time $t_a$. In all our experiments, the annealing time was set to the smallest value allowed by the hardware, $t_a = 20\mu $s. For each problem instance we ran $10^4$ annealing cycles. This was done by performing $10$ programming cycles of $10^3$ runs each. For each programming cycle in which the D-Wave device is programmed with a given problem Hamiltonian, intrinsic control errors (ICE) prevent the physical couplings to be set with a precision better than $\sim 5\%$ of the maximum allowed values $|h_{\max}|,|J_{\max}|$. This error includes both random (low frequency) noise and systematic errors (flux biases). Multiple programming cycles help to average over errors in setting the intended coupling. As an additional precaution designed to average out as much systematic error as possible, we performed random gauge transformations between each programming cycle. A gauge transformation is an invariance transformation of the Ising Hamiltonian Eq.~\eqref{eq:HP} that is achieved by flipping the value of qubit $i$ and by appropriately changing the sign of the couplings:
\beq
  \sigma_i^z \rightarrow - \sigma_i^z \quad \Rightarrow \quad J_{ij} \rightarrow  - J_{ij}  \quad h_{i} \rightarrow  - h_{i}, \quad \forall j \ .
\eeq

We generated $100$ random planted instances per clause density. Below we plot, for each $\alpha$, the mean success probability and its standard error (one $\sigma$ deviation of the mean) for the optimal value of the penalties. Penalties were chosen and optimized according to the uniform strategy described in Sec.~\ref{sec:hom-pen}. The penalties were optimized separately for each $\alpha$. Decoding used the strategy of energy minimization over the broken qubits (we discuss encoding and decoding strategies in more detail in Sec.~\ref{sec:expDS}). 

We stress that all comparisons between the various strategies (i.e., ME, QAC-ME and D) must properly account for the physical resources required by each strategy. Specifically, ME requires twice as many physical qubits as D, while the level-$1$ square code (the only case we tested in this work) implementation of QAC-ME requires four times as many. This implies that one can run two ME or four D embeddings in parallel while consuming the same hardware resources (qubits) as needed for a single QAC-ME embedding. To account for this parallelism we renormalize and report the experimental D and ME ground state (``success'') probabilities as follows:
\bes
\label{eq:Pcorrection}
\begin{align}
P_{\mathrm{D}} &\mapsto 1 - (1-P_{\mathrm{D}})^4 \\
P_{\mathrm{ME}} &\mapsto 1 - (1-P_{\mathrm{ME}})^2\ ,
%&& P_{QAC-ME} \rightarrow  P_{QAC-ME}\,. 
\end{align}
\ees
which is the probability of finding the ground state at least once after four (D) or two (ME) parallel attempts, using the success probability from a single attempt (see  Appendix~\ref{app:data} for a more detailed description of how the quantities above are computed from the raw data).

\subsection{ME \textit{vs} QAC-ME for the 2LG graph}
Figure~\ref{fig:QAC-ME_best-Planted} shows our experimental results for instances with planted solutions as described above and constructed on the entire 2LG graph, i.e., the encoded graph of the square code shown in Fig.~\ref{fig:square-code-a}. Fig.~\ref{fig:QAC-ME_best-Planted} compares the ME [corresponding to Fig.~\ref{fig:square-code-b}] and QAC-ME [corresponding to Fig.~\ref{fig:square-code-c}] implementations on the same set of instances.  It shows that QAC-ME  is effective in boosting the performance of the quantum annealer on the entire range of clause densities. Interestingly, the harder the problems, the larger is the advantage of QAC-ME over ME. We empirically identify the critical clause density as $\alpha_{\mathrm{crit}} \approx 0.94$, where the success probability has a minimum. At this value QAC-ME yields success probabilities that are about one order of magnitude higher than ME's.  
Thus error correction becomes more effective for harder problems, as was also observed in the context of random Ising problems in Ref.~\cite{PAL:14}.

%%%%%%%%%%%%%%%%%%%%%%%%%%%%%%%%%%%%%%%%%
\subsection{D \textit{vs} ME \textit{vs} QAC-ME}
%%%%%%%%%%%%%%%%%%%%%%%%%%%%%%%%%%%%%%%%%
%{\bf embeddable planted.}  
We next compare between D and ME by constructing instances with planted solutions on the 2LG subgraph shown in Fig.~\ref{fig:2LG-sub}. Experimental results are shown in Fig.~\ref{fig:QAC-ME_best-EmbeddablePlanted}. As expected, the ME strategy suffers a steady decay in performance compared to D, until 
%the critical clause density $\alpha_{\mathrm{crit}} \approx 1.17$ is reached (the numerical value depends on the graph).
$\alpha \approx 1.17$ is reached. At this value, the normalized success probabilities for ME are an order of magnitude smaller than those for the D case. QAC-ME completely reverses the performance loss due to ME; the QAC-ME success probabilities are comparable to those obtained with the D implementation. This again demonstrates that QAC acts as a successful error correction strategy. 

We note that there is no clear evidence of a critical clause density (or hardness peak) in this case. 
% We can estimate the critical clause density to be at $\alpha \approx 1.17$. 
The ME strategy exhibits a plateau after $\alpha \approx 1.17$, while the D and QAC-ME success probabilities continue to decrease, with QAC-ME decreasing more slowly. Up to this value, the increase in hardness is intrinsic and can be related to the increase in frustration. 
It may be possible to associate the further increase in hardness for larger clause densities to a precision effect, e.g., an increase in the range (the ratio between the largest and smallest couplings in the problem Hamiltonians; see Appendix~\ref{sec:rang&frust} for a quantification of this effect). This adversely affects the performance of the DW2 because a large coupling range means that some couplings are set to smaller values than can be physically implemented (the DW2 has a range of about 4 bits), due to the well-studied effect of ICE that leads to the specification of the wrong problem Hamiltonian \cite{q108,speedup,Venturelli:2014nx,King:2014uq,King:2015zr}. The saturation of the ME success probabilities can be understood as being due to having arrived at this precision limit before D and QAC-ME, i.e., for $\alpha > 1.17$ the ME results are dominated by ICE, and the improvement due to QAC is at least in part due to countering ICE. The D implementation is, of course, also susceptible to ICE, but is impacted less than ME due to the absence of broken qubits in D.

\begin{figure*}[t]
\begin{center}
\subfigure[\ Planted (full 2LG graph)]{\includegraphics[width=0.45\textwidth]{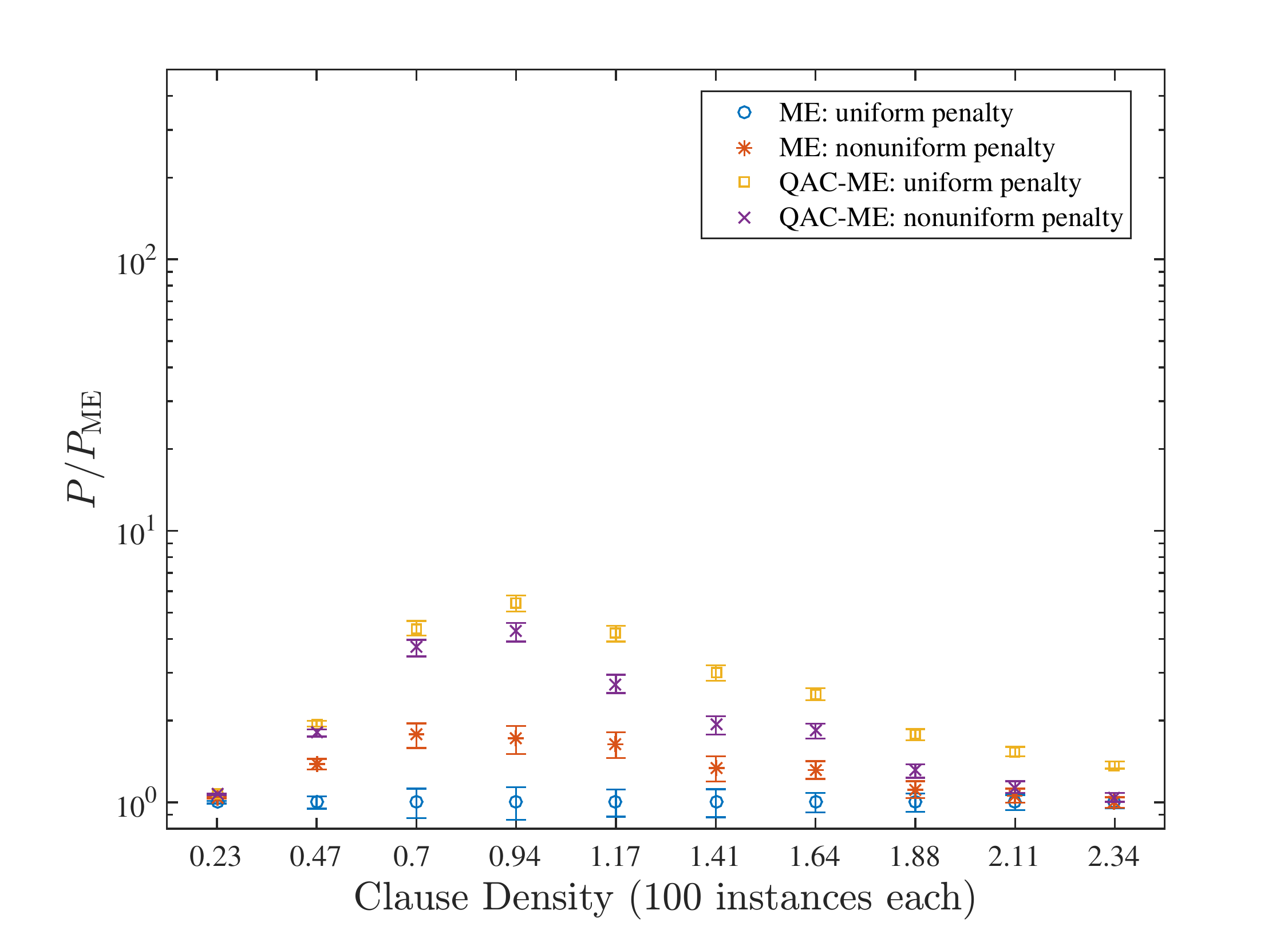}\label{fig:homvsnonhom1(a)}}
\subfigure[\ Weighted planted (full 2LG graph)]{\includegraphics[width=0.45\textwidth]{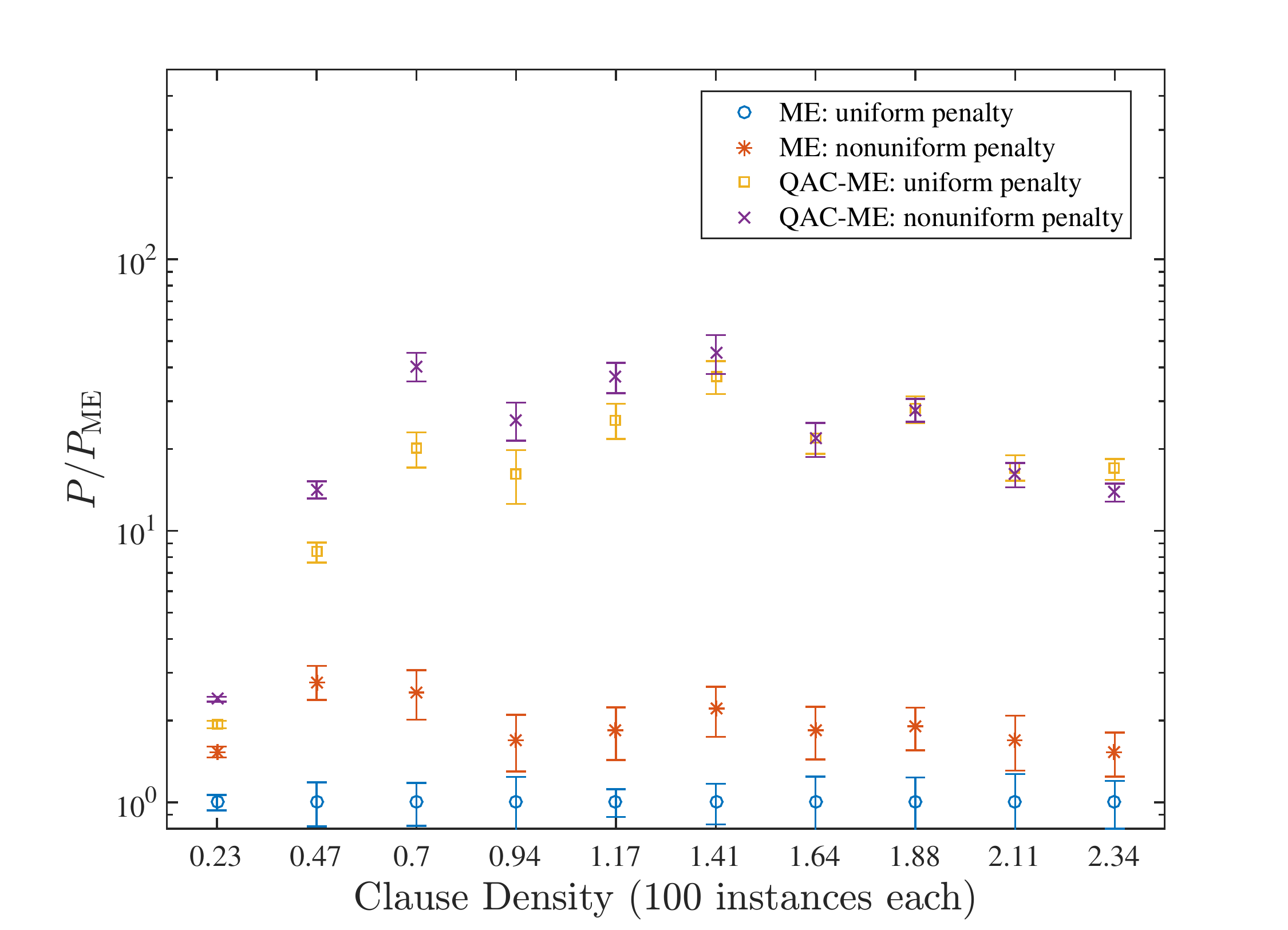}\label{fig:homvsnonhom1(b)}}
\subfigure[\ Embeddable planted (full 2LG graph)]{\includegraphics[width=0.45\textwidth]{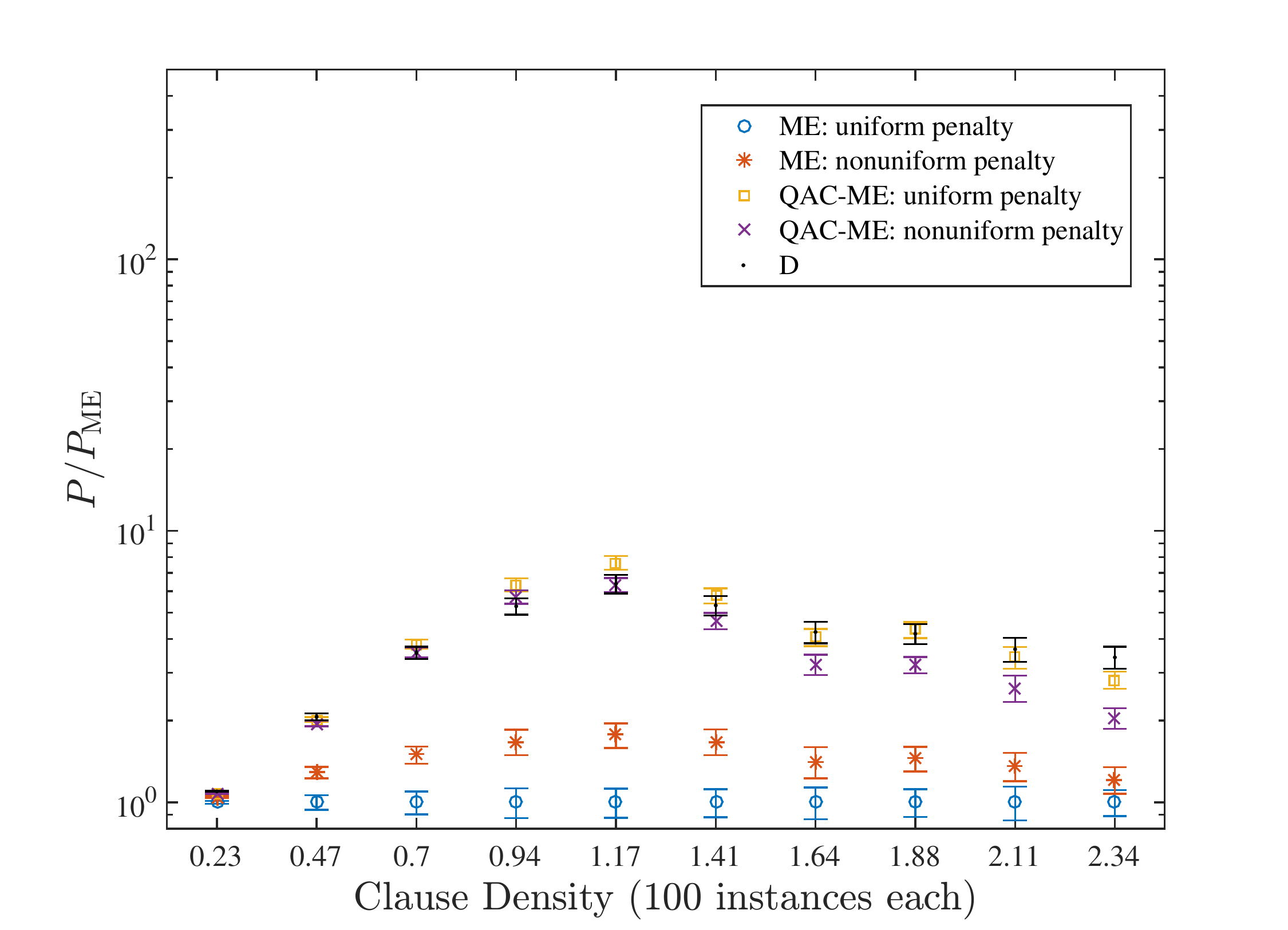}\label{fig:homvsnonhom1(c)}}
\subfigure[\ Deformed embeddable (embeddable 2LG subgraph)]{\includegraphics[width=0.45\textwidth]{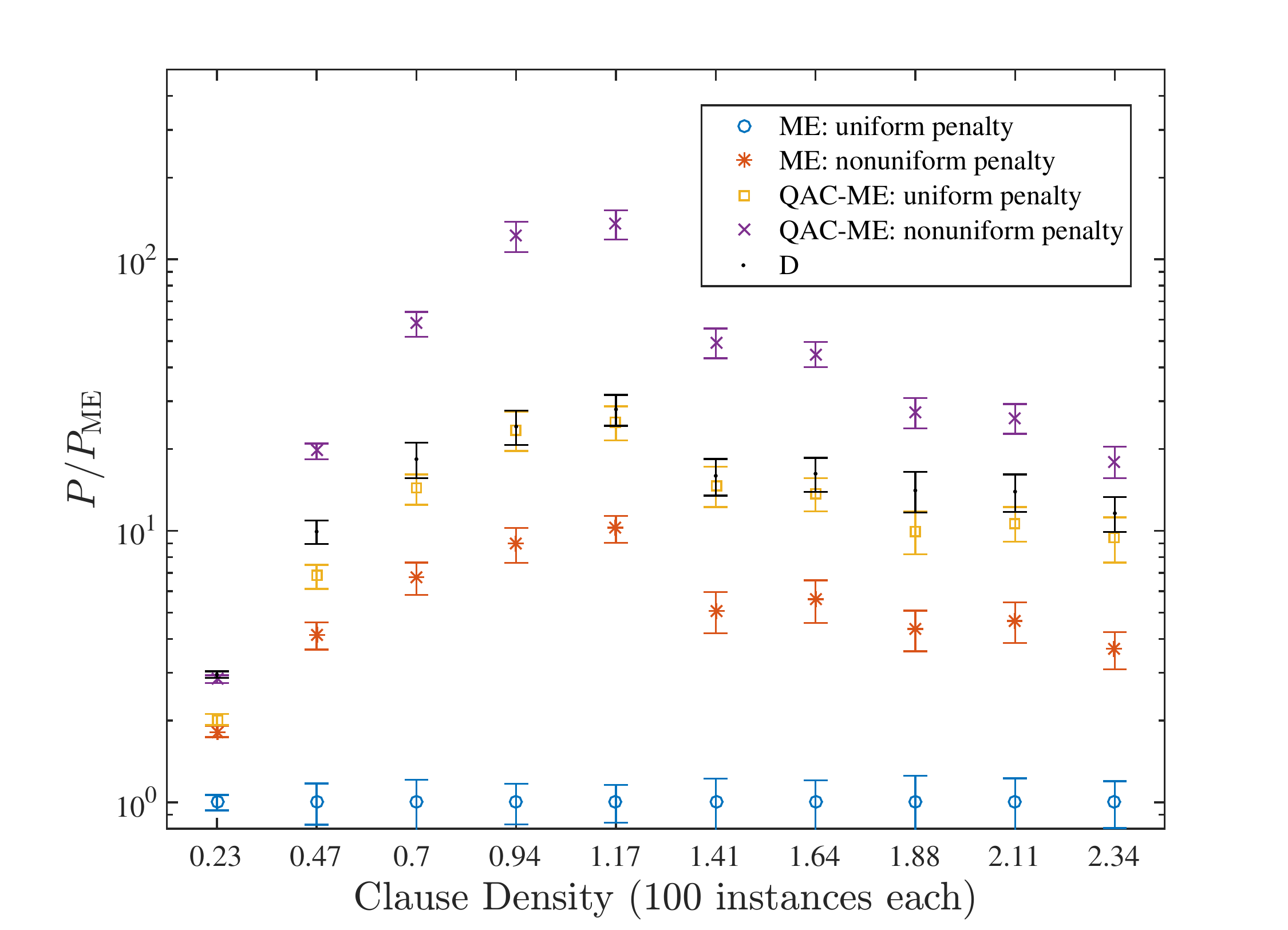}\label{fig:homvsnonhom1(d)}}
\caption{Success probabilities of ME and QAC-ME for uniform and nonuniform penalties for the four types of problem instances, normalized to the success probability of ME with uniform penalties. Decoding used the strategy of energy minimization. In all cases ME with uniform penalties performs worst, across the entire range of clause densities. ME with nonuniform penalties improves the ME performance somewhat, but always less than QAC-ME with uniform penalties. QAC-ME with nonuniform penalties does worse than QAC-ME with uniform penalties when the instances have uniform couplings [(a),(c)], but almost always does better than QAC-ME with uniform penalties when the instances have non-uniform couplings, as in (b) and (d). In the deformed embeddable case (d) QAC-ME with nonuniform couplings does significantly better even than the direct embedding.} 
\label{fig:homvsnonhom1}
\end{center}
\end{figure*}

%%%%%%%%%%%%%%%%%%%%%%%%%%%%%%%%%%%%%%%%%
\subsection{Optimized uniform \textit{vs} nonuniform penalties}
%%%%%%%%%%%%%%%%%%%%%%%%%%%%%%%%%%%%%%%%%
In Sec.~\ref{sec:ES} we discussed optimizing the energy penalties and proposed a new nonuniform strategy. In this section we experimentally test the effectiveness of this strategy and compare it to the previously proposed uniform strategy. Specifically, we compare the performance of the square code when penalties are chosen according to either Eq.~\eqref{eq:pen-uni} (uniform case) or Eq.~\eqref{eq:pen-nonuni} (nonuniform case), where the overall penalty energy scale $\gamma$ is a parameter that must be optimized independently. 

Intuitively, a nonuniform choice of penalties should perform best when the logical couplings are distributed nonuniformly on the logical graph. The problem instances we have considered so far were generated by adding frustrated loops that are uniformly distributed over the logical graph, resulting in similarly uniformly distributed couplings. We thus consider two additional sets of nonuniform instances.

%%%%%%%
\begin{figure*}[t]
\begin{center}
\subfigure[\ Planted, uniform]{\includegraphics[width=0.45\textwidth]{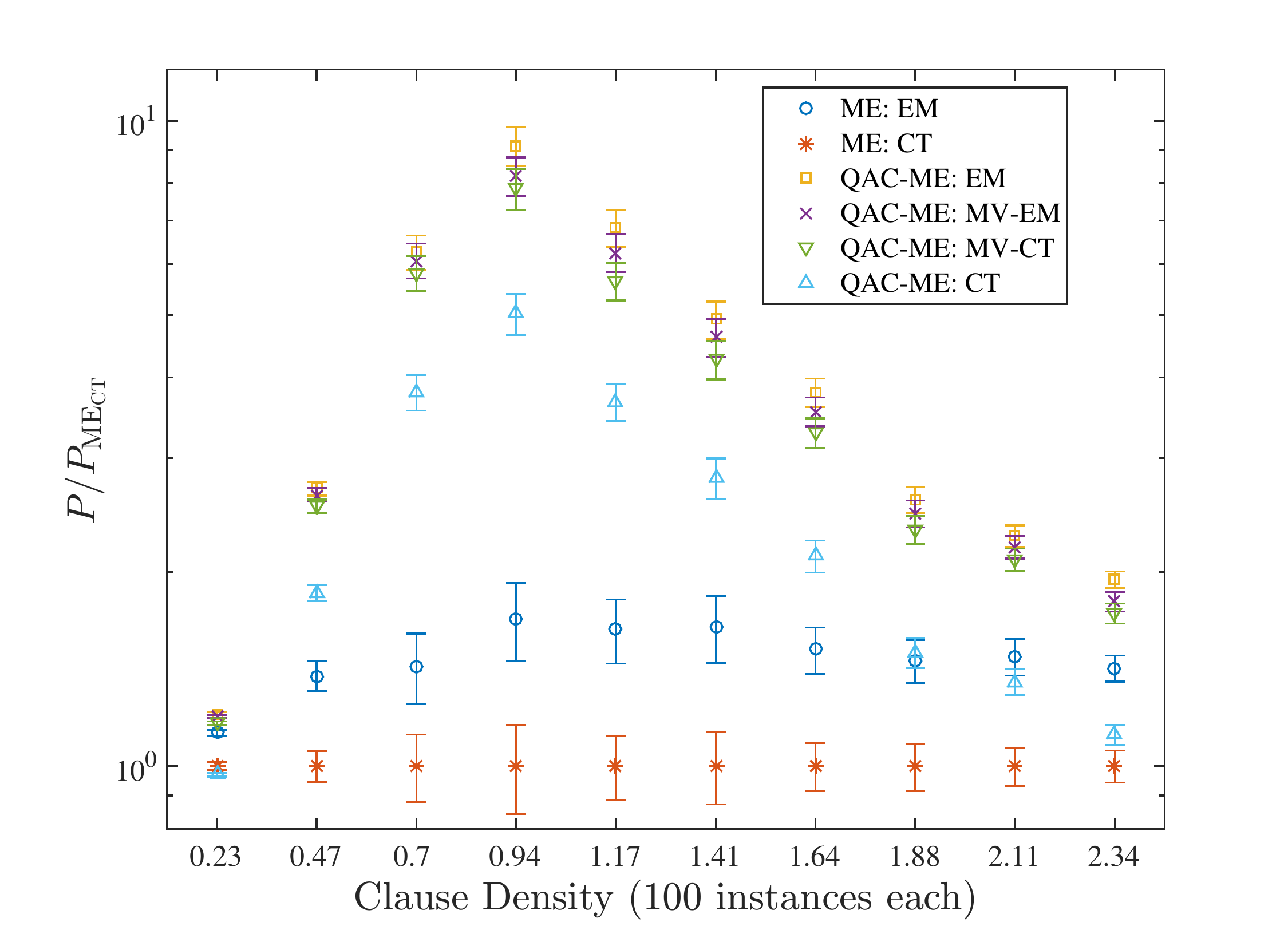}\label{fig:majvsmin(a)}}
\subfigure[\ Planted, nonuniform]{\includegraphics[width=0.45\textwidth]{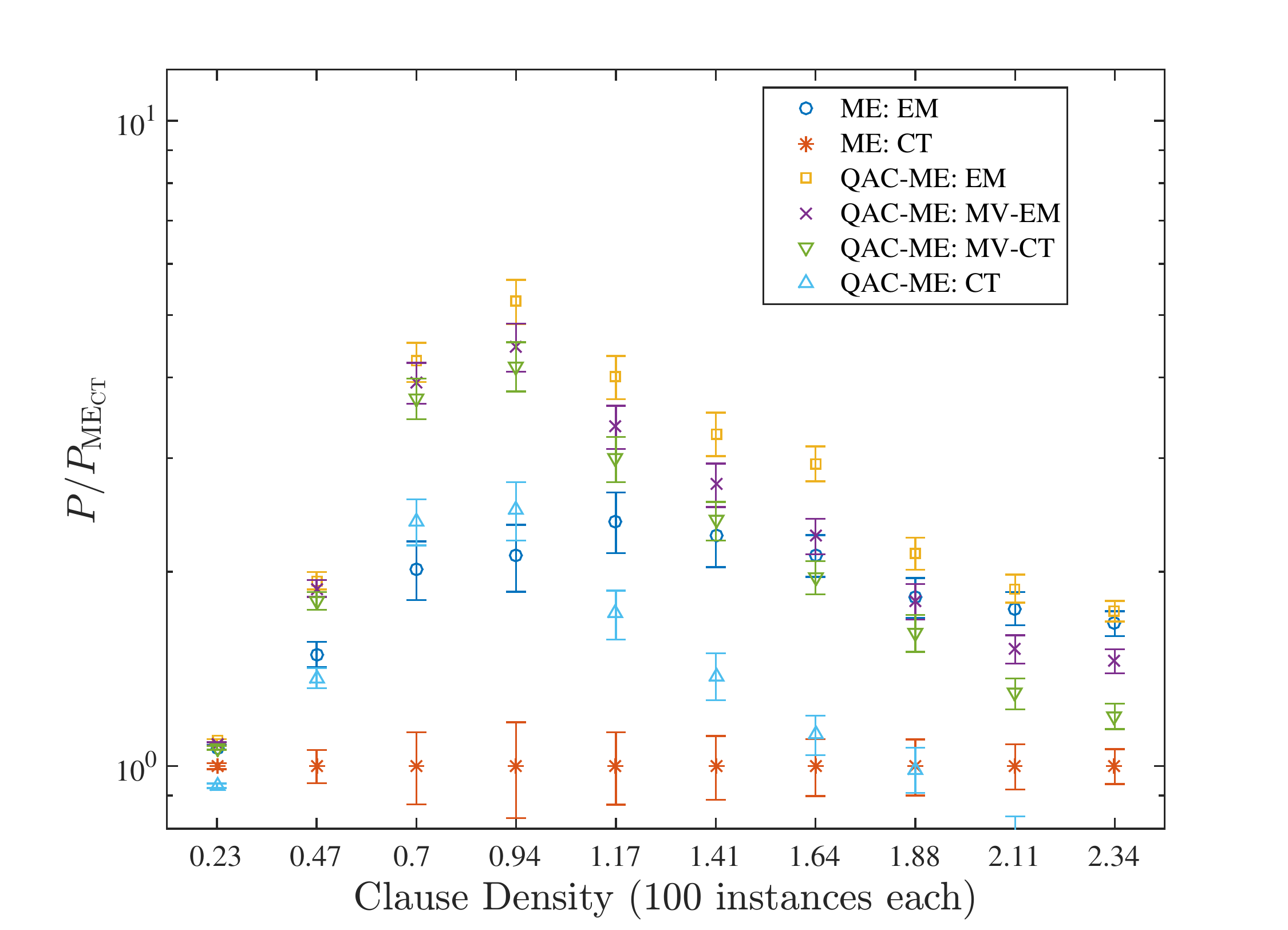}\label{fig:majvsmin(b)}}
%\caption{
%Comparison between different decoding strategies for the ME and QAC-ME implementations, for planted instances. Panel (a) [(b)] is for the case of uniform [nonuniform] penalties. 
%The success probabilities plotted are normalized to the ME case with random decoding ($P_{\mathrm{ME}_{\mathrm{CT}}}$). Both panels demonstrate the improvement of the decoded results when going from CT to MV and EM. In the ME case, the EM strategy resoundingly beats CT. In the QAC-ME case the ordering is EM$>$MV-EM$>$MV-CT$>$CT, with the advantage of the best strategy (EM) being more significant in the nonuniform case.}
%\label{fig:majvsmin}
%\end{center}
%\end{figure*}
%%%%%%%%
%
%%%%%%%%
%\begin{figure*}[t]
%\begin{center}
\subfigure[\ Weighted planted, uniform]{\includegraphics[width=0.45\textwidth]{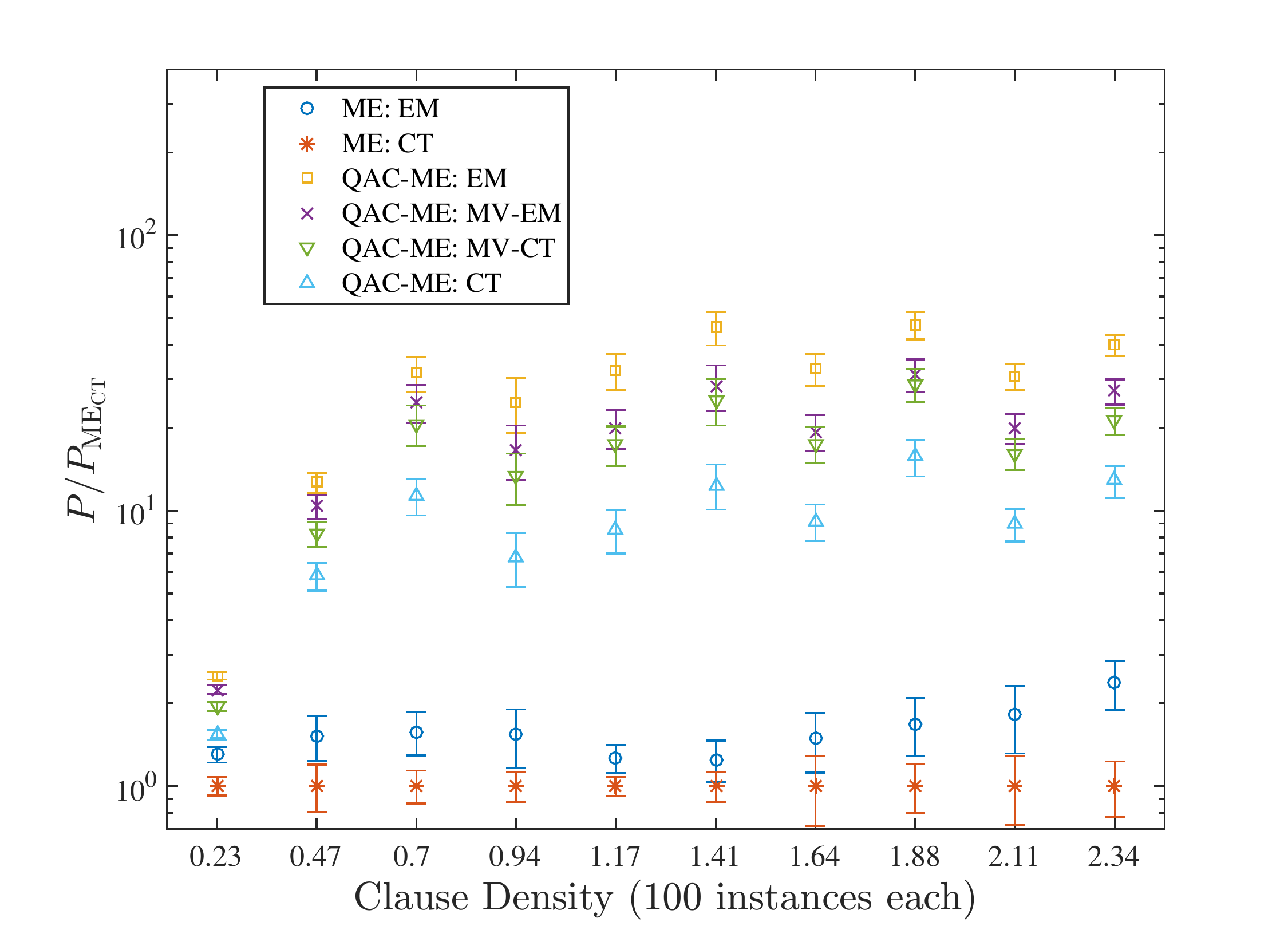}\label{fig:majvsmin(d)}}
\subfigure[\ Weighted planted, nonuniform]{\includegraphics[width=0.45\textwidth]{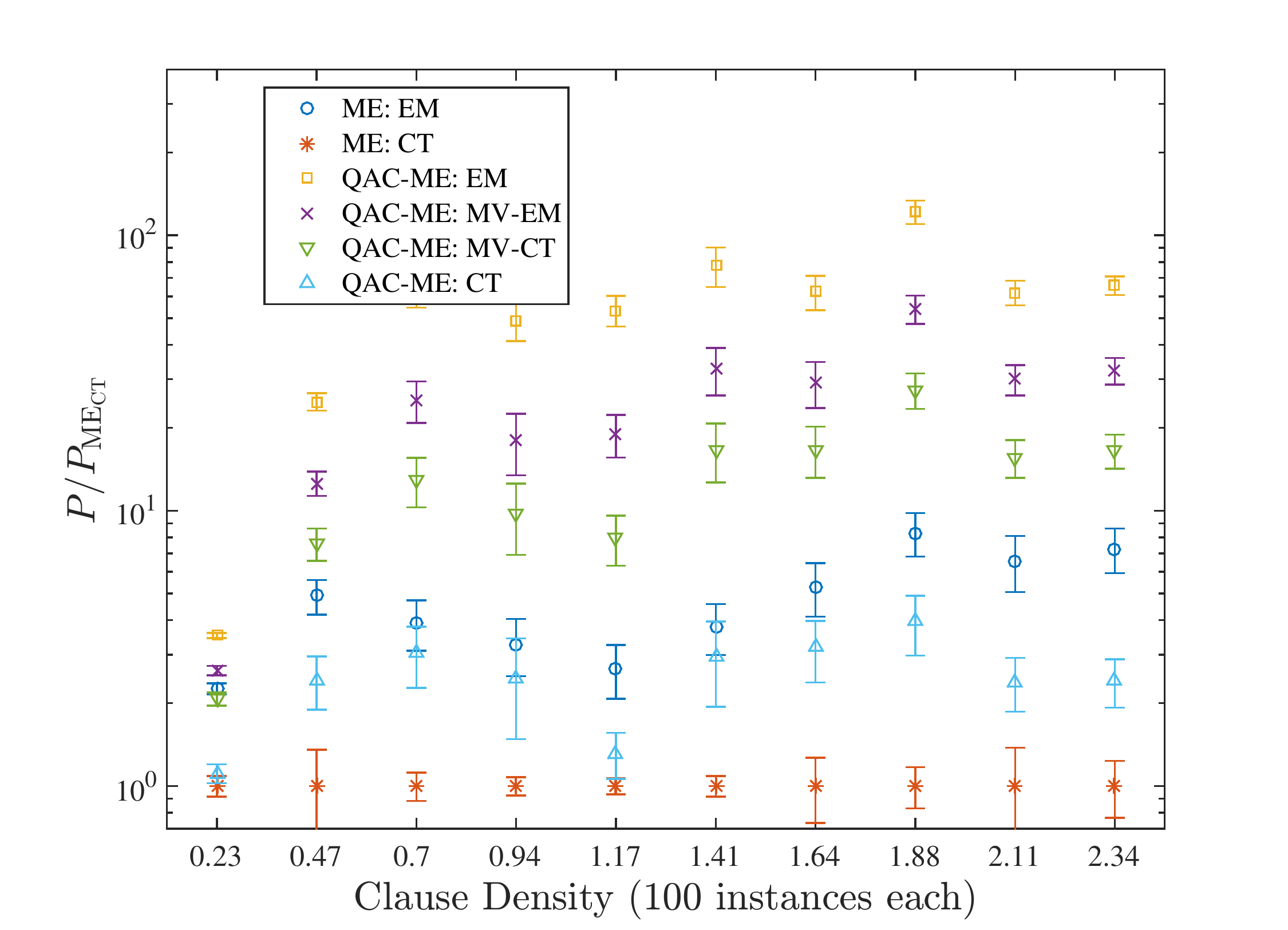}\label{fig:majvsmin(e)}}
%\caption{As in Fig.~\ref{fig:majvsmin}, for weighted planted instances. The ordering of the decoding strategies is also as in Fig.~\ref{fig:majvsmin}, with the case of nonuniform penalties again exhibiting a larger gain due to the EM strategy than the uniform case.} 
\caption{
Comparison between different decoding strategies for the ME and QAC-ME implementations, for planted [(a) and (b)] and weighted planted [(c) and (d)] instances. Panel (a) and (c)  [(b) and (d)] is for the case of uniform [nonuniform] penalties. 
The success probabilities plotted are normalized to the ME case with random decoding ($P_{\mathrm{ME}_{\mathrm{CT}}}$). Panels demonstrate the improvement of the decoded results when going from CT to MV and EM. In the ME case, the EM strategy resoundingly beats CT. In the QAC-ME case the ordering is EM$>$MV-EM$>$MV-CT$>$CT, with the advantage of the best strategy (EM) being more significant in the nonuniform case.}
\label{fig:majvsmin}
\end{center}
\end{figure*}
%%%%%%%

\subsubsection{Weighted planted frustrated-loop instances}  
The first set of instances is a simple modification of the construction using frustration-free Hamiltonians described above. The only difference compared to Eq.~\eqref{eq:planting} is that here we consider a weighted sum of the same set of frustrated loops as used to generate Fig.~\ref{fig:planted}, with the weights varying nonuniformly on the logical graph:
\beq
H_{\mathrm{P}} = \sum_{i}^L w_i H_i^{\mathrm{loop}}\ ,
\label{eq:weightedplanting}
\eeq 
We choose the $w_i$ as follows. We define $\{\bar{x}_i,\bar{y}_i, \bar{z}_i\}$
%[ H_i^{\mathrm{loop}}]_{\bar x,\bar y, \bar z}$
to be the ``center'' of loop $i$ on the 2LG (defined as the integer part of the mean of the coordinates of the spins comprising the loop), 
%We thus have $0 \le [ H_i^{\mathrm{loop}}]_{\bar I,\bar J} \le 8$ and $0 \le [ H_i^{\mathrm{loop}}]_{\bar K} \le 2$. 
%The $w_i$ are then chosen as follows
%
%\beq
%w_i = [ H_i^{\mathrm{loop}}]_{\bar I}\,.
%\eeq 
%
and set $w_i = \bar{x}_i$. This increases the typical strength of the couplings by a factor of $8$ between the two opposite sides of the $8\times 8\times 2$ 2LG logical graph embedded into the DW2. This introduces a spatial non-uniformity and also reduces the magnitude of the smallest couplings of this set of instances (see Appendix~\ref{sec:rang&frust}), thus making the instances significantly harder, as we shall see below. 
Note that the same planted solution as before ($\ket{\Psi_{0}}$) applies, independently of the choice of the weights. 

%%%%%%%%%%%%%%%%%%%%%%%%%
\subsubsection{Deformed embeddable instances} 
The second set are nonuniform instances we generate by ``deformation'' of the set of instances with planted solutions defined on the embeddable subgraph [Fig.~\ref{fig:2LG-sub}], as follows. For each instance we randomly pick a subset $\bar I$ of encoded qubits (vertices of $\bar{G}$) and apply a linear transformation (shifting and compression) to the encoded couplings such that:
\bes
\begin{align}
\label{eq:34a}
  J_{\bar i,\bar j} & \rightarrow    J_{\bar i,\bar j} /3   \quad \quad \quad \forall \bar i \in \bar I, \forall  j \not \in \bar I  \\
 \label{eq:34b}
  J_{\bar i,\bar j} & \rightarrow   (2 \, {\rm sign} ( J_{\bar i,\bar j} ) +  J_{\bar i,\bar j} )/3 \quad \quad  \forall  \bar i,\bar j  \not \in \bar I\ .
\end{align}
\ees
This partitions the encoded qubits into three sets $\bar I, \bar J, \bar K$, where the set $\bar I$  (the one picked) is connected only through coupling that have been transformed according to Eq.~\eqref{eq:34a} (small couplings), $\bar K$ is connected through couplings that have been transformed according to Eq.~\eqref{eq:34b} (large couplings), and $\bar J$ is connected through couplings that have been transformed according to both Eqs.~\eqref{eq:34a} and \eqref{eq:34b}.

The average of the encoded couplings is $\sim 0.5$. Using this value in Eqs.~\eqref{eq:34a} and \eqref{eq:34b} yields average coupling strengths of $1/6$, $1/2$ and $5/6$ for the sets $\bar I, \bar J, \bar K$ respectively. We randomly pick $35$ qubits out of the available $120$  for the $\bar I$ set, which yields three similarly sized sets. Under this deformation, the originally planted configuration is no longer necessarily the ground state. In this case, we use the exact solver that uses a bucket tree elimination strategy (BTE) and is included in the D-Wave user libraries to determine the exact ground state energy. This exact solver is specifically  designed to solve small, Chimera-structured problems.  
%\mg{Updated figures are coming soon. I expect a small increase (if any) in the advantage that QAC-ME has over ME.}

Figure~\ref{fig:homvsnonhom1} shows the experimental results for ME and QAC-ME, with both uniform and nonuniform penalties. The panels show the four sets of instances: planted and weighted planted (both defined on the full 2LG), embeddable planted and deformed embeddable (defined on the embeddable subgraph).  In all panels of Fig.~\ref{fig:homvsnonhom1} we see that QAC-ME outperforms ME. The choice of nonuniform penalties improves the success probabilities of both the ME and QAC-ME  in the case of nonuniform instances [panels (b) and (d)]. Moreover, we see that the nonuniform choice of penalties improves the ME implementation in the uniform case too, while it does not improve the QAC-ME  scheme, suggesting that QAC-ME with a uniform penalty is already close to optimal in the case of uniform instances.  (The dependence of the optimal penalty strength on clause density is shown in Appendix~\ref{app:penalties}.)

As a final remark of this section, note the large boost of nearly two orders in magnitude in the success probabilities that QAC-ME achieves over ME in the weighted case shown in Fig.~\ref{fig:homvsnonhom1(b)}. In this case QAC-ME is particularly successful. A similar advantage of QAC-ME over ME is observed in Fig.~\ref{fig:homvsnonhom1(d)}, where the advantage of QAC-ME with nonuniform penalties is the most pronounced, easily outperforming the direct embedding as well [in contrast to the tie in the embeddable planted case, seen in Fig.~\ref{fig:QAC-ME_best-EmbeddablePlanted}]. 

%We summarize this section noticing that a nonuniform choice of penalties is typically not worse than a uniform one and  has the potential, in the case of non- uniform instances,  to  significantly improve the performance of Vesuvius on minor embedded optimization problems. We also stress  the huge boost of performance achieved on the weighted and deformed types of instances by QAC-ME. These instances are very hard for Vesuvius because of the combined effects of large range and frustration, both very detrimental to the performance of Vesuvius.  

%%%%%%%%%%%%%%%%%%%%%%%%%%%%%%%%%%%%%%%%%
\section{Impact of the decoding strategy}
\label{sec:dec-sc}
%%%%%%%%%%%%%%%%%%%%%%%%%%%%%%%%%%%%%%%%%

We now consider the impact of various decoding strategies defined in Sec.~\ref{sec:DS} on performance as measured in terms of success probabilities. As already discussed in Sec.~\ref{sec:MV}, because both the ME and QAC-ME implementations of the square code involve an even number of physical qubits (two and four respectively), the possibility of ties means that broken qubits can be undecodable by majority vote. In the ME case, in particular, every broken qubit is indeed a tie. In the QAC-ME case there are two possibilities: ``partially" broken qubits (decodable through majority vote) and ties.

We thus consider the following decoding strategies:
\begin{enumerate}
\item {ME:}
	\begin{enumerate}
	\item Decode ties randomly via coin tossing (CT);
	\item Decode ties via energy minimization (EM).
	\end{enumerate}
\item {QAC-ME:}
	\begin{enumerate}
	\item Decode all broken qubits via CT;
	\item Decode all broken qubits via energy minimization (EM);
	\item Decode partially broken qubits via MV and ties via CT (MV-CT);
	\item Decode partially broken qubits via MV and ties via EM (MV-EM);
	\item Decode partially broken qubits via MV and ties via EM. Randomly select a number of unbroken qubits equal to the number of partially broken qubits and decode via EM (MV-EM(R)).
	\end{enumerate}
\end{enumerate}

%%%%%%%%%%%%%%%%
\subsection{Performance of the various decoding strategies}
\label{sec:expDS}
%%%%%%%%%%%%%%%%

%%%%%%%
\begin{figure*}[t]
\begin{center}
\subfigure[\ Planted, nonuniform]{\includegraphics[width=0.45\textwidth]{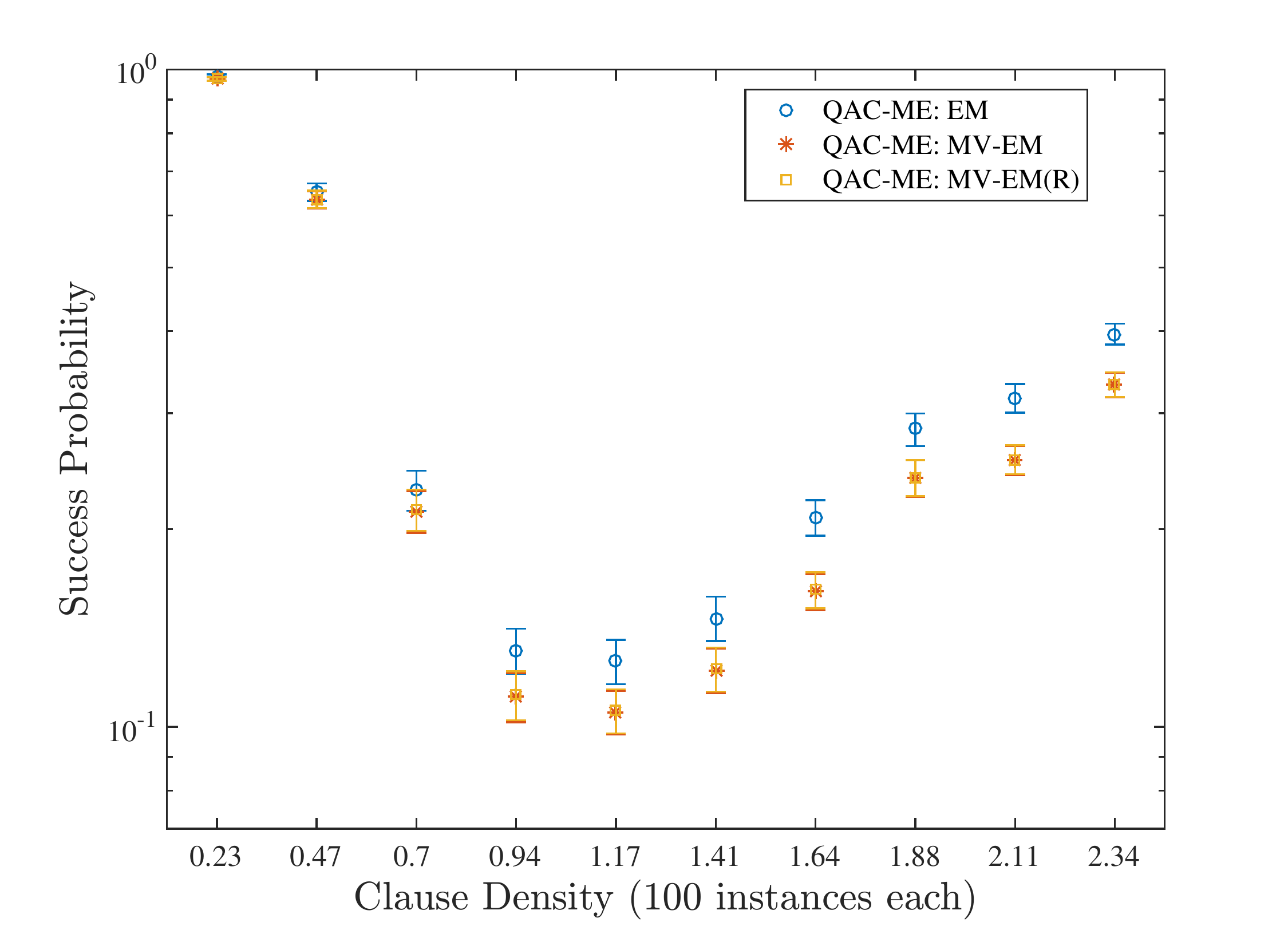}\label{fig:majvsmin(c)}}
\subfigure[\ Weighted planted, nonuniform]{\includegraphics[width=0.45\textwidth]{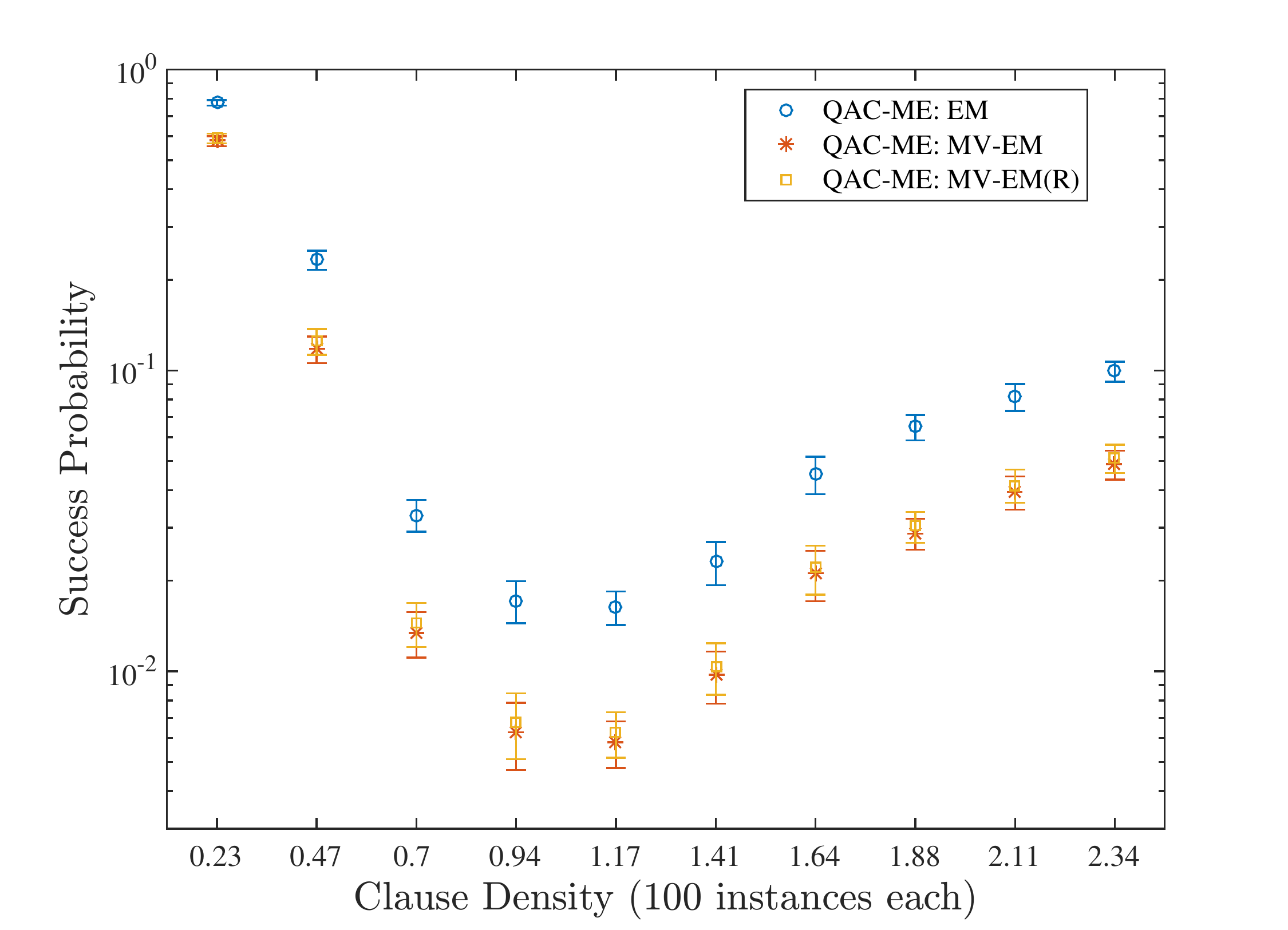}\label{fig:majvsmin(f)}}
\caption{Comparison between different decoding strategies for the QAC-ME implementation with nonuniform penalties, for (a) planted and (b) weighted planted instances. The MV-EM(R) strategy, where a random set of unbroken qubits is subjected to energy minimization, is shown along with MV-EM and EM. It can be seen that MV-EM(R) does no better than MV-EM. This implies that logical errors are mostly concentrated on broken qubits. The EM data shown here is the same as for the QAC-ME case in Figs.~\ref{fig:majvsmin(b)} and \ref{fig:majvsmin(e)}.} 
\label{fig:majvsmin3}
\end{center}
\end{figure*}
%%%%%%%

%%%%%%%
\begin{figure}[t]
\begin{center}
{\includegraphics[width=0.45\textwidth]{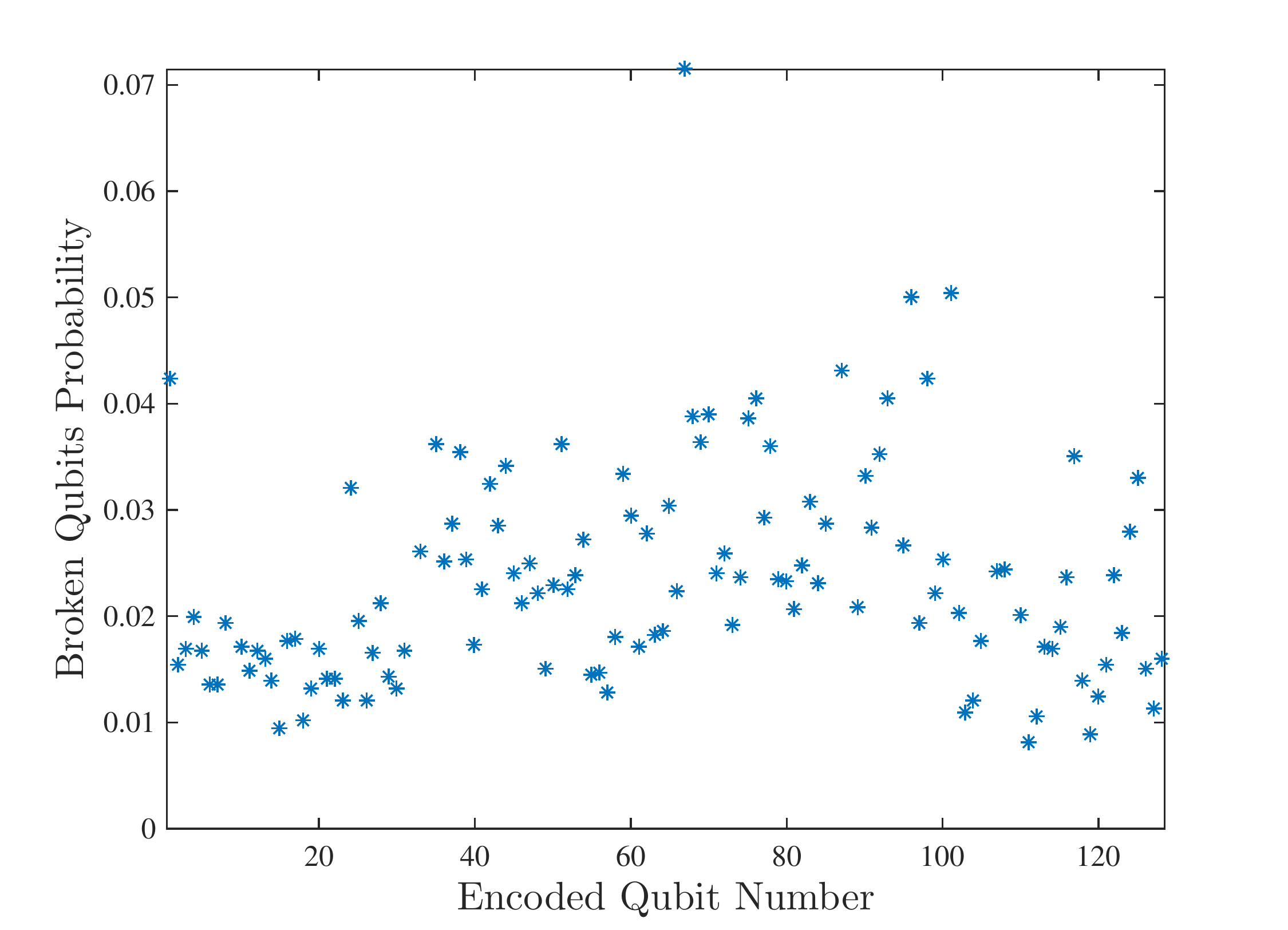}}
\caption{Broken qubit median probabilities for the ``planted'' instance set at the critical clause density $\alpha = 0.94$. Results are shown for the full set of $100$ instances implemented with the QAC-ME scheme, for the optimal value of the (uniform) penalty $\gamma = 0.2$. The abscissa is the index of the encoded qubits on the 2LG graph shown in Fig.~\ref{fig:square-code-a}. 
%\mg{Using the labeling of Fig.~\ref{fig:square-code-a}, qubit 127 is used, while, for example, qubit 8 is unused.}
}
\label{fig:BQ_density}
\end{center}
\end{figure}
%%%%%%%

%%%%%%%%%%%%%%%%
\begin{figure*}[ht]
\begin{center}
\subfigure[\ Planted, uniform]{\includegraphics[width=0.45\textwidth]{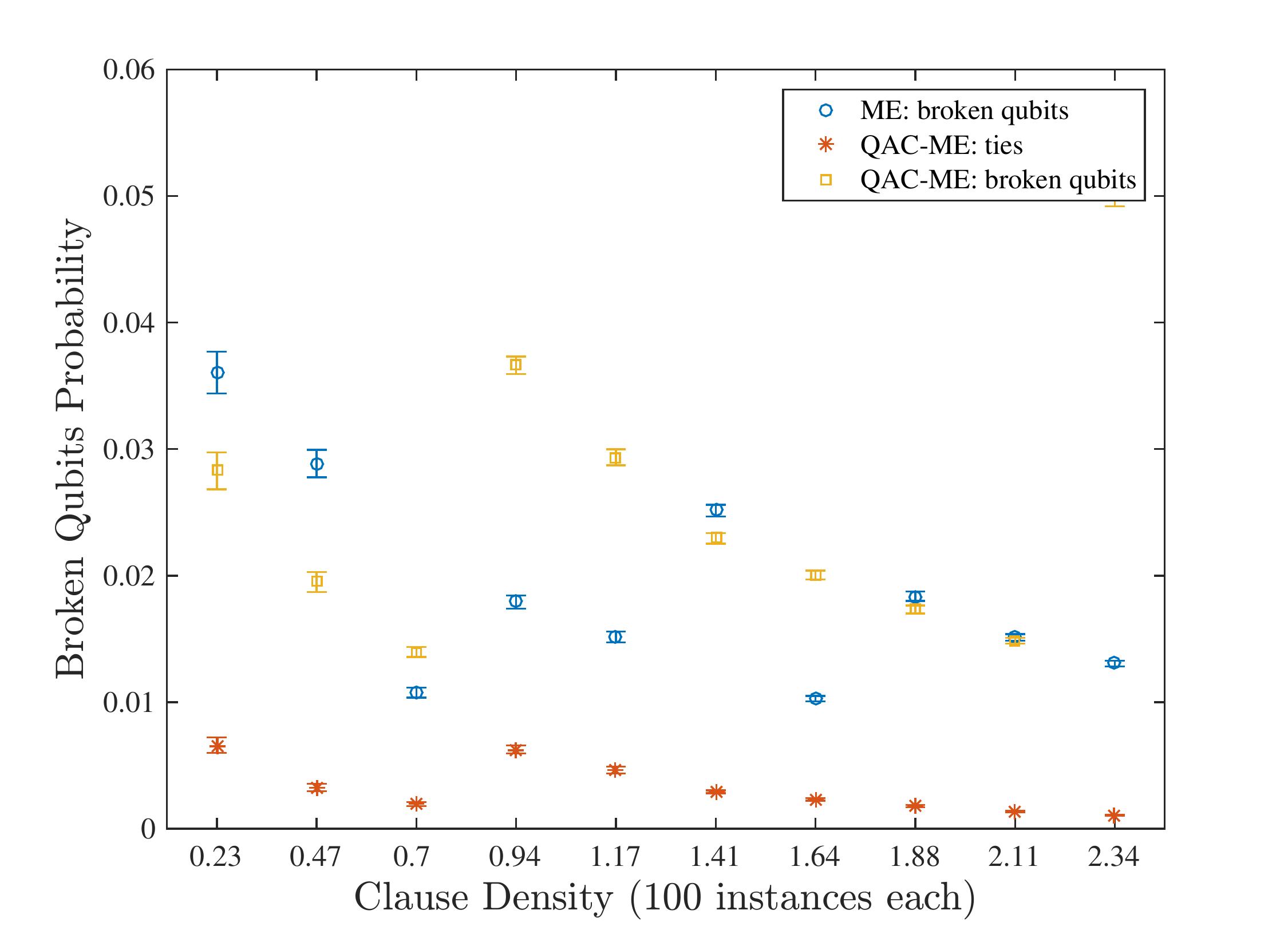}}
\subfigure[\ Planted, nonuniform]{\includegraphics[width=0.45\textwidth]{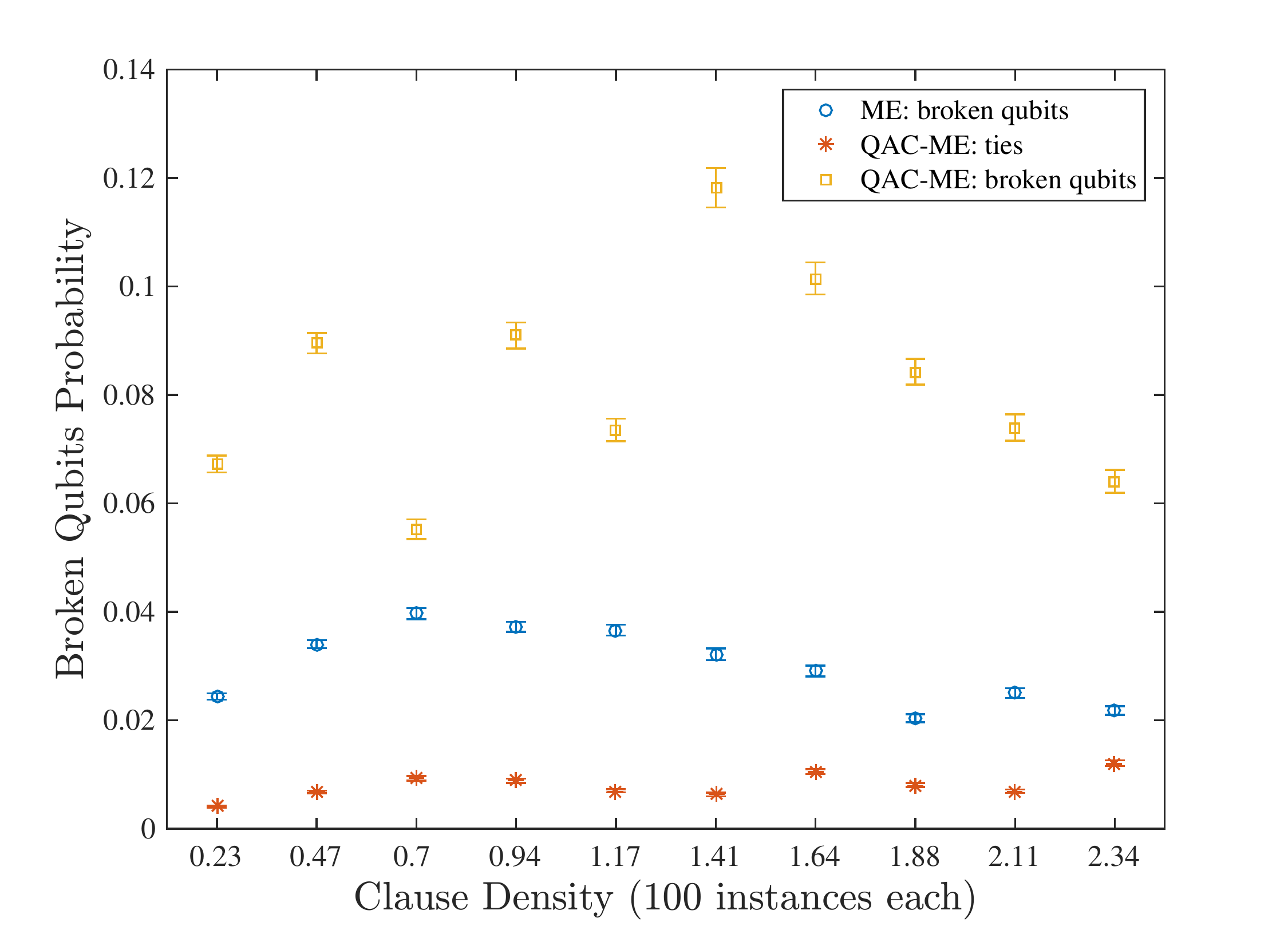}}
\subfigure[\ Weighted planted, uniform]{\includegraphics[width=0.45\textwidth]{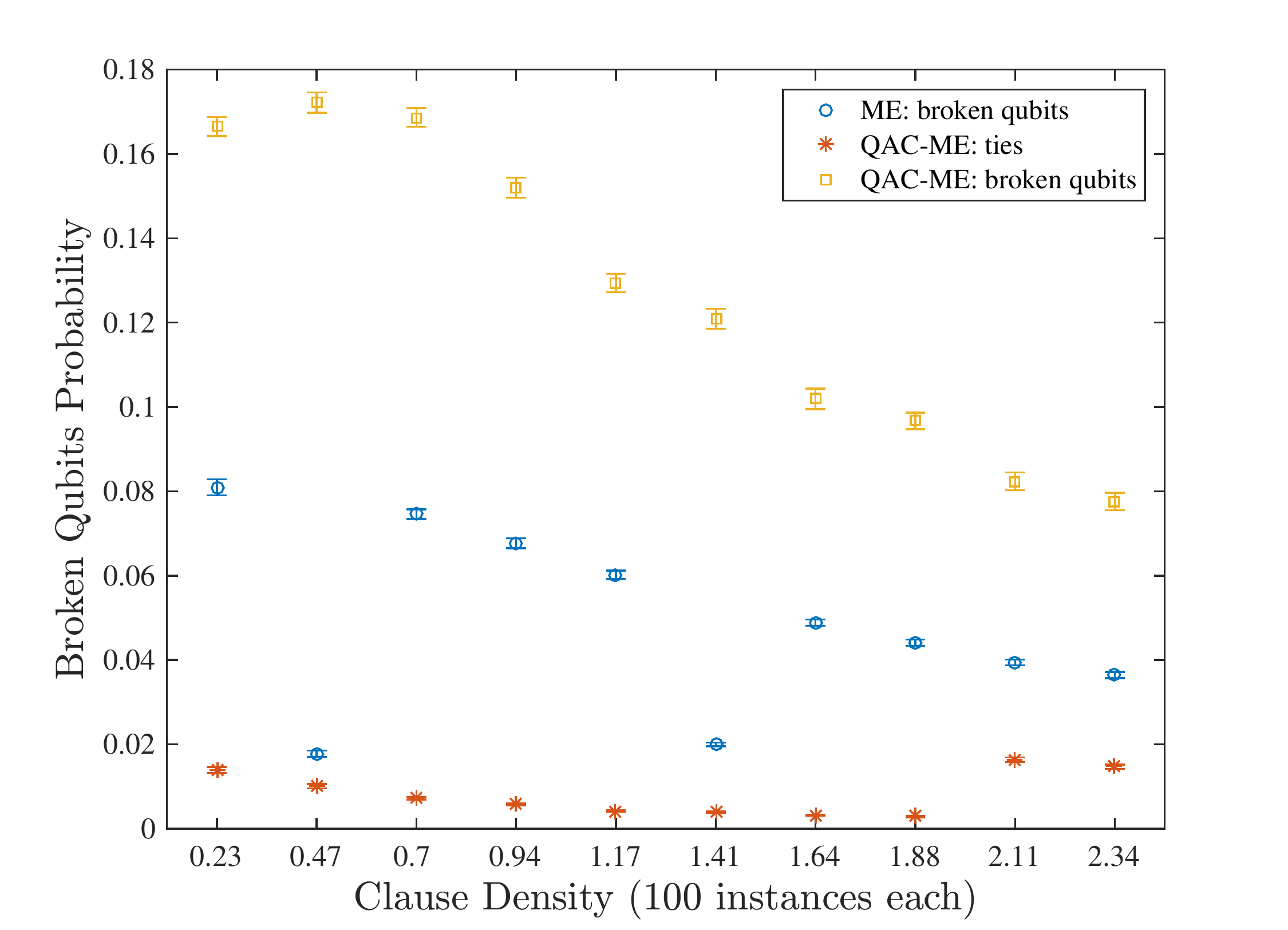}}
\subfigure[\ Weighted planted, nonuniform]{ \includegraphics[width=0.45\textwidth]{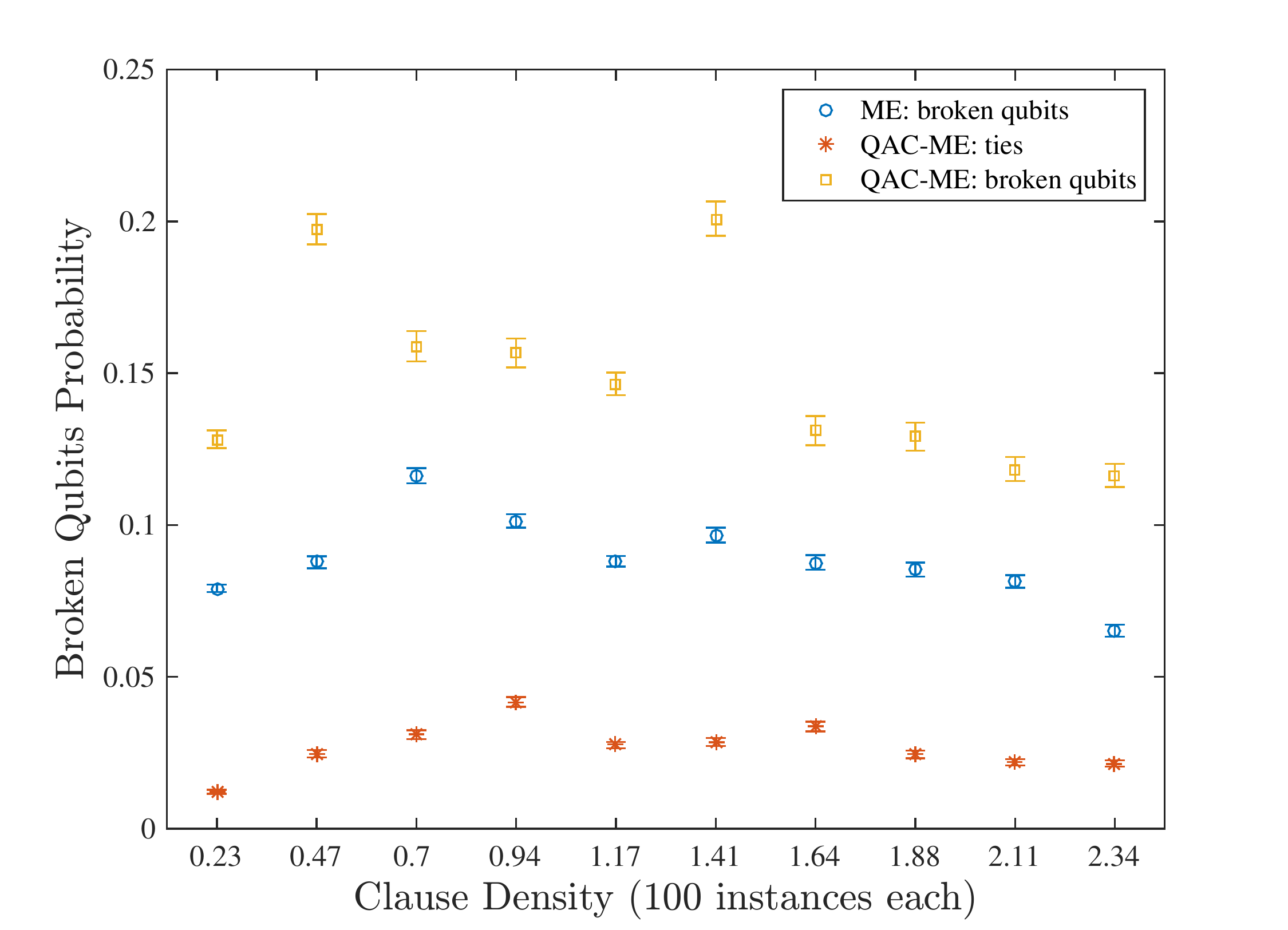}}
\caption{Probability $p_{\mathrm{\mathrm{BQ}}}$ of broken qubits in ME and QAC-ME, as a function of the clause density, for planted (top) and weighted planted (bottom) problem instances defined on the full 2LG graph. Left: uniform penalties. Right: nonuniform penalties. ME almost always has a lower $p_{\mathrm{\mathrm{BQ}}}$ than QAC-ME. Also shown is the probability of tied qubits in the QAC-ME case (which is already included in the broken qubits probability).} 
\label{fig:exptie}
\end{center}
\end{figure*}
%%%%%%%%%%%%%%%%

While we do not expect random decoding (the CT strategy) to work well, it sets a useful baseline against which we can compare our other decoding strategies. 
The comparison between all strategies in the case of instances defined on the full 2LG graph is shown in Figs.~\ref{fig:majvsmin}~and~\ref{fig:majvsmin3}. Figure~\ref{fig:majvsmin} is for the planted and weighted planted instances and reports results for both the ME and QAC-ME cases.
Figure~\ref{fig:majvsmin3} focuses on the MV-EM(R) decoding strategy (QAC-ME only). Complementary results for the embeddable planted and the deformed embeddable cases are shown in Appendix~\ref{app:dec}.

As expected, CT is the least effective decoding strategy, and is substantially outperformed by MV. The optimal choice is always the EM strategy (recall Sec.~\ref{sec:EM}), which yields improvements over both MV and MV-EM. The benefit of EM is larger in the case of nonuniform penalties and weighted instances. In  Fig.~\ref{fig:majvsmin(e)} we see that EM decoding improves the success probabilities by a factor of  about $3$ as compared to the second best decoding strategy, MV-EM. The advantage of EM over MV and CT is directly related to the number of broken qubits at the optimal penalty value. The larger this number, the more bit flip errors we expect to correct during the decoding process by doing energy minimization. As we shall see in the next section, the typical number of broken qubits is usually larger for the nonuniform penalties and for the nonuniform instances we have constructed.

%%DL: I have trouble making sense of the (already heavily edited) paragraph below:
%It is instructive to provide a further comparison between the best (EM) and second best (MV-EM) decoding strategies. EM involves an optimization over a strictly larger number of logical qubits than MV-EM and hence it must give a strictly better result. However, this gain comes with the price of a concomitant increased computational effort. The difference between EM and MV-EM is the optimization over partially broken qubits done in EM (thus completely discarding the MV step). This extra effort can be justified only if we can confirm the intuitive idea that partially broken qubits are more likely to contain logical errors than unbroken qubits (where a logical error could arise only if all physical qubits had flipped, a scenario that cannot be decoded by MV). The MV-EM(R) decoding strategy is designed to probe this, by singling out a fraction of the unbroken qubits and applying the EM strategy to them. We thus compare EM and MV-EM to the MV-EM(R) decoding. This comparison is  shown in Fig.~\ref{fig:majvsmin3}, again for the planted and weighted planted instances defined on the full 2LG graph, with nonuniform penalties. We find that the MV-EM(R) strategy offers no advantage over MV-EM, leaving EM as the superior strategy. This is an experimental indication that unbroken qubits typically contain fewer errors than partially broken qubits. It thus seems justified  to apply EM to partially broken qubits, as they are more likely to contain errors that cannot be correctly decoded by MV.

It is interesting to check whether unbroken qubits (where all physical qubits agree), which cannot be decoded via MV, can be further improved by EM. The MV-EM(R) decoding strategy is designed to probe this, by singling out a fraction of the unbroken qubits and applying the EM strategy to them. We thus compare MV-EM(R) to MV-EM and EM decoding in Fig.~\ref{fig:majvsmin3}, again for the planted and weighted planted instances defined on the full 2LG graph, with nonuniform penalties. We find that the MV-EM(R) strategy offers no advantage over MV-EM, leaving EM as the superior strategy. Since MV-EM(R) is essentially indistinguishable from MV-EM, we may conclude that the unbroken qubits make a negligible contribution to the error rate. 

%\mg{We can check an independent errors model by studying the probability distribution of $k$ errors per logical qubit ($k=1,2,3,4$) and show that it declines with increasing $k$}

%%%%%%%%%%%%%%%%%%%%%%%%%%%%%%%%%%%%%%%%%
 \subsection{Broken qubit probability and the percolation threshold for efficient decoding}
%%%%%%%%%%%%%%%%%%%%%%%%%%%%%%%%%%%%%%%%%
\begin{figure*}[ht]
\begin{center}
\subfigure[\ $T = 2$, $\chi = 5\%$   ]{\includegraphics[width=0.45\textwidth]{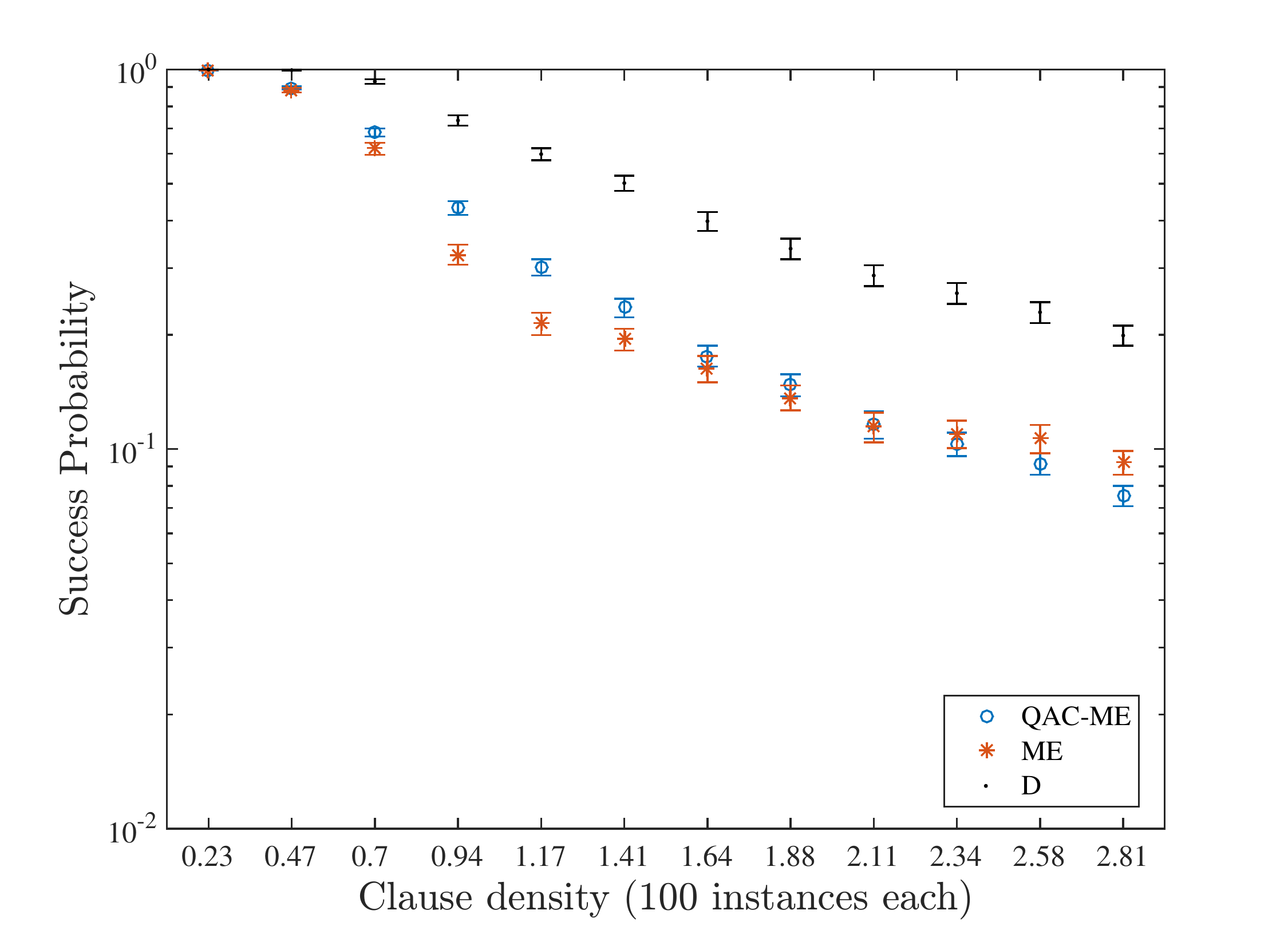} }
\subfigure[\ $T = 2$, $\chi = 10\%$    ]{\includegraphics[width=0.45\textwidth]{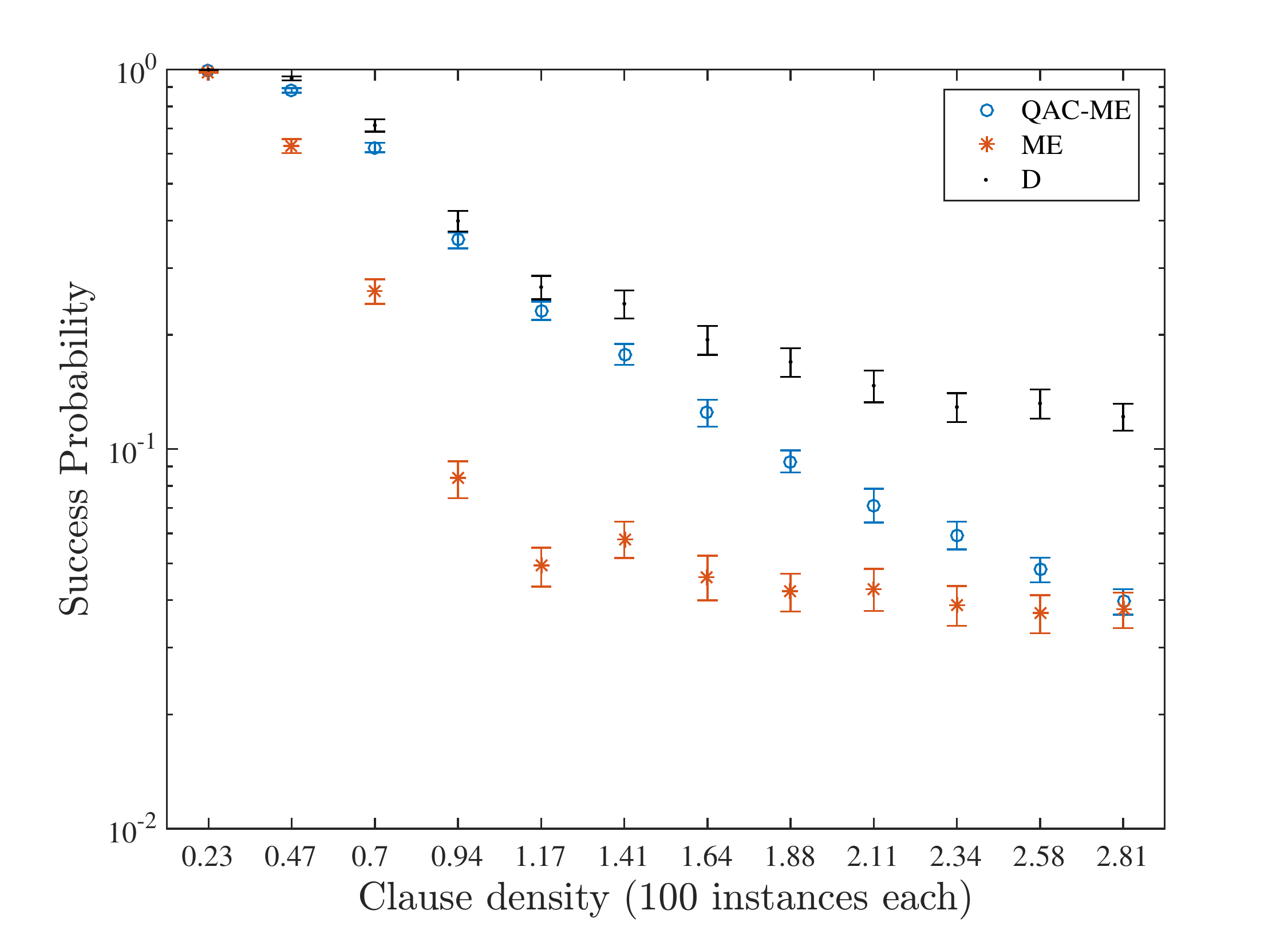} }
\subfigure[\  $T = 4$, $\chi = 5\%$    ]{\includegraphics[width=0.45\textwidth]{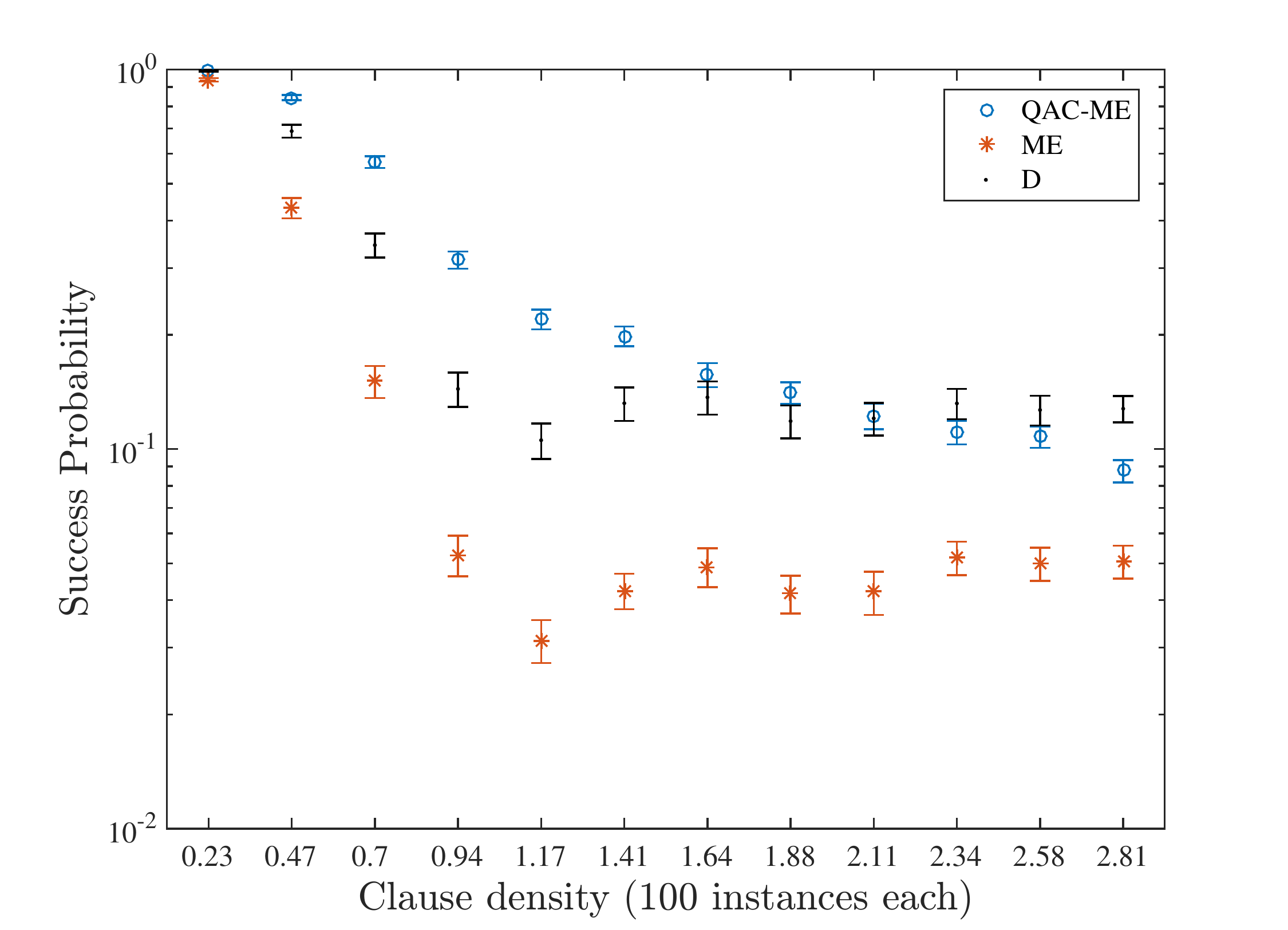} }
\caption{SQA simulation results for the ``embeddable planted'' instances class. (a) Relatively low temperature $T$ and low noise strength $\chi$: QAC-ME does not provide an advantage over the uncorrected ME approach. Note how success probabilities are typically larger for SQA in this case with respect to the experimental results [Fig.~\ref{fig:QAC-ME_best-EmbeddablePlanted}]. (b): Relatively low temperature and high noise strength: QAC-ME gains a clear advantage over ME but is still inferior to D. (c) Relatively high temperature and low noise strength: QAC-ME gives a consistent advantage over ME, and it is competitive with D. This situation is the most similar to the experimental results. We used $20,\!000$ sweeps in all cases shown. Temperature is in units of $B(t_f)/k_B$.} 
\label{fig:num_SQA}
\end{center}
\end{figure*}

Following the discussion in Sec.~\ref{sec:EOD}, efficient optimal decoding via the EM strategy is possible if the probability $p_{\mathrm{\mathrm{BQ}}}$ to find a broken encoded qubit is smaller than the percolation threshold $p_{\textrm{2LG}}$ of the 2LG. This ensures that no large clusters of broken qubits are formed that would make energy minimization too computationally costly. While we do not know the value of the percolation threshold of the (infinite) 2LG, it must certainly be bounded between the thresholds $p_{\textrm{cubic}}$ and $p_{\textrm{square}}$ for cubic and square lattices, respectively \cite{levinshtein1975relation,1992JPhA25.5867G}:
\beq
0.3116  = p_{\textrm{cubic}} < p_{\textrm{2LG}} < p_{\textrm{square}} = 0.5927\ .
\eeq
We may further expect $p_{\textrm{2LG}}$ to be closer to $p_{\textrm{square}}$ than to $p_{\textrm{cubic}}$ since the 2LG is geometrically more similar to a square lattice.
Recall that the 2LG is the encoded graph of the square code, and hence we identify the latter's percolation threshold with $p_{\textrm{2LG}}$.

We first check that the experimental probability $p_{\mathrm{BQ}}$ to find broken qubits in the square code depends only weakly on the position of the encoded qubits. This is confirmed (for the QAC-ME case) in Fig.~\ref{fig:BQ_density} at the optimal penalty value for a fixed clause density, and justifies the assumption underlying our percolation threshold argument about the efficiency of the EM decoding strategy in Sec.~\ref{sec:EOD}.

Figure~\ref{fig:exptie} shows the probability of a broken qubit as a function of the clause density, for both ME and QAC-ME, in the case of instances with planted and weighted planted solutions on the full 2LG, for both uniform and nonuniform penalties. The most important finding is that in all cases, the experimental probability of a broken logical or encoded qubit is well below $p_{\textrm{cubic}}$, and hence below the percolation threshold of the square code. This suggests that scalable and efficient decoding of the square code through energy minimization is possible, at least for the (frustrated) class of instances considered here. Additional supporting arguments are given in Appendix~\ref{app:worstcase} using a worst-case analysis.

There are a few other interesting observations we can make about Fig.~\ref{fig:exptie}. First, the very fact that the observed $p_{\mathrm{\mathrm{BQ}}}$ is non-vanishing is the reason that a decoding of ME and QAC-ME is necessary. Second, note that in both the ME and QAC-ME cases, the optimal penalty values tend to keep the probability $p_{\mathrm{\mathrm{BQ}}}$ of broken qubits fairly constant over the range of clause densities $\alpha$ we have studied. The optimal penalty values can thus be converted into an optimal broken qubits probability that depends on the set of instances chosen and on the implementation (ME or QAC-ME) considered. Third, note that $p_{\mathrm{\mathrm{BQ}}}$ is typically smaller for ME than for QAC-ME, which is unsurprising given that the QAC step in QAC-ME adds a second way for qubits to break, in addition to ME. Fourth, note that $p_{\mathrm{\mathrm{BQ}}}$ is typically larger with nonuniform penalties or weighted instances. The intuitive reason for this is that nonuniform penalties are by design weaker for qubits that are less strongly coupled. This facilitates the appearance of more broken qubits. This is consistent with the experimental finding of the previous section that EM decoding results in a larger advantage in these cases, simply because there are more broken qubits to be optimized. 
Finally, note that the differences in performance of the various decoding strategies considered in Sec.~\ref{sec:expDS} can be related to $p_{\mathrm{\mathrm{BQ}}}$.

%%%%%%%%%%%%%%%%%%%%%%%%%%%%%%%%%%%%%%%%%
\section{Testing temperature and noise dependence via simulated quantum annealing}
\label{sec:NS}
%%%%%%%%%%%%%%%%%%%%%%%%%%%%%%%%%%%%%%%%%

To gain a better understanding of the mechanisms that are most responsible for the experimental success of QAC, we conducted numerical simulations that illuminate the role of temperature and ICE. A master equation simulation treating the device as an open quantum system \cite{ABLZ:12-SI} is unfeasible due to the large number of qubits involved in our experiments. Hence we resort to a semi-dynamical approach based on quantum Monte Carlo. The approach, called simulated quantum annealing (SQA) \cite{sqa1,Santoro,q108}, attempts to mimic the evolution of an open quantum system where stochastic processes (typical of a noisy regime) dominate the dynamics, and essentially provides the instantaneous Gibbs state along the annealing evolution (see Appendix~\ref{app:SQA} for details). This approach has successfully reproduced the success probability distribution of the D-Wave devices for random Ising instances \cite{q108}, although it does not provide a complete description when one also accounts for ground state degeneracy and excited states \cite{Albash:2014if}, or open system dynamics \cite{Albash:2015d}.

Our strategy is not to use SQA as a faithful model for the DW2 device. We thus do not attempt to tune the SQA parameters to find a best fit for the experimental results. Rather, we use SQA as a numerical tool to explore the response of our QAC-ME scheme to changes of various important quantities that cannot be controlled experimentally, specifically temperature and noise. The noise level $\chi$ is the standard deviation of statistically independent Gaussian noise added to all physical couplings $J_{ij}$ and local fields $h_i$, and is introduced to reproduce ICE. We performed simulations for different values of the number of sweeps (a sweep is a single update of all spins), but found that the results have a very weak dependence on this number, so it is not a crucial parameter in our analysis. On the other hand, the temperature and the noise level heavily affect our numerical results. Figure~\ref{fig:num_SQA} summarizes our numerical results for the ``embeddable planted'' instances [compare with the experimental results shown in Fig.~\ref{fig:QAC-ME_best-EmbeddablePlanted}]. It shows that QAC-ME becomes competitive with respect to ME and D only for sufficiently large temperature and/or noise. We regard this result as strong evidence that the QAC schemes are particularly efficient in reducing the amount of errors due to thermal noise and imprecisions in parameter settings.  

%%%%%%%%%%%%%%%%%%%%%%%%%%%%%%%%%%%%%%%%%
\section{Conclusions and Discussion}
\label{sec:CD}
%%%%%%%%%%%%%%%%%%%%%%%%%%%%%%%%%%%%%%%%%

In order to exploit the potential advantages of quantum annealing over classical solvers in solving ``non-native'' optimization problems they must accommodate a minor embedding (ME) step that maps the given problem to the given device hardware graph. In this work we introduced a new quantum annealing correction (QAC) scheme that is compatible with the minor embedding step---the ``square code''---and studied its efficacy experimentally using the DW2 device. We demonstrated that this hybrid QAC-ME scheme significantly boosts performance by applying it to frustrated Ising problems with known (``planted'') ground states. In the construction of such instances, a controllable clause density parameter $\alpha$ tunes both hardness and frustration. The largest performance boost (measured in terms of the success probability of finding the ground state) via QAC-ME occurs at the critical clause density $\alpha_{\mathrm{c}}$, where  hardness 
%and frustration both peak
peaks. 
%\mg{I don't see that $\alpha_{\mathrm{c}}$ is also the location of the frustration peak; see Fig.~\ref{fig:frust-range}, where for planted and weighted planted the peak is $0.7$, for embeddable planted at $0.47$. This disagrees with the peak location in Fig.~\ref{fig:planted}. Therefore we can't say that ``This is consistent with the intuitive expectation that the performance of quantum annealers after ME is especially sensitive to frustration."}
%This is consistent with the intuitive expectation that the performance of quantum annealers after ME is especially sensitive to frustration.

We have also considered problem instances that are not only embeddable into the square code but also directly into the physical ``Chimera'' hardware graph of the DW2 device. This allows for a comparison of the direct, ME and QAC-ME embeddings for the same problem instance. As expected ME behaves poorly with respect to the direct embedding. The performance improvement achieved with QAC-ME is particular significant at $\alpha_{\mathrm{c}}$ and suffices to match that of the direct embedding. This is a proof-of-concept that QAC-ME can (and should) be used to overcome the limitations of minor embedding on quantum annealers. QAC-ME thus has the potential to extend the competitiveness of these devices beyond natively embeddable optimization problems.

Several questions remain to be addressed in future work, requiring the use of next-generations quantum annealers. The most important of these is arguably the scaling of error-corrected quantum annealing. To properly address this question we need to be able to embed larger logical problems (the square code considered in this work implements up to $120$ encoded qubits) on devices with a larger number of physical qubits. 
An important aspect for any scaling study is the determination of the ``optimal" implementation. In the case of a direct embedding, this boils down to the question of determining the optimal annealing time (see, e.g., Ref.~\cite{Hen:2015rt}). 
Optimality in minor embedded quantum annealing is a broader question. This is due to the freedom in choosing the minor embedding and, once this is given, the exact map between the logical and the physical couplings. In particular, the performance of minor embedded quantum annealing is sensitive to the choice of energy penalties, and determining the optimal choice of penalties is important both for determining the scaling of physical quantum annealing and for optimizing the performance on practical applications. 
%The choice of energy penalties and the decoding step both play important roles in minor embedded quantum annealing. 
We have shown that 
%a significant improvement results when these procedures are chosen carefully. In particular, 
choosing nonuniform penalties that depend on the local strength of the logical couplings can give a significant advantage with respect to a uniform choice. 

The optimal choice of the decoding procedure is another important question. We found global energy minimization over broken qubits to be the optimal choice for our problem instances. We showed that in order for this decoding technique to be performed at a negligible computational cost, the typical probability to have broken logical or encoded qubits should be lower than the per-site percolation threshold of the logical or encoded graph. It turns out that the square code, for the frustrated problems we have considered, is efficiently and optimally decodable through energy minimization even when its size is scaled.
Our results demonstrate that the square code on the tested DW2 lies in a ``decodable" regime, with a sufficiently small number of broken qubits when the penalties are optimized. It is important to study whether optimal decoding can be performed efficiently in other ME and QAC-ME schemes. Whenever this is not possible, efficient, albeit suboptimal, strategies should be developed. To this end, it is likely that techniques developed in the decoding of Low Density Parity Check (LDPC) codes \cite{Gallager:1962vn} will find important applications in quantum annealing too. In particular, suboptimal decoding through belief propagation \cite{MacKay:book} is a promising approach to successfully decode ME quantum annealing when the broken qubit probability is larger than the decoding threshold.

More generally, it is important to understand how the performance of minor embedded quantum annealing degrades with the increase of the size of the logical or encoded qubits. It is likely that the larger the logical qubits required for a given minor embedding are (this will be the case for increasingly larger logical problems) the more important QAC-ME will be as a tool to prevent the degradation of the performance of quantum annealers on minor embedded implementations. 

While our experimental conclusions are naturally limited to the particular set of problem instances we studied, we may extrapolate that our results more generally support the effectiveness of the QAC-ME strategy. Confirming this can have important implications in scaling quantum annealing to solve larger and harder optimization problems of practical interest, and we expect QAC-ME techniques to play a central role in achieving this goal.

\section{Acknowledgements}

We thank Anurag Mishra and Leonid Pryadko for helpful discussions. Access to the D-Wave Two was made available by the USC-Lockheed Martin Quantum Computing Center. This research used resources of the Oak Ridge Leadership Computing Facility at the Oak Ridge National Laboratory, which is supported by the Office of Science of the U.S. Department of Energy under Contract No. DE-AC05-00OR22725. Part of the computing resources were provided by the USC Center for High Performance Computing and Communications. 
I.H. and D.A.L. acknowledge
support under ARO grant number W911NF-12-1-0523. The work of W.V., T.A. and D.A.L. was supported under ARO MURI Grant No. W911NF-11-1-0268.  

%\bibliography{refs}
\bibliography{refs}

\newpage 
\clearpage

%\newpage
%\onecolumngrid

\appendix

\begin{figure*}[b]
\begin{center}
\subfigure[\ Planted, uniform]{\includegraphics[width=0.48\textwidth]{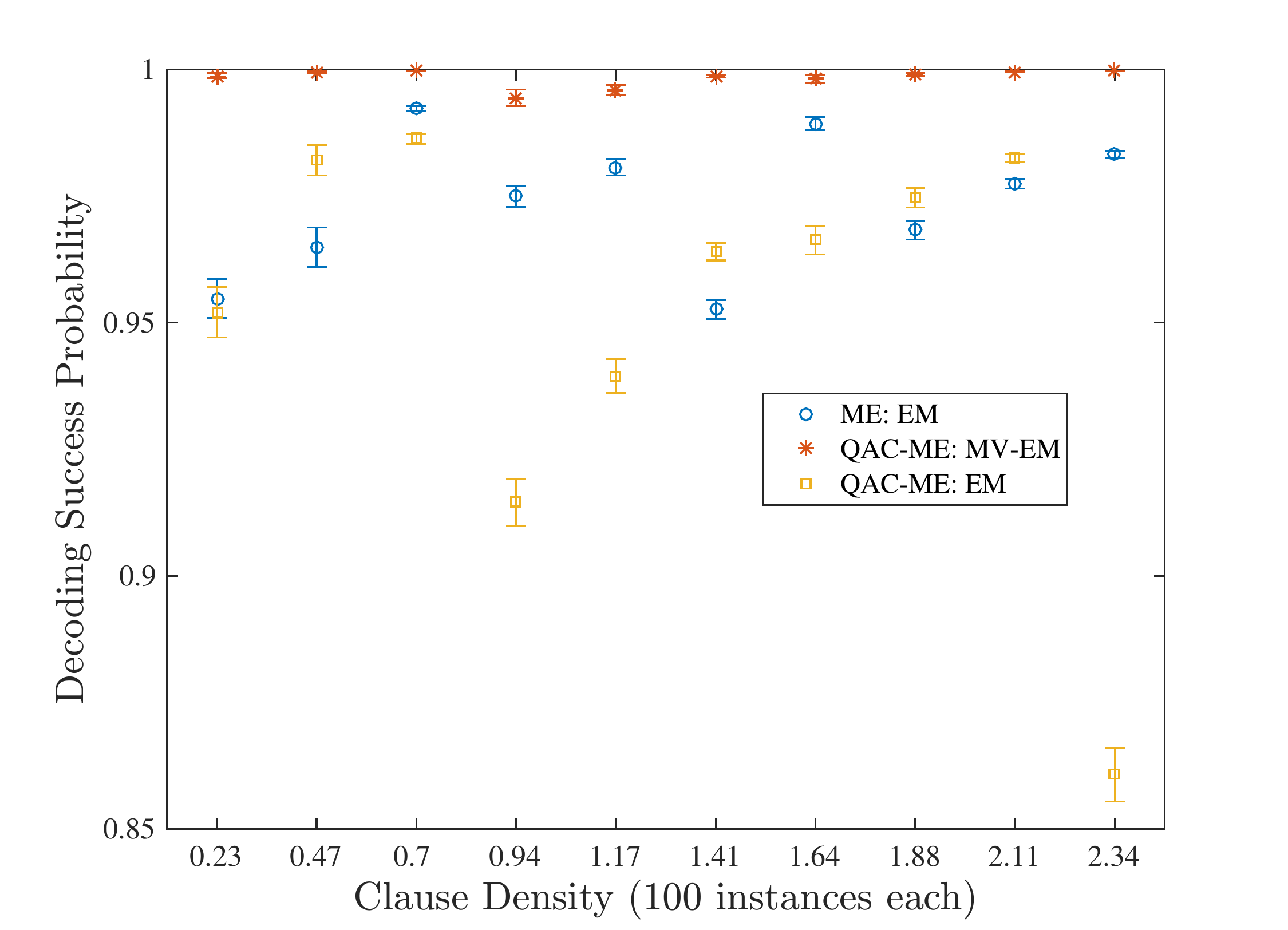}}
\subfigure[\ Planted, nonuniform]{\includegraphics[width=0.48\textwidth]{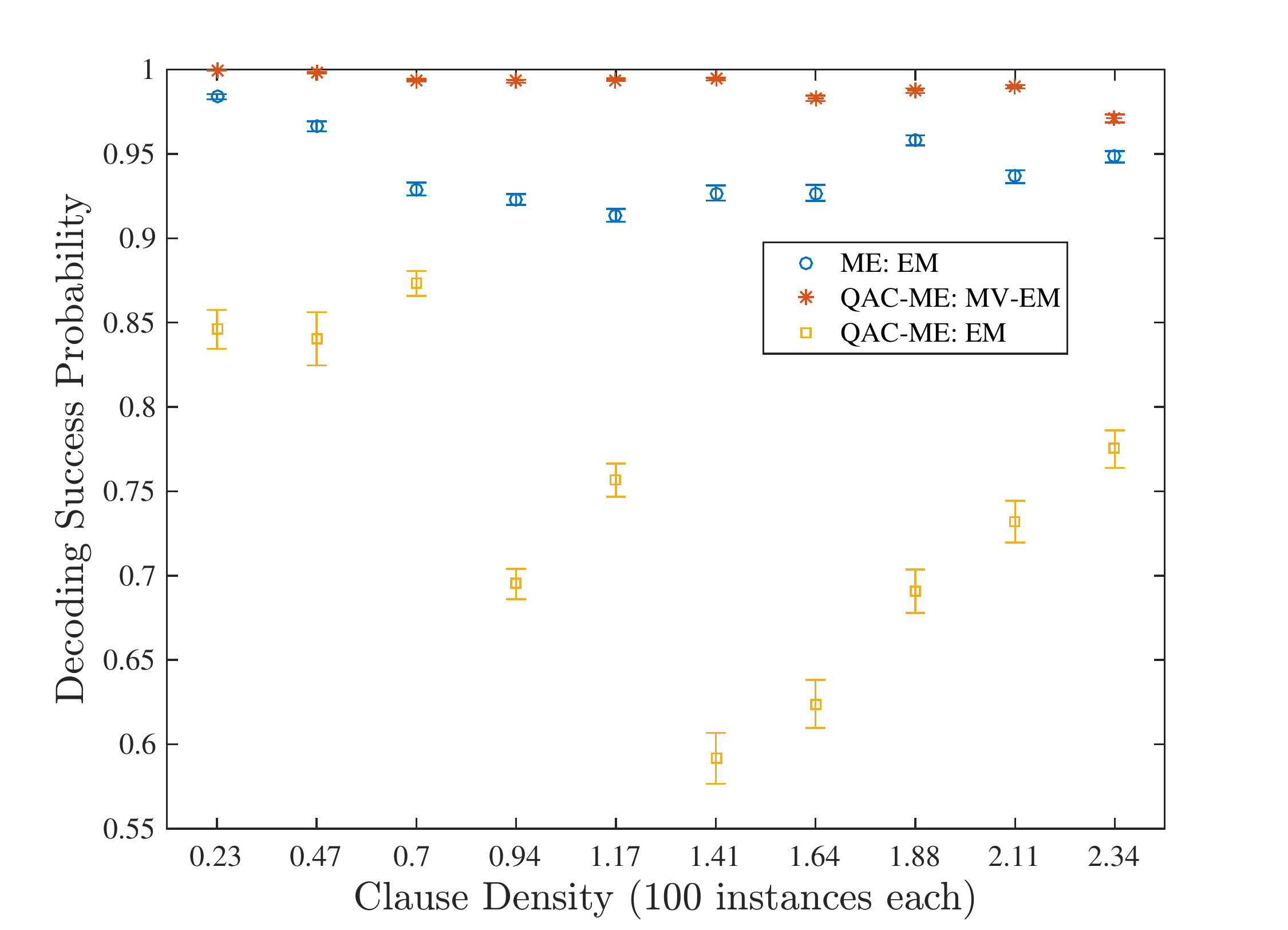}}
\subfigure[\ Weighted planted, uniform]{\includegraphics[width=0.48\textwidth]{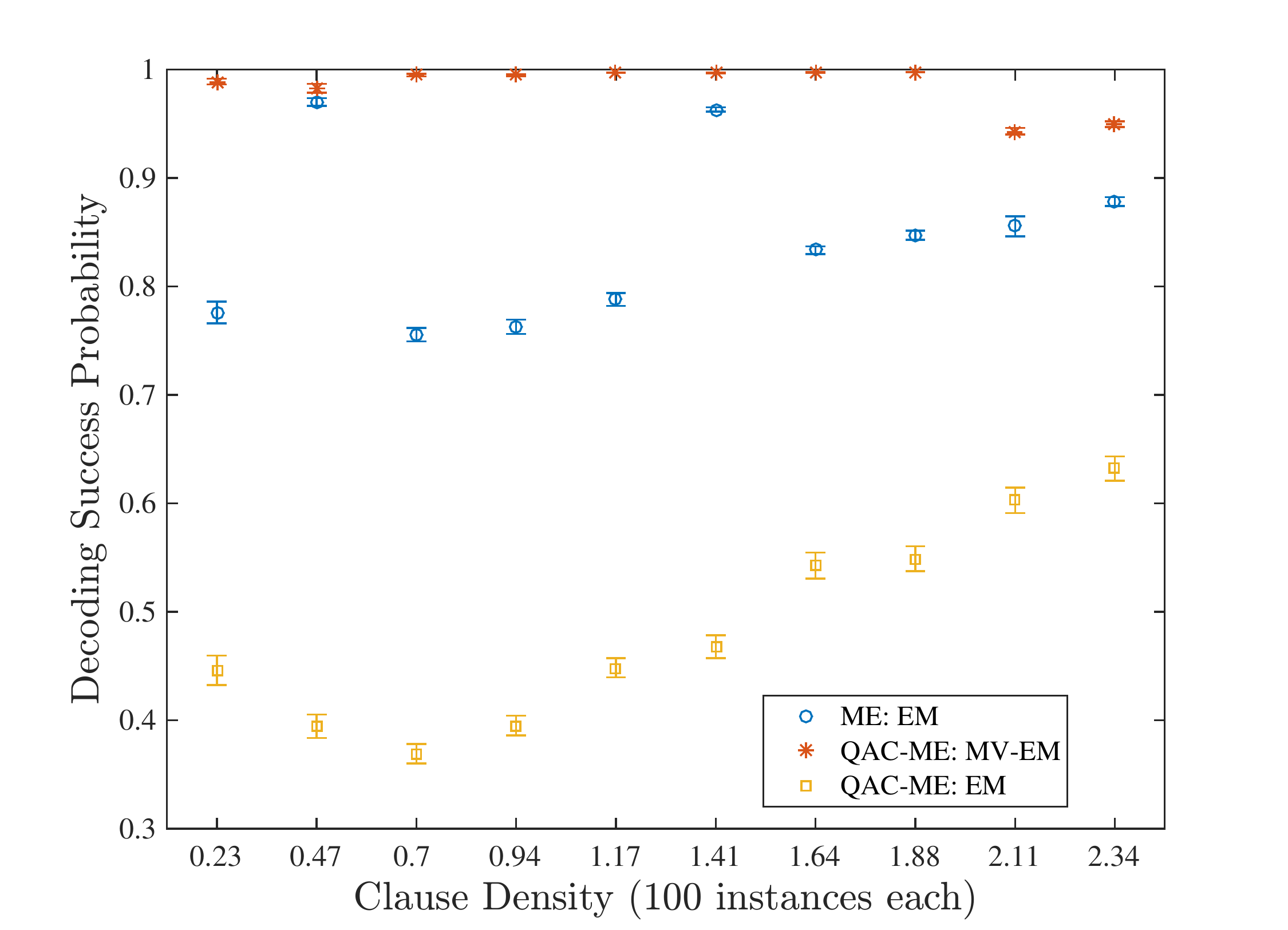}}
\subfigure[\ Weighted planted, nonuniform]{\includegraphics[width=0.48\textwidth]{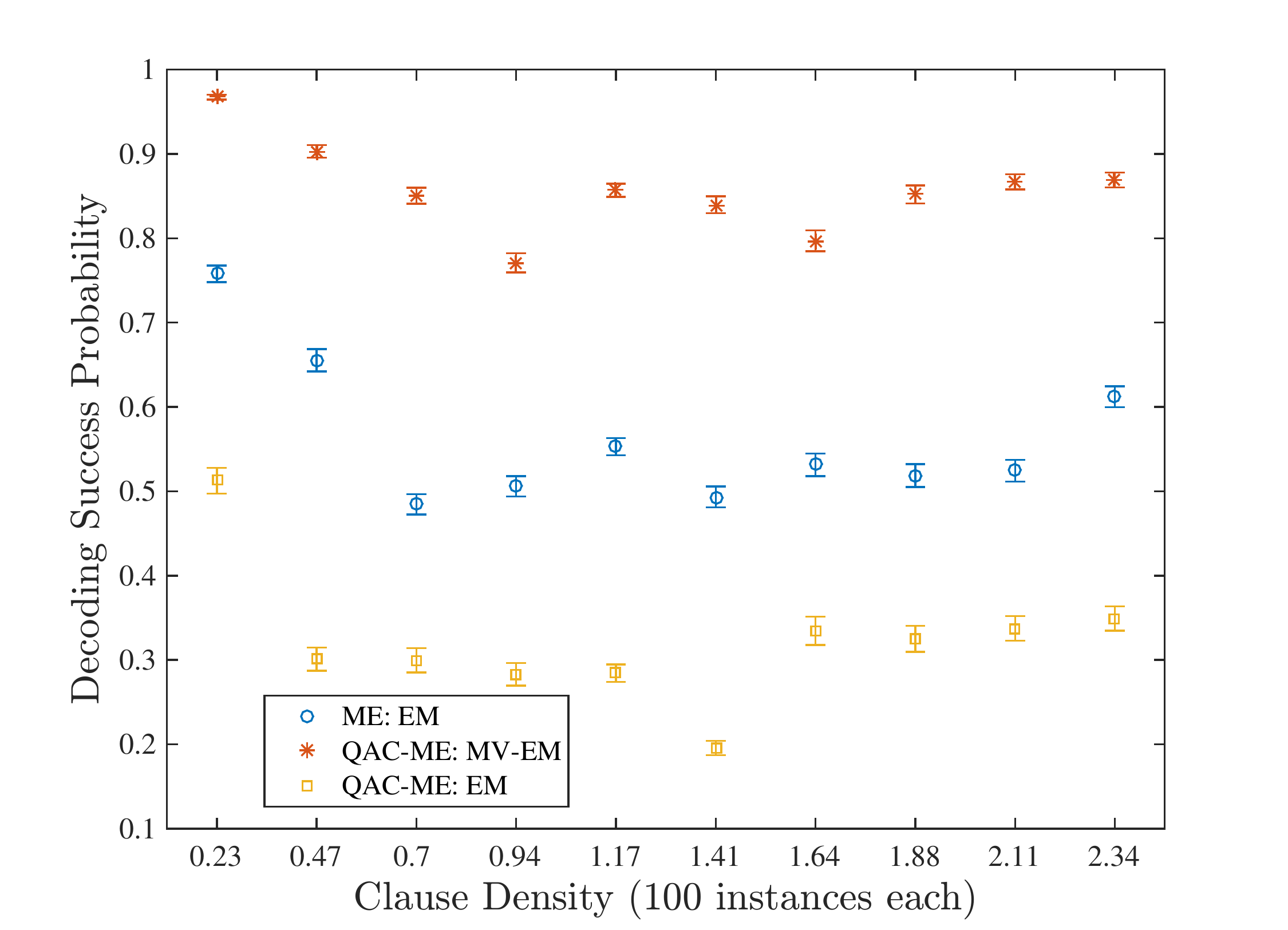}}
\caption{Energy minimization success probabilities of decoding using simulated annealing.} 
\label{fig:deceff}
\end{center}
\end{figure*}
%%%%%%%%%%%%%%%%%%%%

\begin{figure*}[t]
\begin{center}
\subfigure[\ Planted, uniform]{\includegraphics[width=0.48\textwidth]{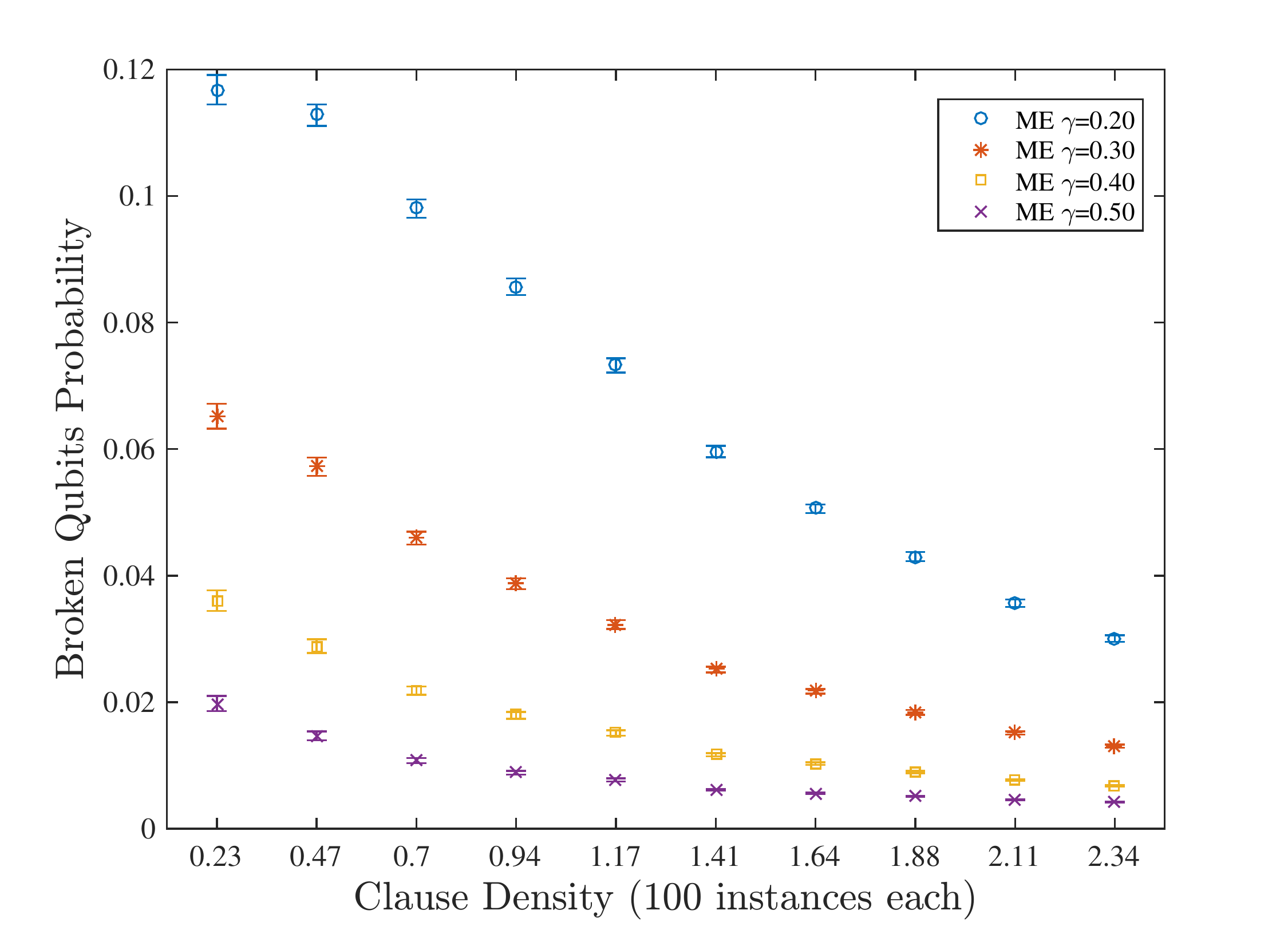}}
\subfigure[\ Planted, uniform]{\includegraphics[width=0.48\textwidth]{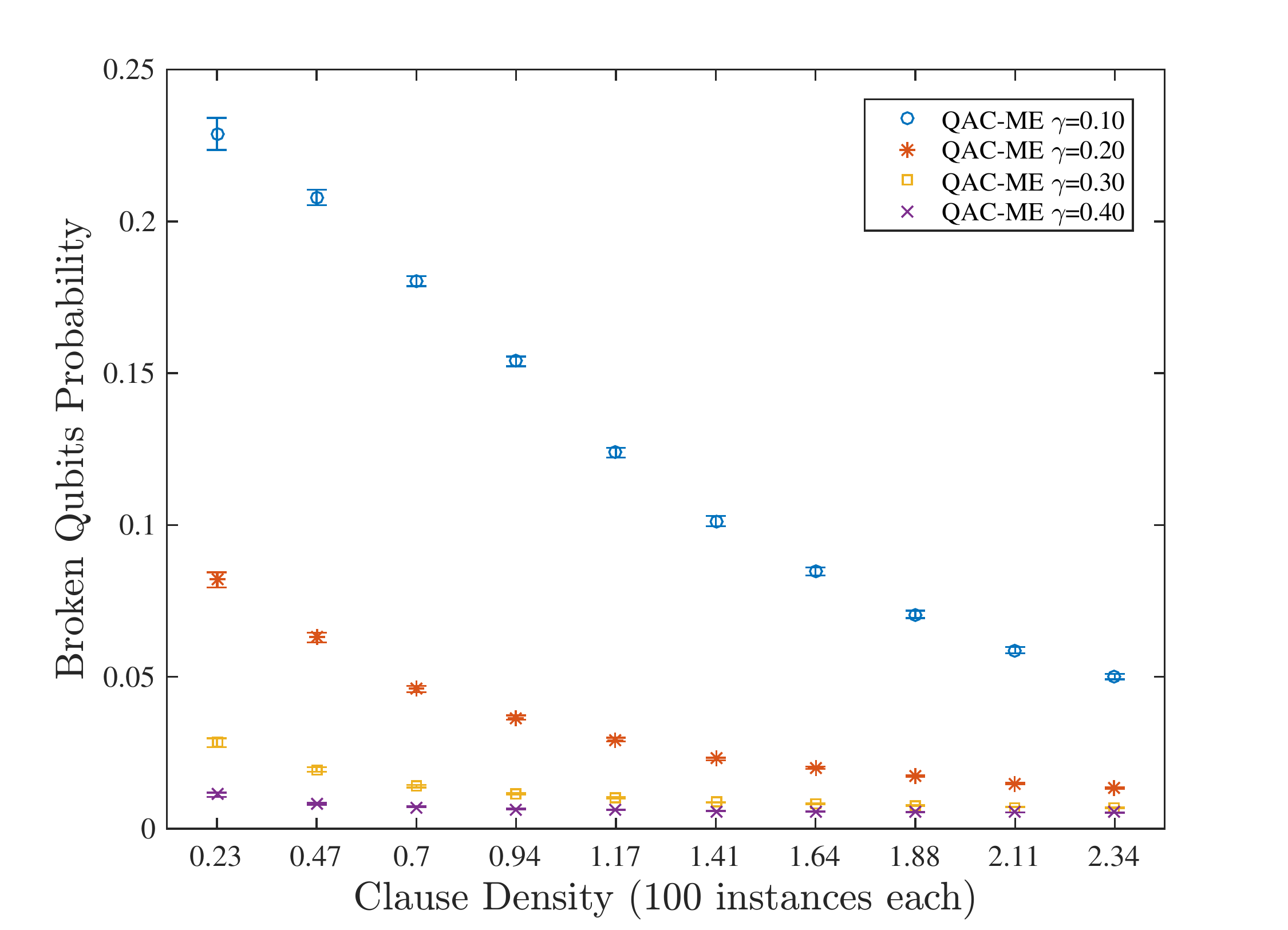}}
\subfigure[\ Planted, nonuniform]{\includegraphics[width=0.48\textwidth]{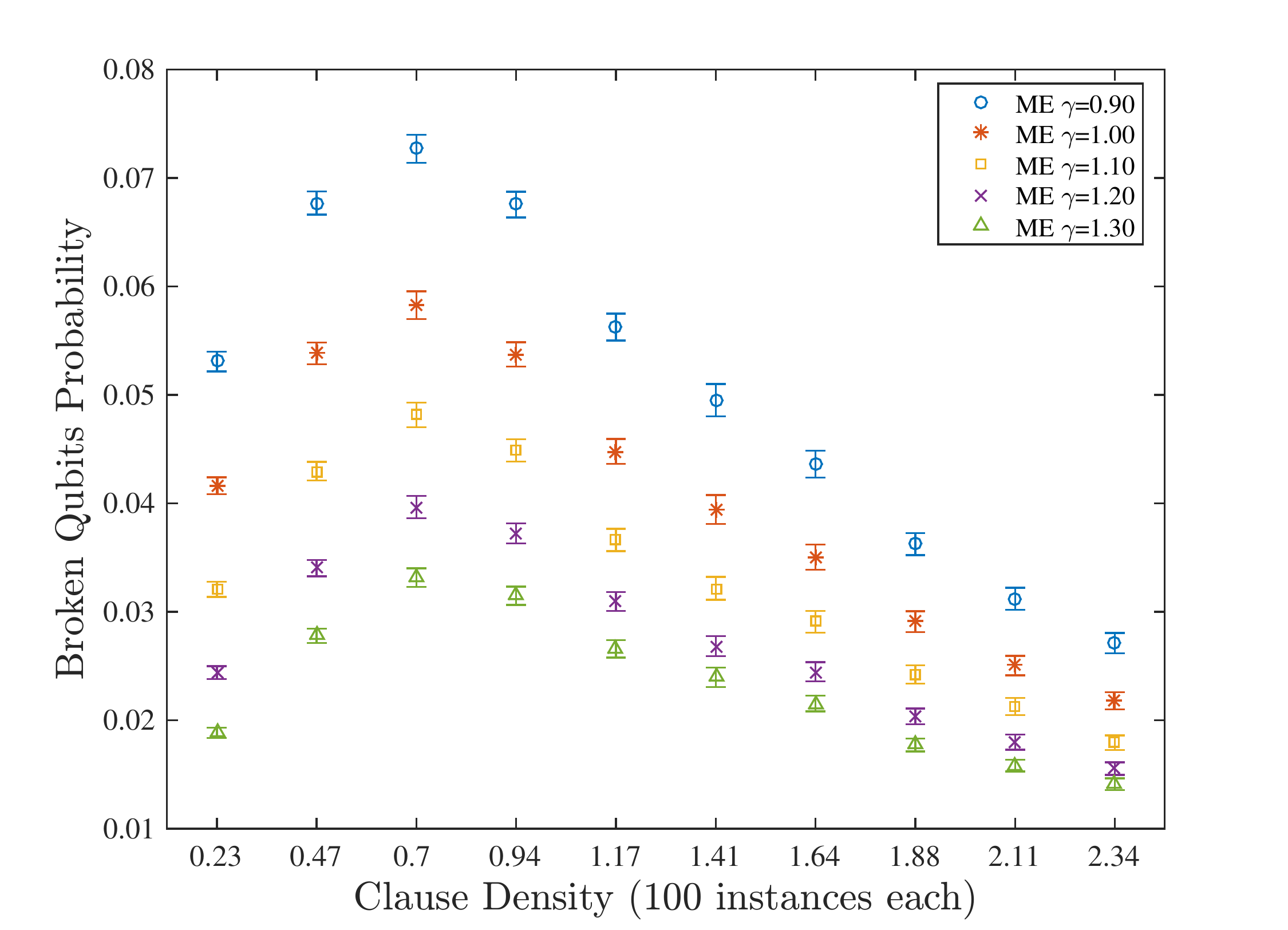}}
\subfigure[\ Planted, nonuniform]{ \includegraphics[width=0.48\textwidth]{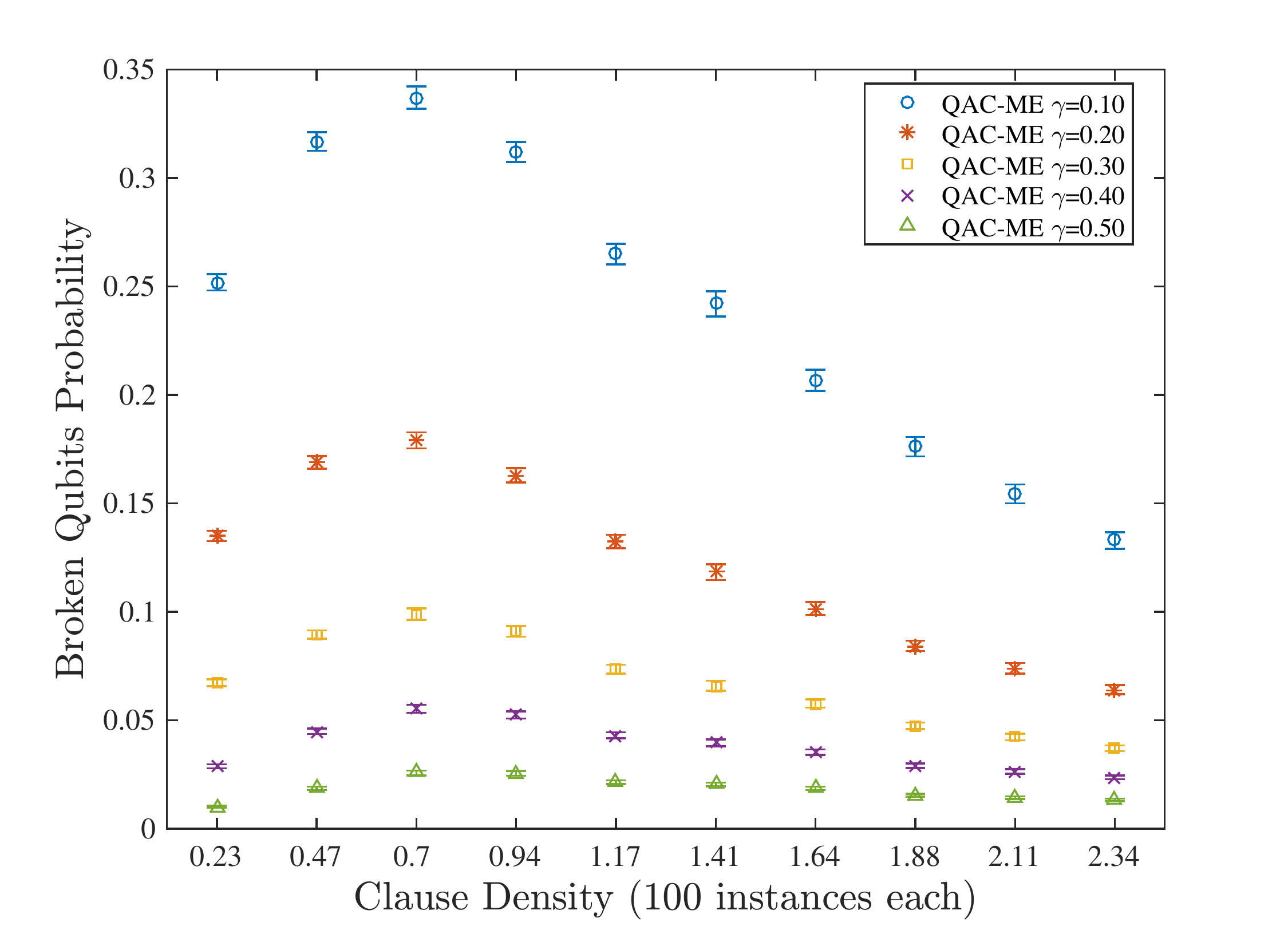}}
\caption{Probability of broken qubits for ME (left) and QAC-ME (right) at different penalty values, for the planted and embeddable planted instances. Top row: uniform. Bottom row: nonuniform. The dependence on the clause density is monotonic in the uniform case and exhibits a maximum in the nonuniform case.}
\label{fig:plantedfull3}
\end{center}
\end{figure*}

\begin{figure*}[t]
\begin{center}
\subfigure[\ Embeddable planted, uniform]{\includegraphics[width=0.48\textwidth]{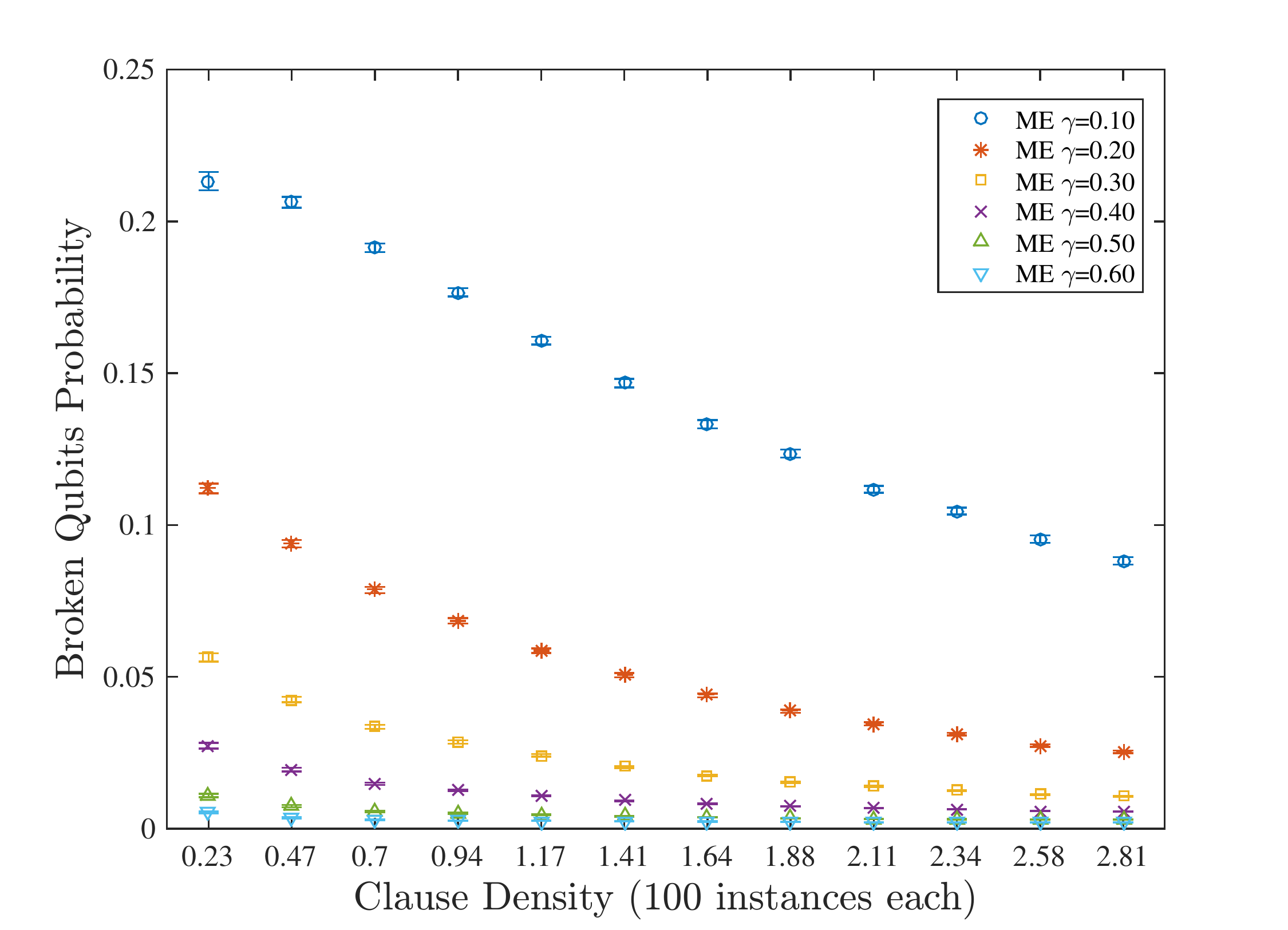}}
\subfigure[\ Embeddable planted, uniform]{\includegraphics[width=0.48\textwidth]{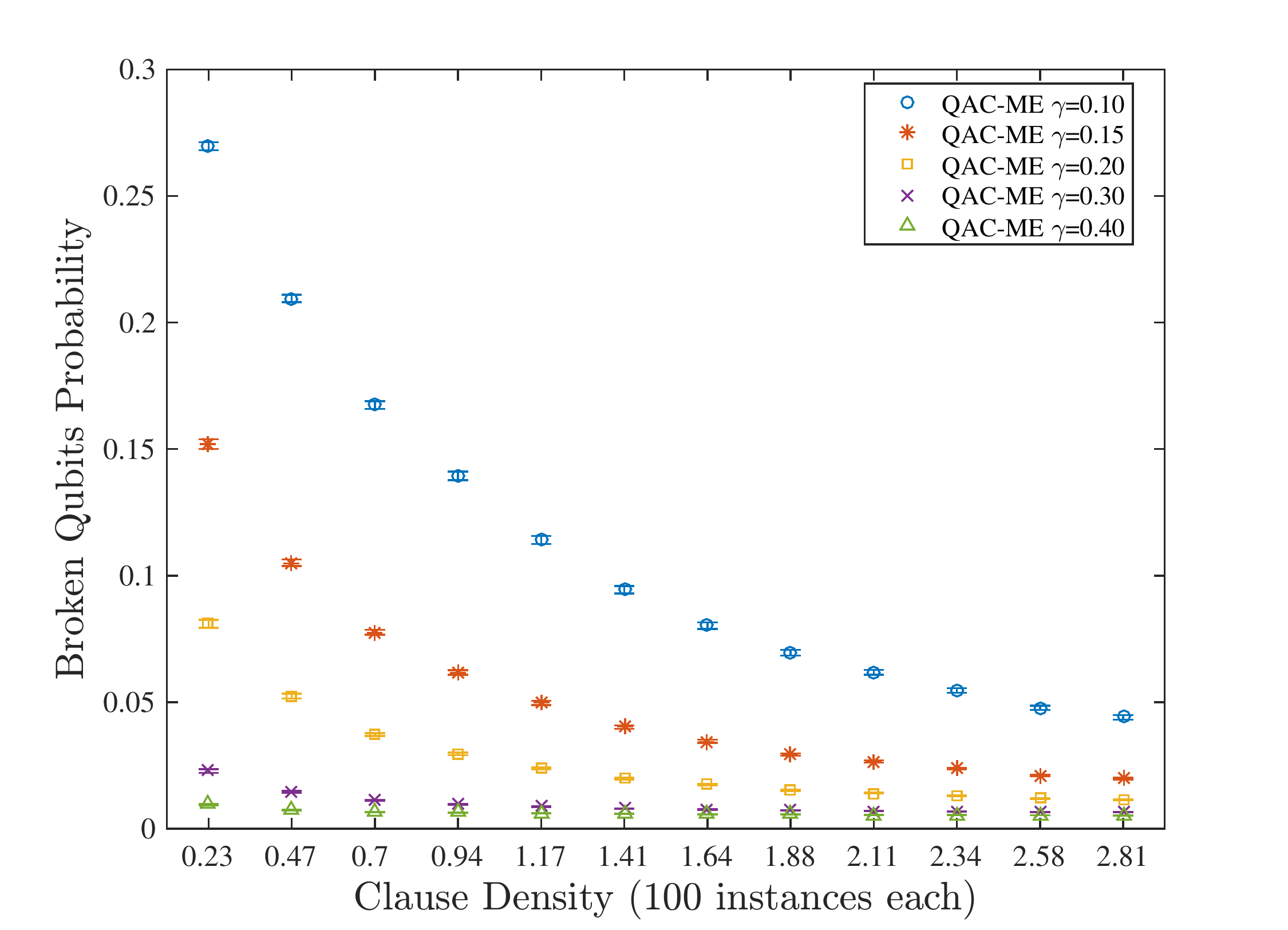}}
\subfigure[\ Embeddable planted, nonuniform]{\includegraphics[width=0.48\textwidth]{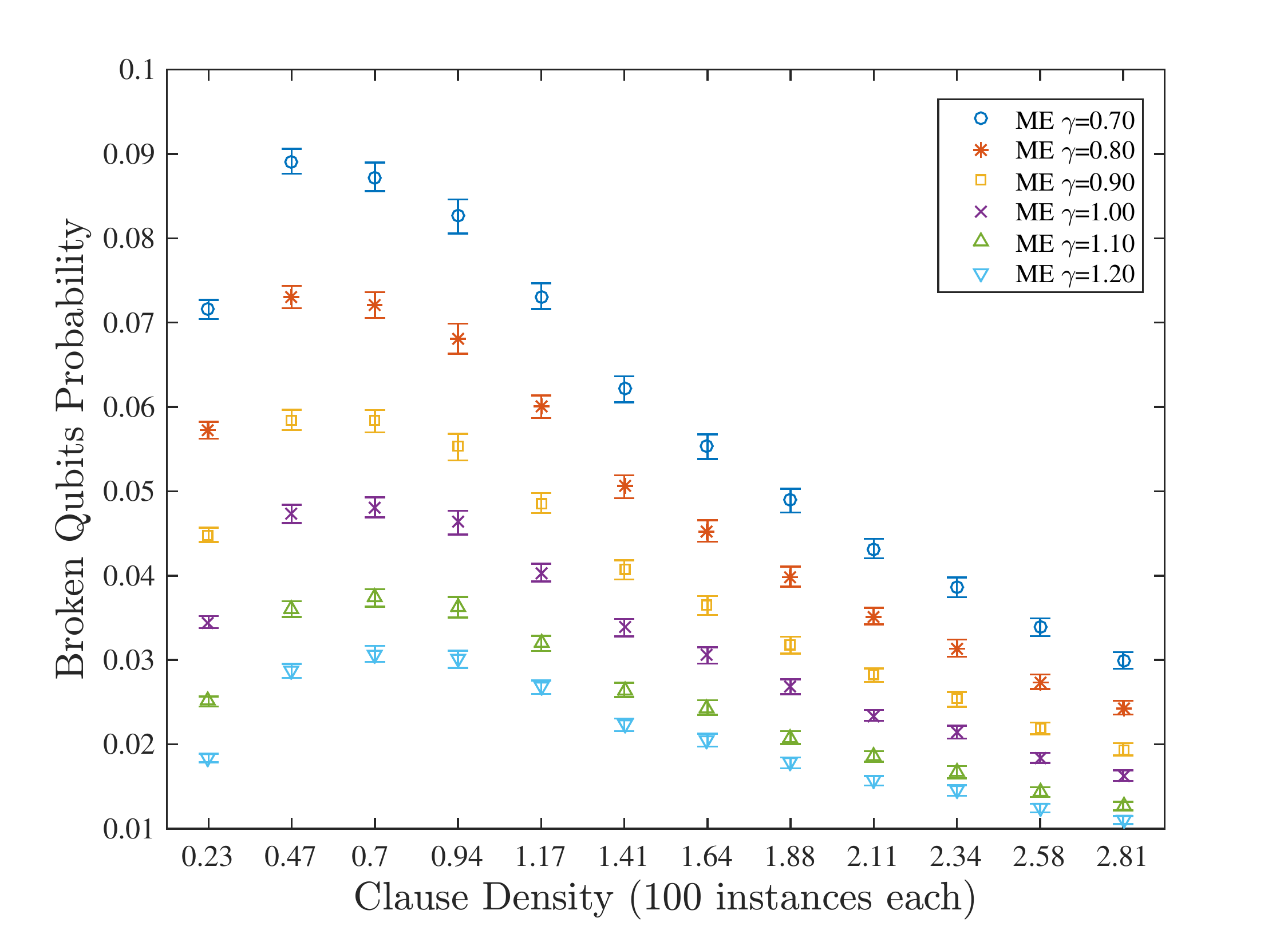}}
\subfigure[\ Embeddable planted, nonuniform]{ \includegraphics[width=0.48\textwidth]{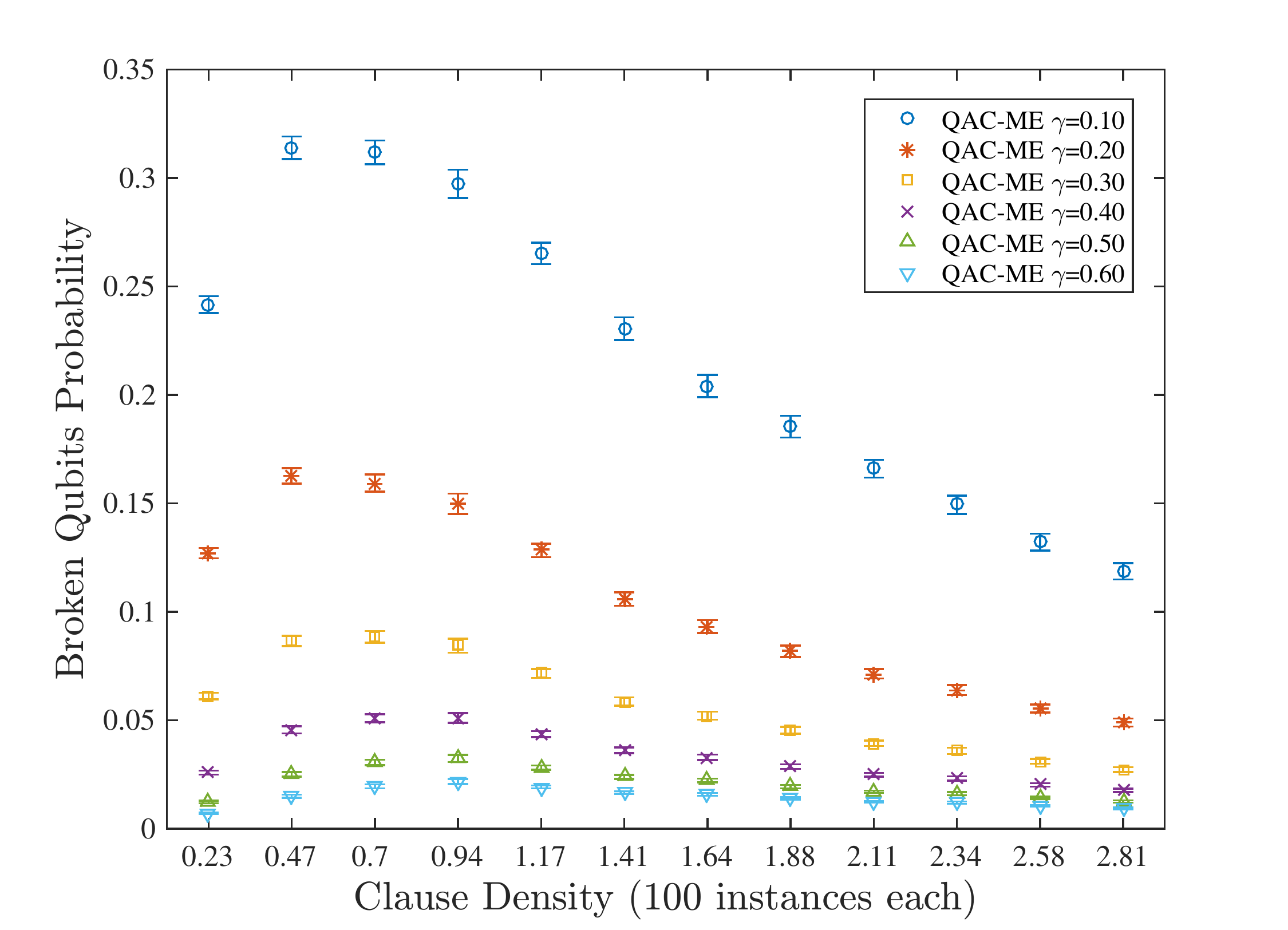}}
\caption{Probability of broken qubits for ME (left) and QAC-ME (right) at different penalty values, for the embeddable planted instances. Top row: uniform. Bottom row: nonuniform. The dependence on the clause density is monotonic in the uniform case and exhibits a maximum in the nonuniform case.} \label{fig:embeddableplantedfull3}
\end{center}
\end{figure*}

\begin{figure*}[t]
\begin{center}
\subfigure[\ Weighted planted, uniform]{\includegraphics[width=0.48\textwidth]{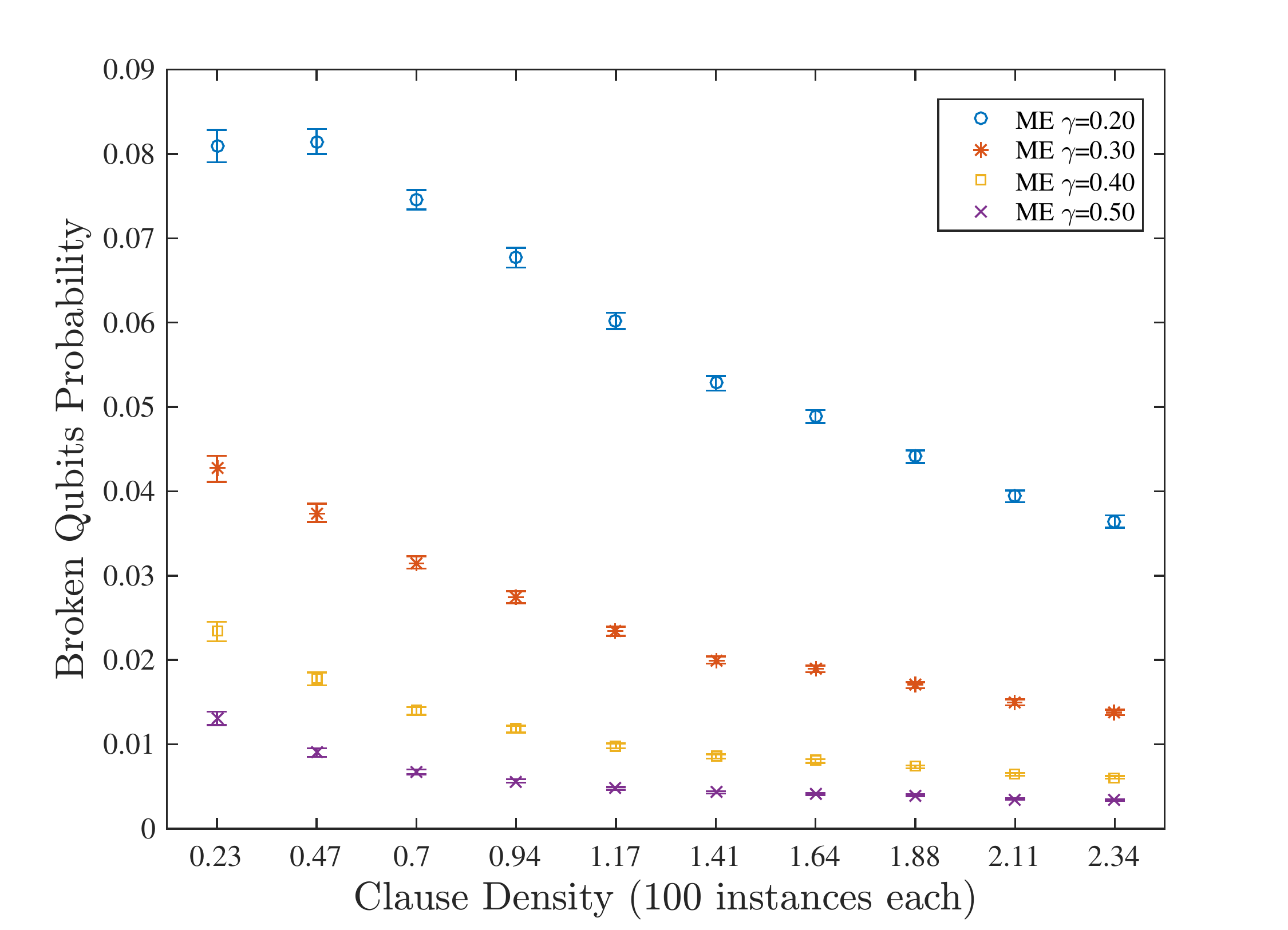}}
\subfigure[\ Weighted planted, uniform]{\includegraphics[width=0.48\textwidth]{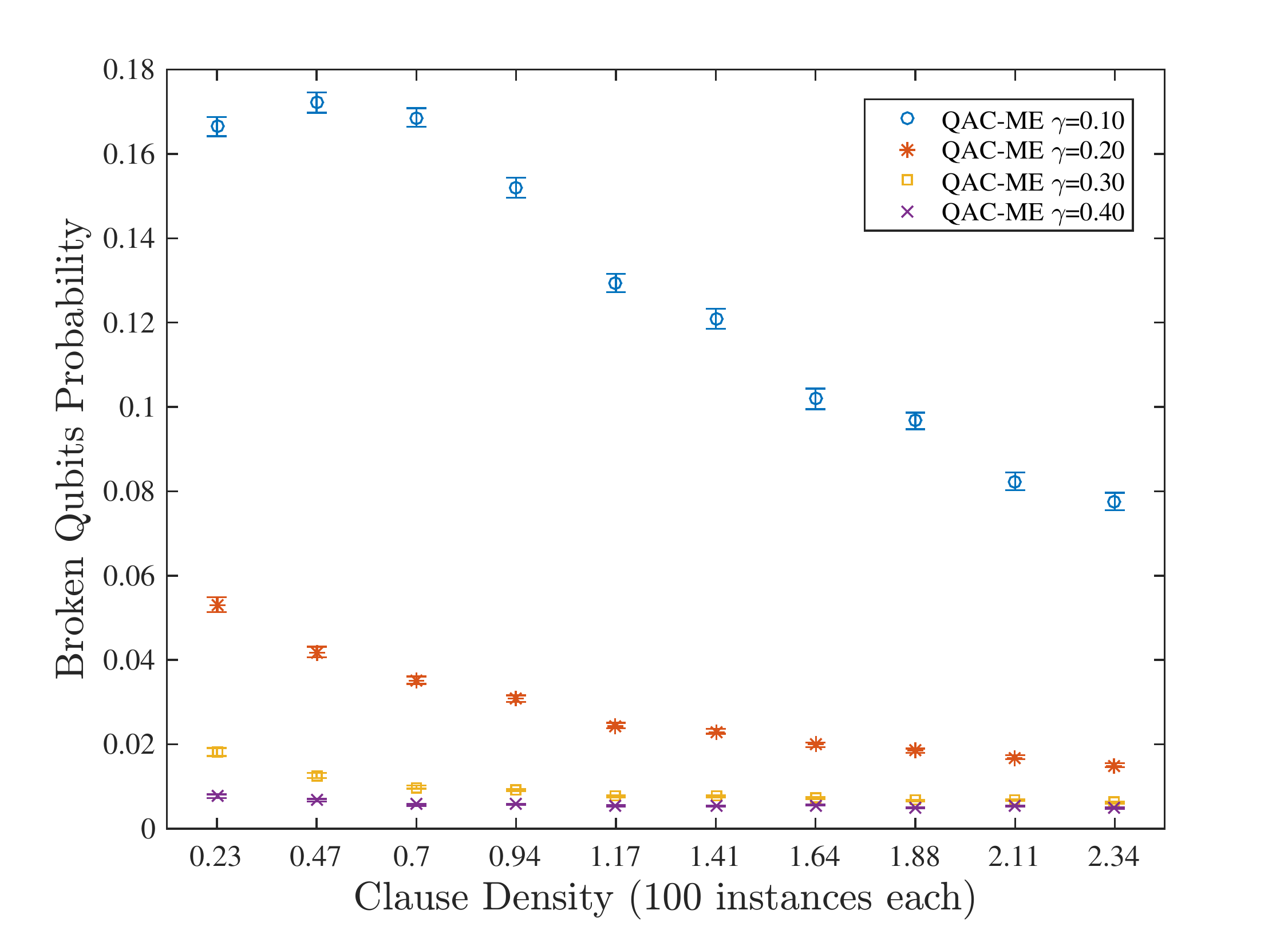}}
\subfigure[\ Weighted planted, nonuniform]{\includegraphics[width=0.48\textwidth]{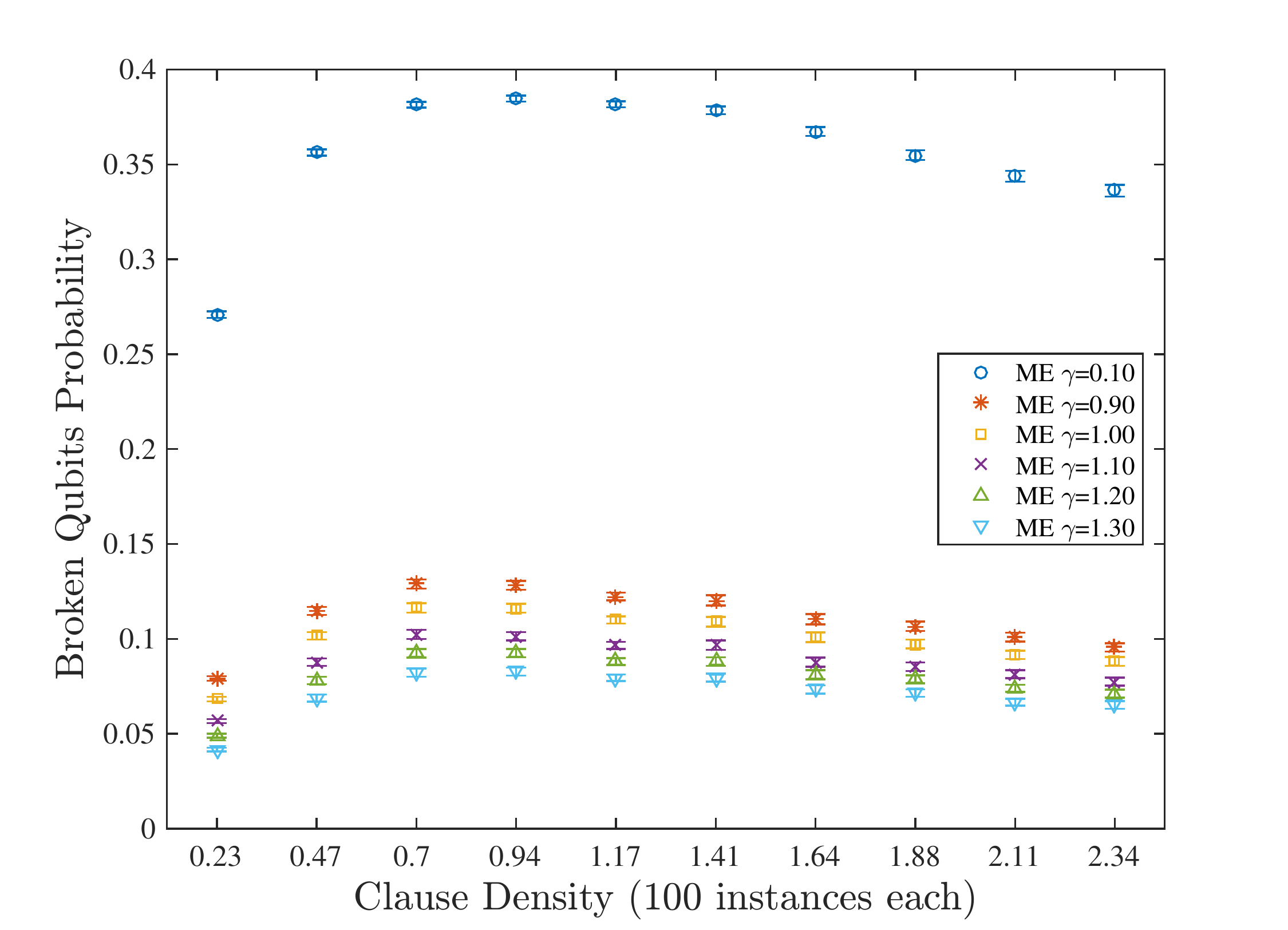}}
\subfigure[\ Weighted  planted, nonuniform]{ \includegraphics[width=0.48\textwidth]{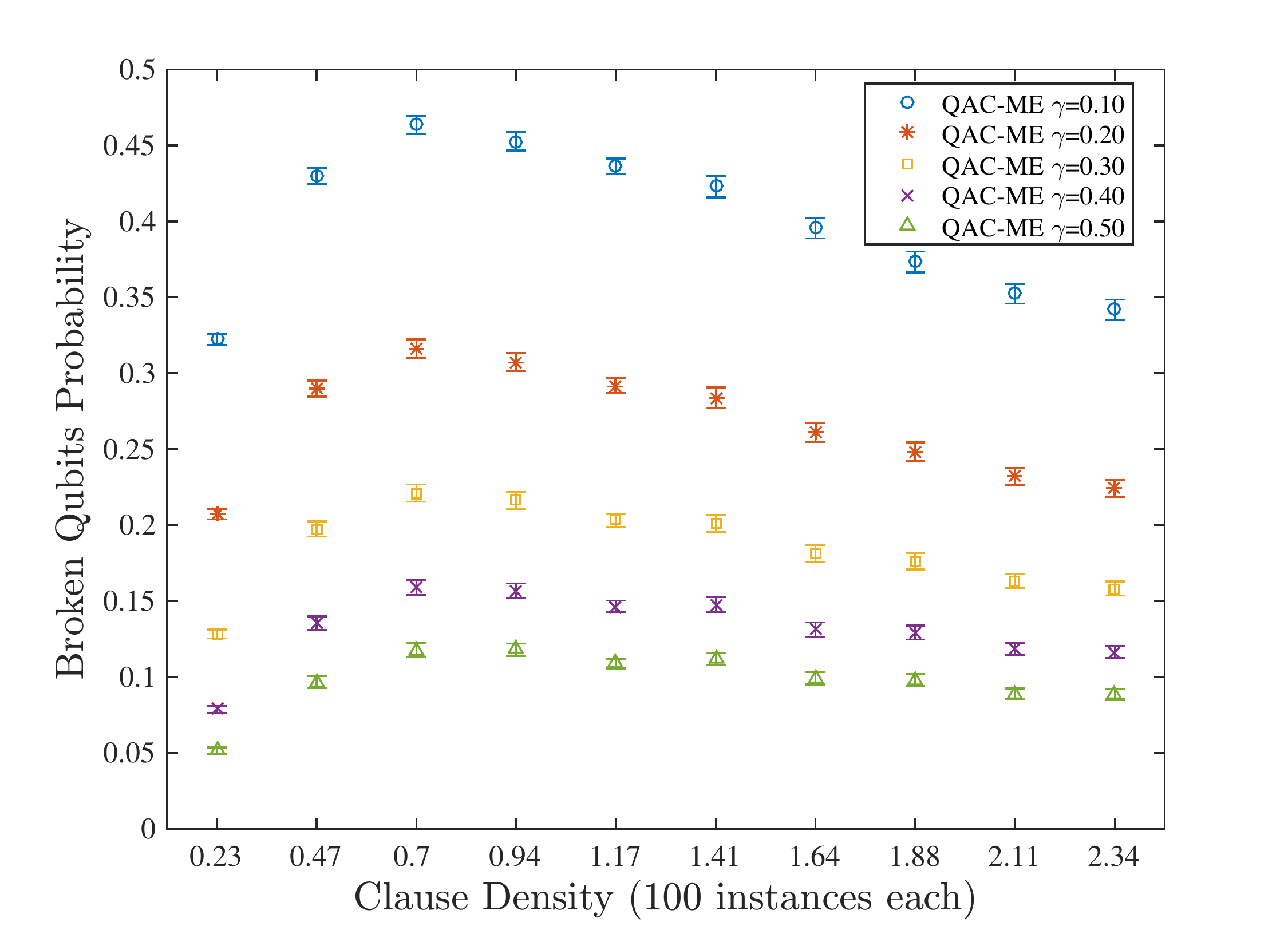}}
\caption{Probability of broken qubits for ME (left) and QAC-ME (right) at different penalty values, for the weighted planted and deformed embeddable instances. Top row: uniform. Bottom row: nonuniform. The dependence on the clause density is monotonic in the uniform case and exhibits a maximum in the nonuniform case.} 
\label{fig:weightedplantedfull3}
\end{center}
\end{figure*}

\begin{figure*}[t]
\begin{center}
\subfigure[\ Deformed embeddable, uniform]{\includegraphics[width=0.48\textwidth]{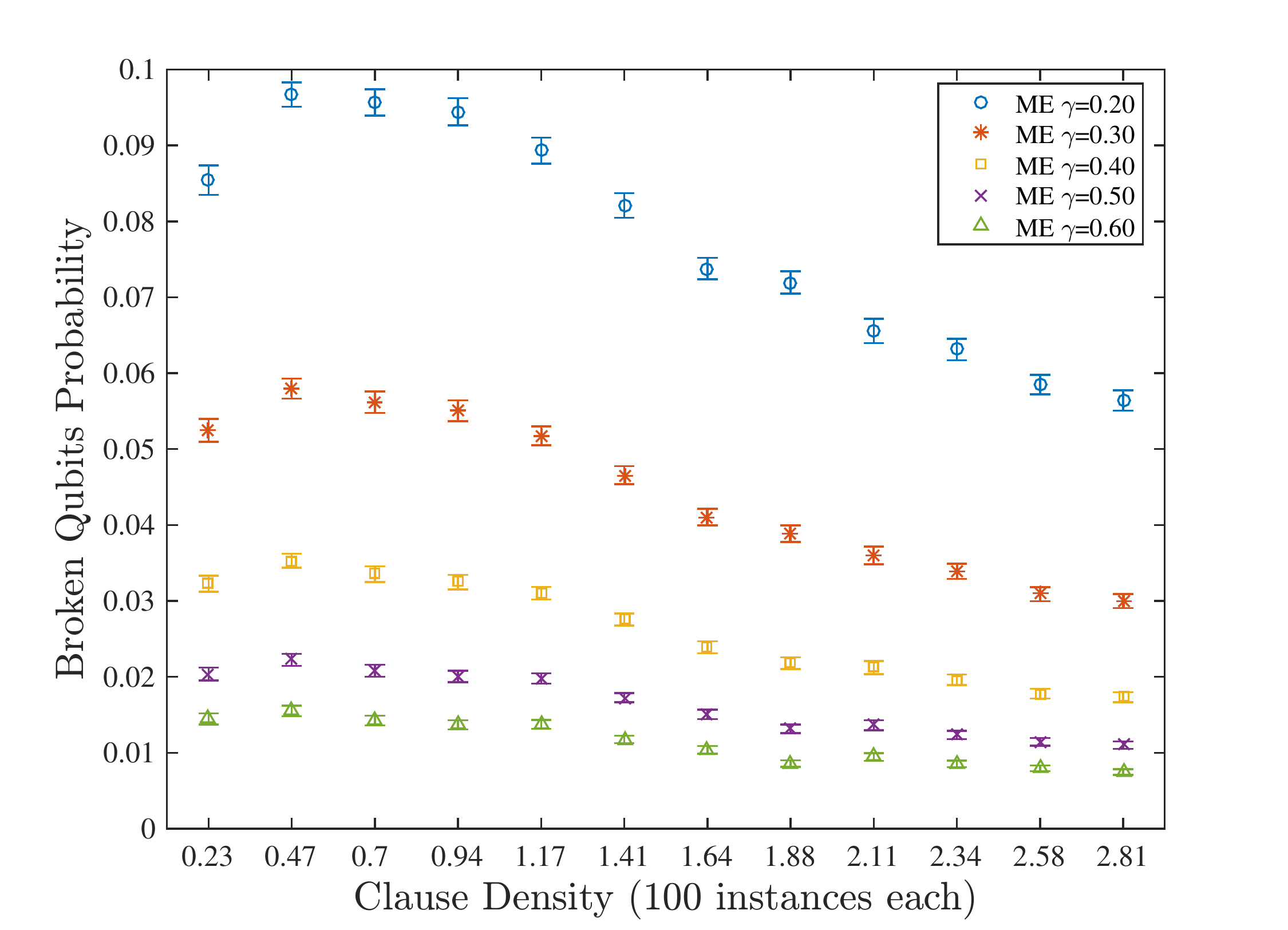}}
\subfigure[\ Deformed embeddable, uniform]{\includegraphics[width=0.48\textwidth]{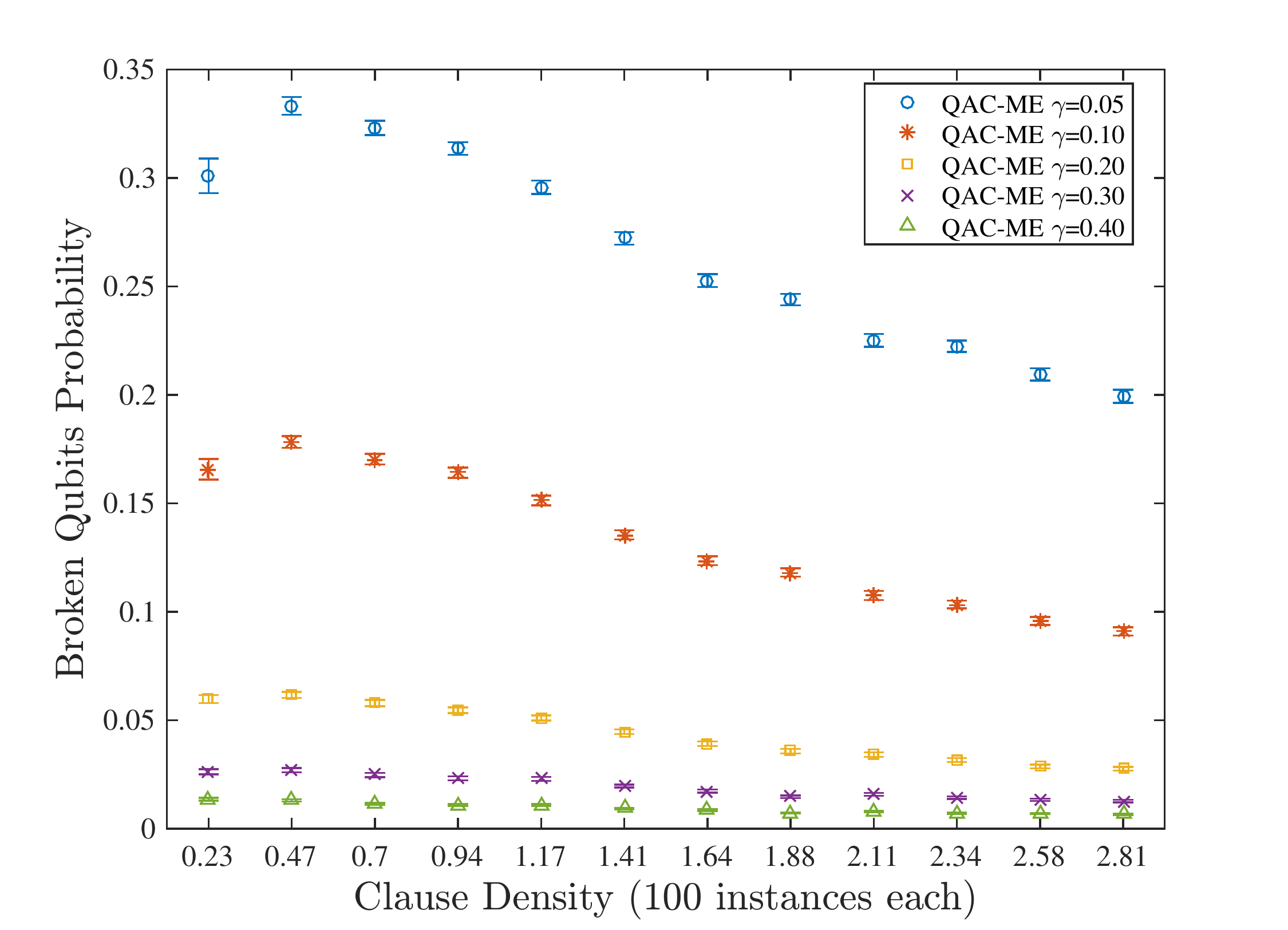}}
\subfigure[\ Deformed embeddable, nonuniform]{\includegraphics[width=0.48\textwidth]{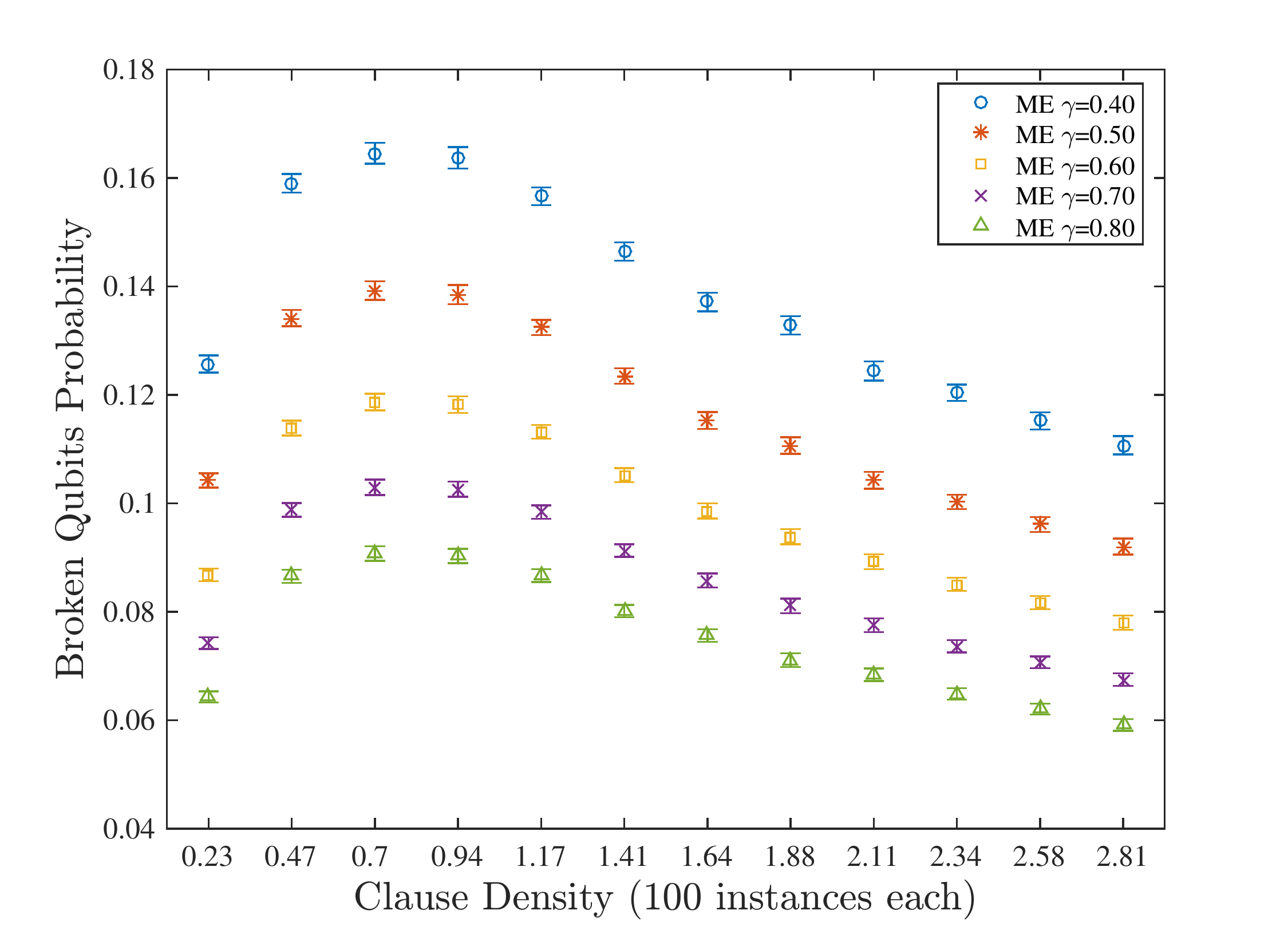}}
\subfigure[\ Deformed embeddable, nonuniform]{ \includegraphics[width=0.48\textwidth]{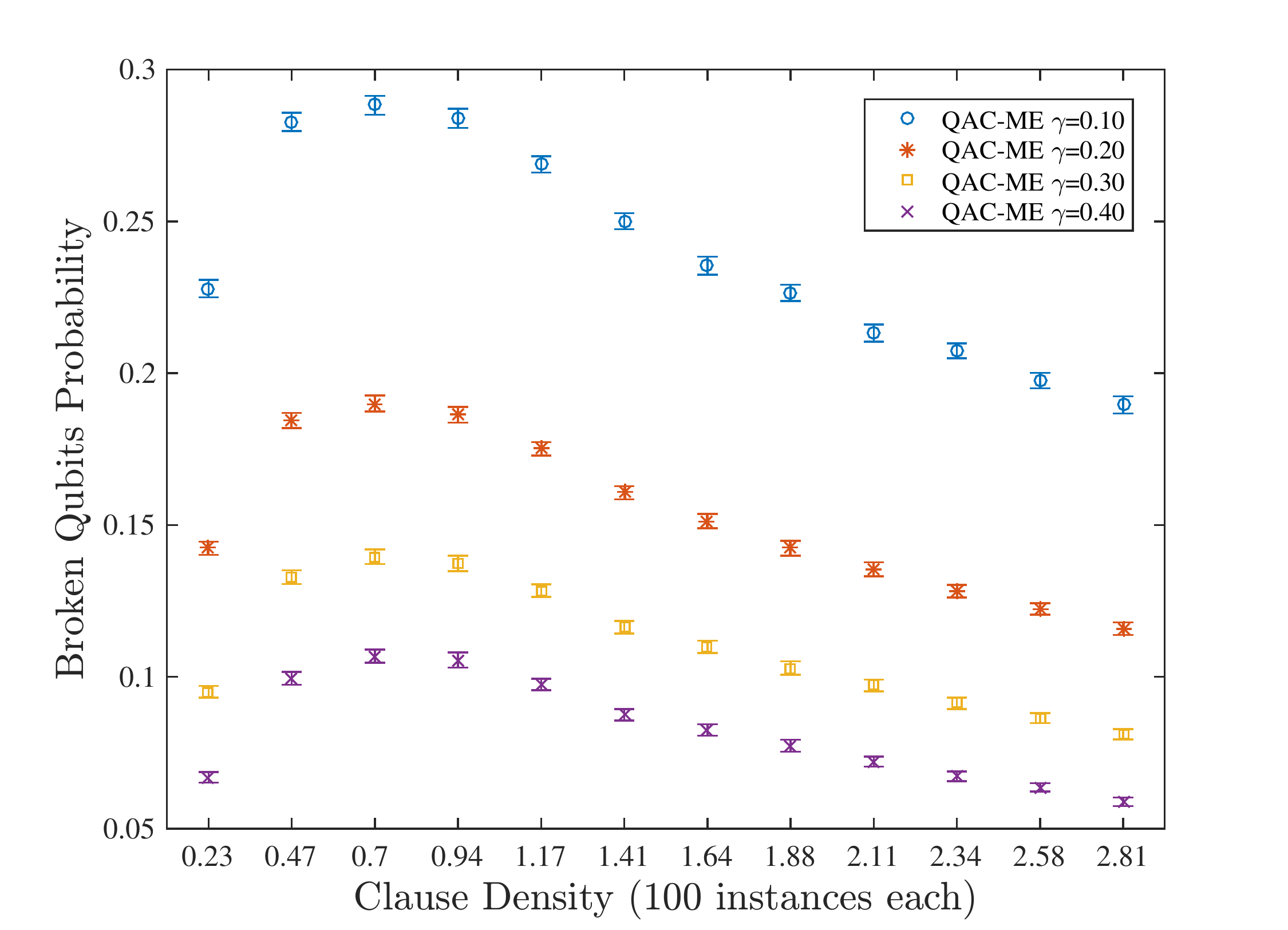}}
\caption{Probability of broken qubits for ME (left) and QAC-ME (right) at different penalty values, for the deformed embeddable instances. Top row: uniform. Bottom row: nonuniform. The dependence on the clause density is monotonic in the uniform case and exhibits a maximum in the nonuniform case.} 
\label{fig:deformedembeddableplantedfull3}
\end{center}
\end{figure*}

%%%%%%%%%%%%%%%%%%%%

\begin{figure*}[t]
\begin{center}
\subfigure[\ Planted]{\includegraphics[width=0.32\textwidth]{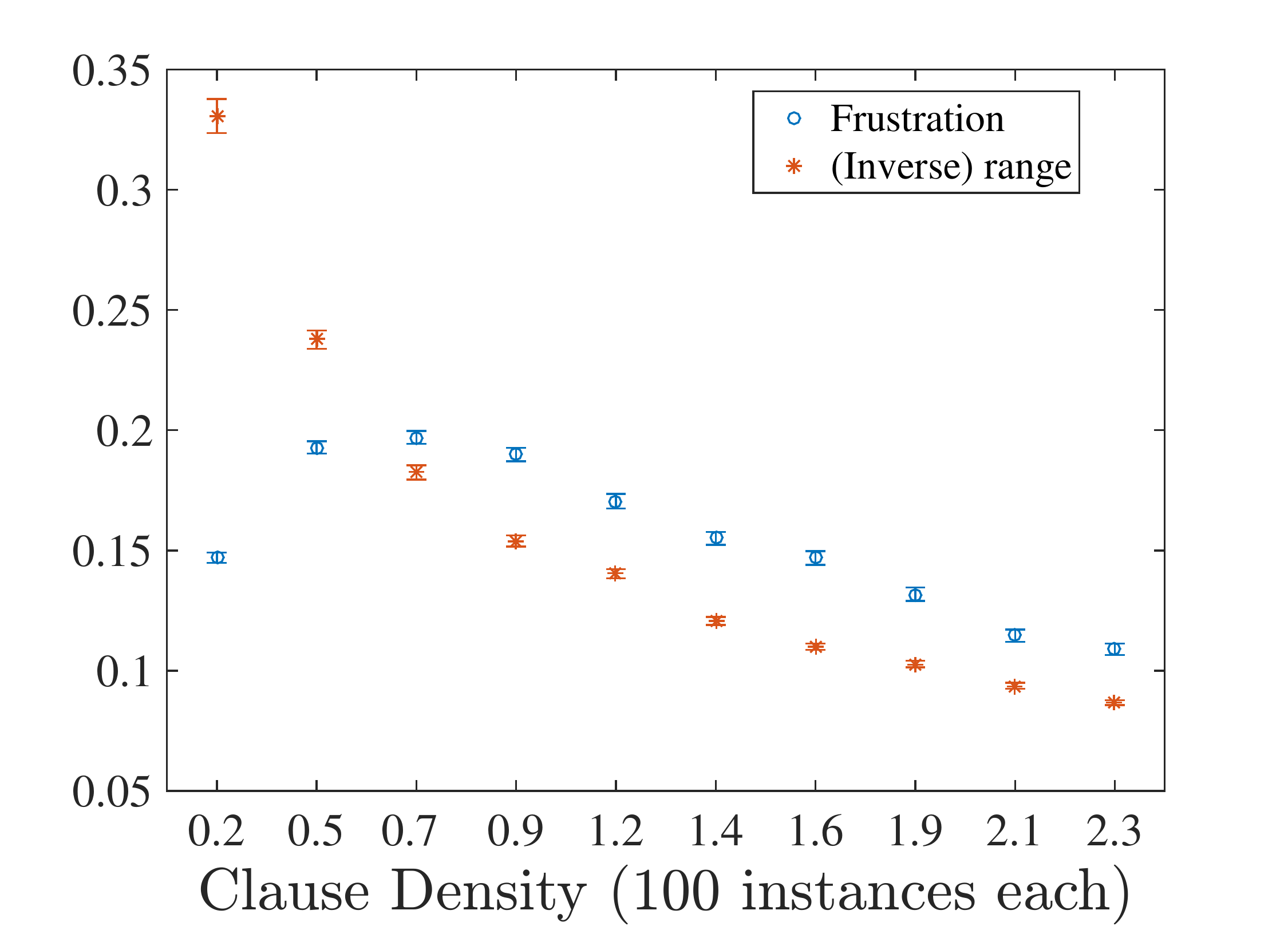} }
\subfigure[\ Embeddable planted]{\includegraphics[width=0.32\textwidth]{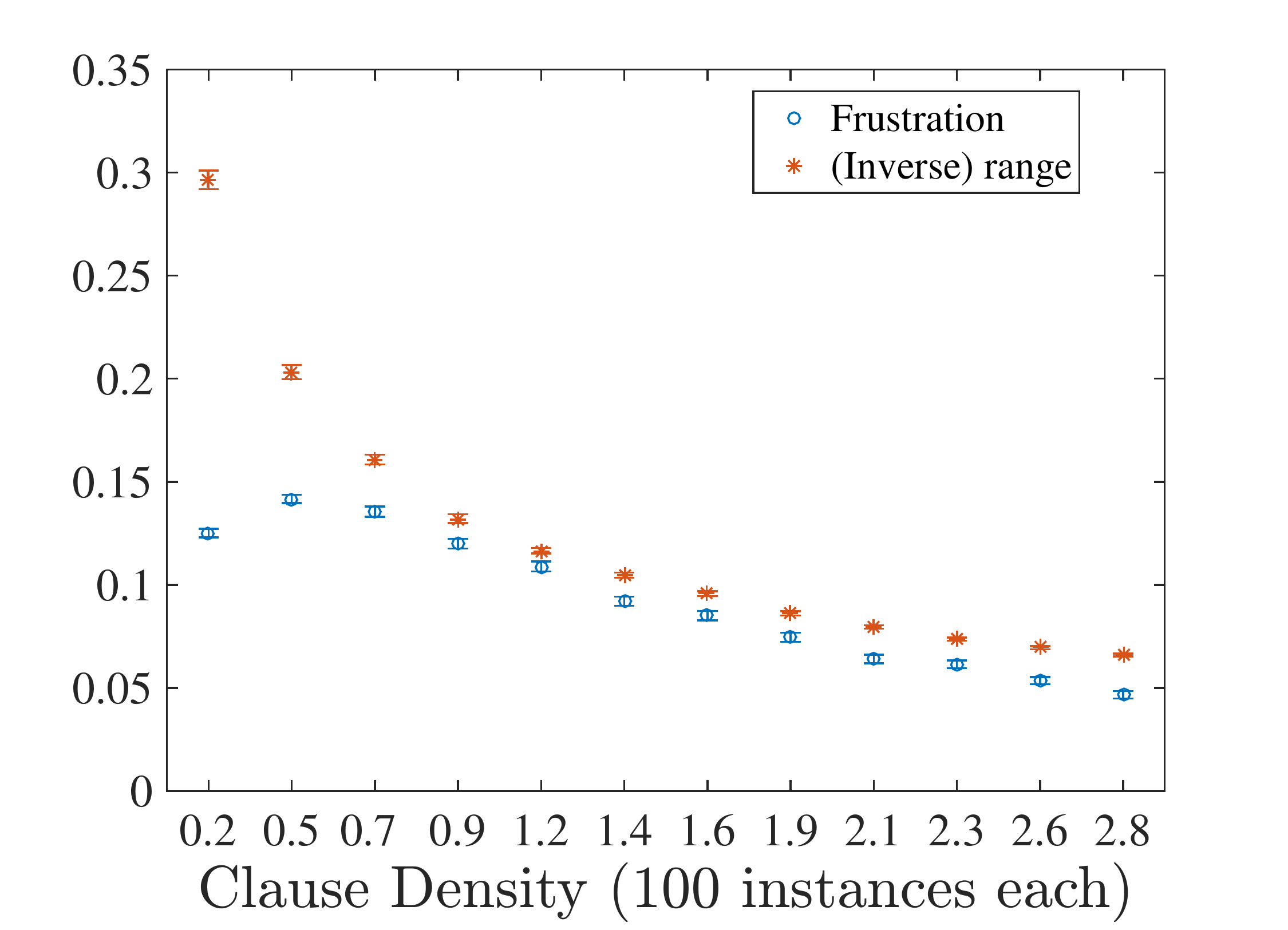} }
\subfigure[\ Weighted planted]{\includegraphics[width=0.32\textwidth]{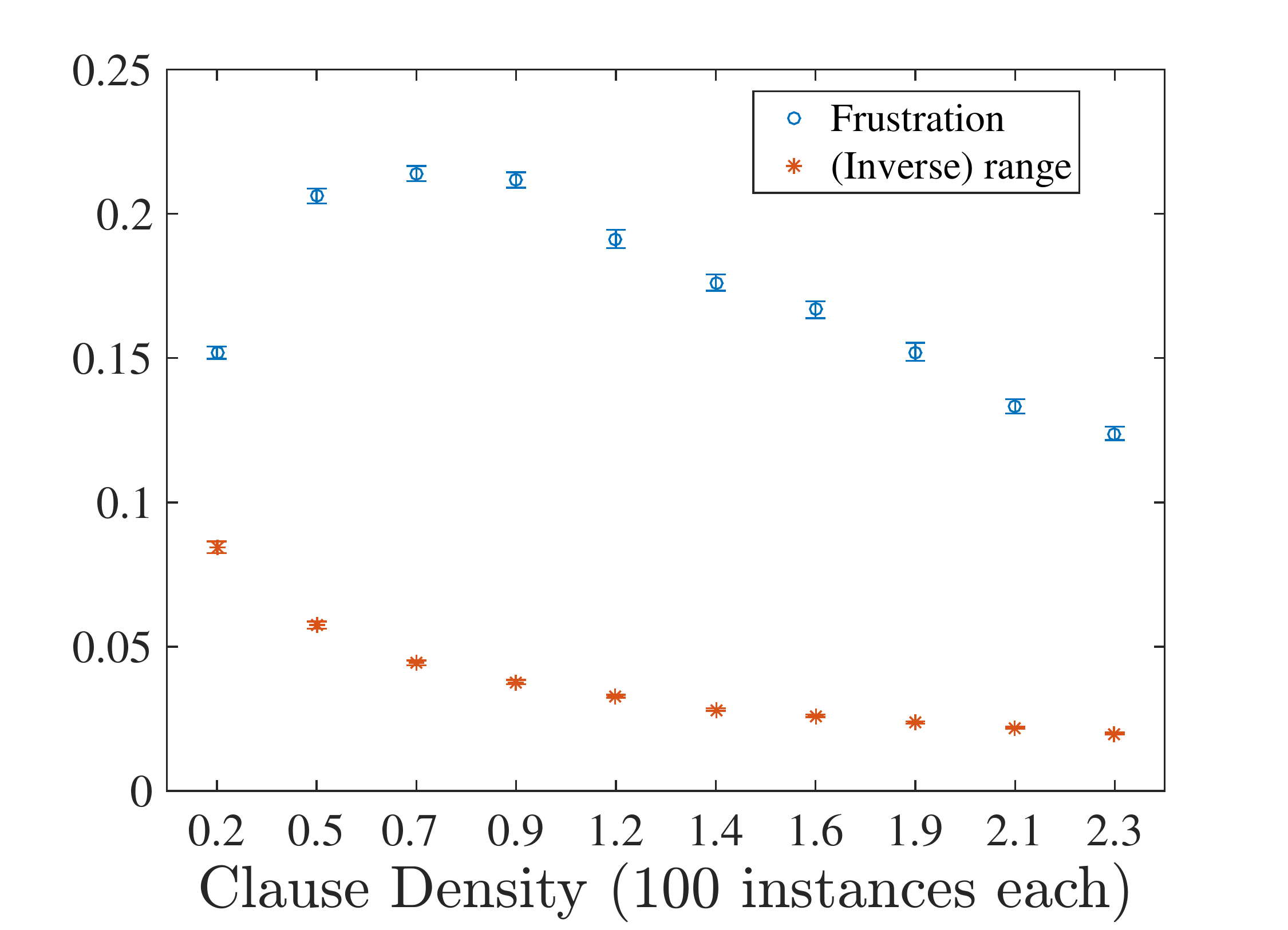} }
%\subfigure[\ Deformed Embeddable]{\includegraphics[width=0.45\textwidth]{Figures/Frust-Rang-NonHomEmbeddablePlanted-} }
\caption{Range and frustration as a function of the clause density. Note the correlation of frustration (range) to the number of broken qubits obtained with the nonuniform (uniform) encoding choices of Figs.~\ref{fig:plantedfull3}-\ref{fig:deformedembeddableplantedfull3}.} 
\label{fig:frust-range}
\end{center}
\end{figure*}
%%%%%%%%%%%%%%%%%%%%

\begin{figure*}[t]
\begin{center}
\subfigure[\ Planted]{\includegraphics[width=0.48\textwidth]{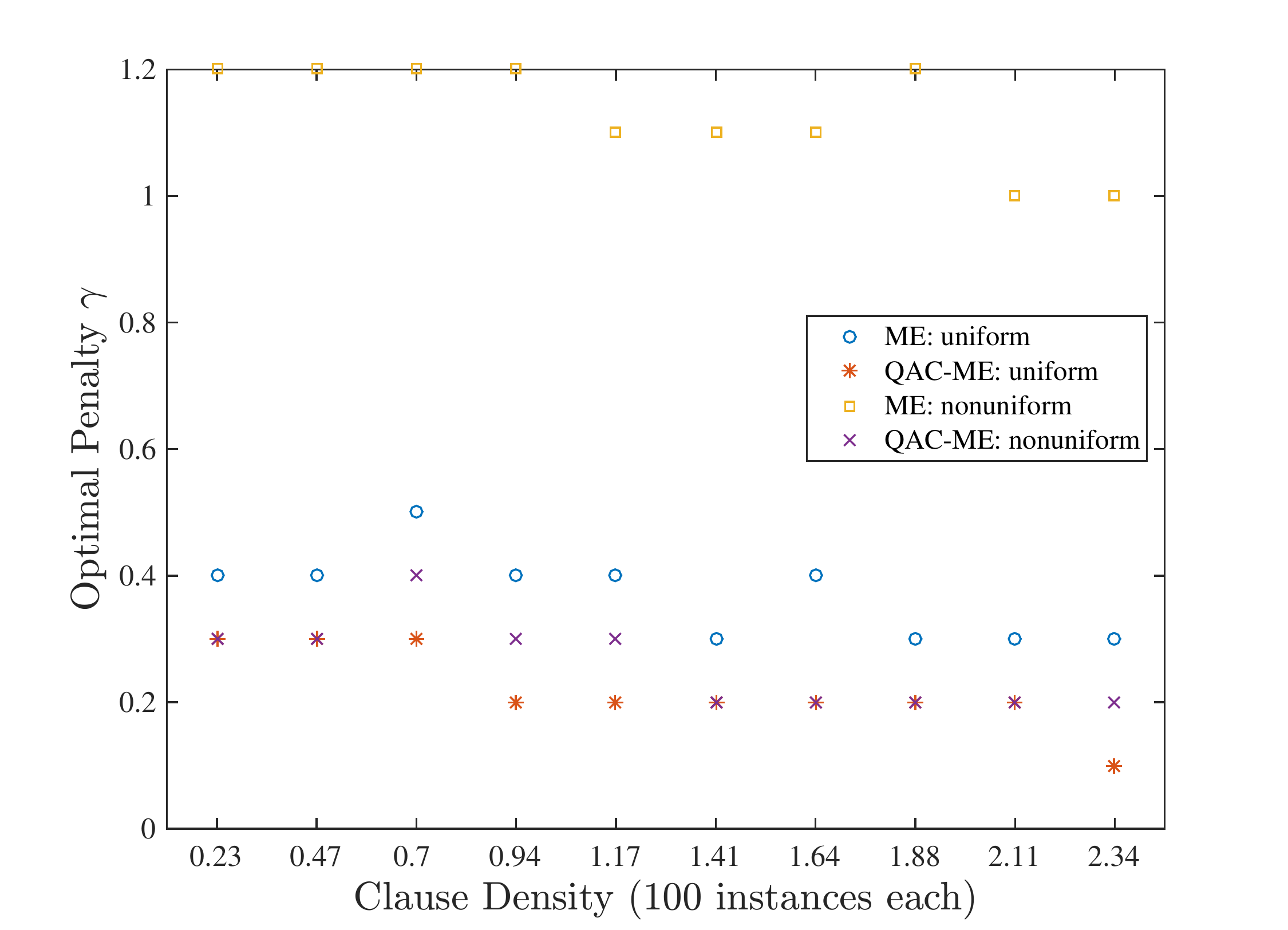}}
\subfigure[\ Weighted planted]{\includegraphics[width=0.48\textwidth]{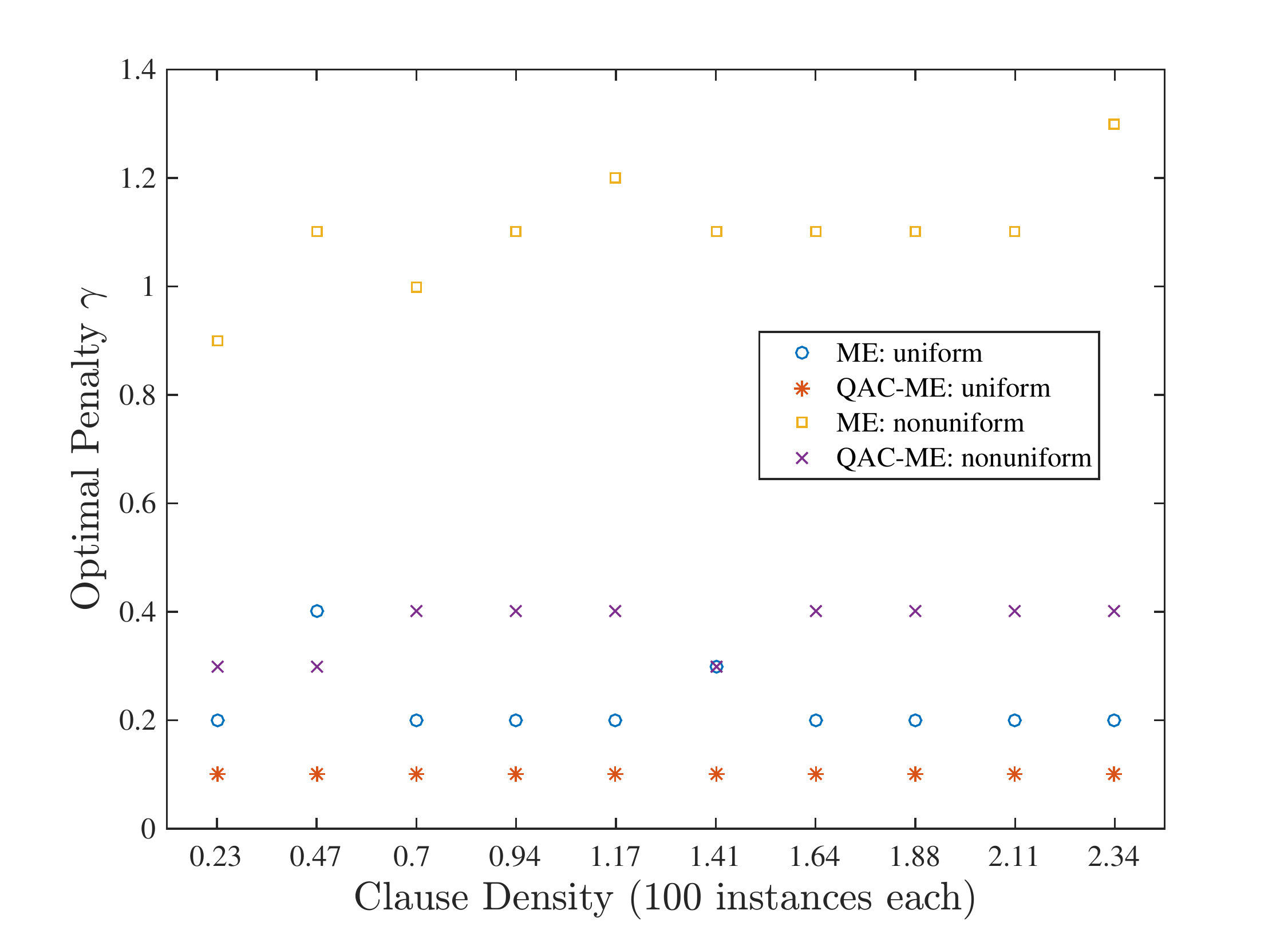}}
\subfigure[\ Embeddable planted]{\includegraphics[width=0.48\textwidth]{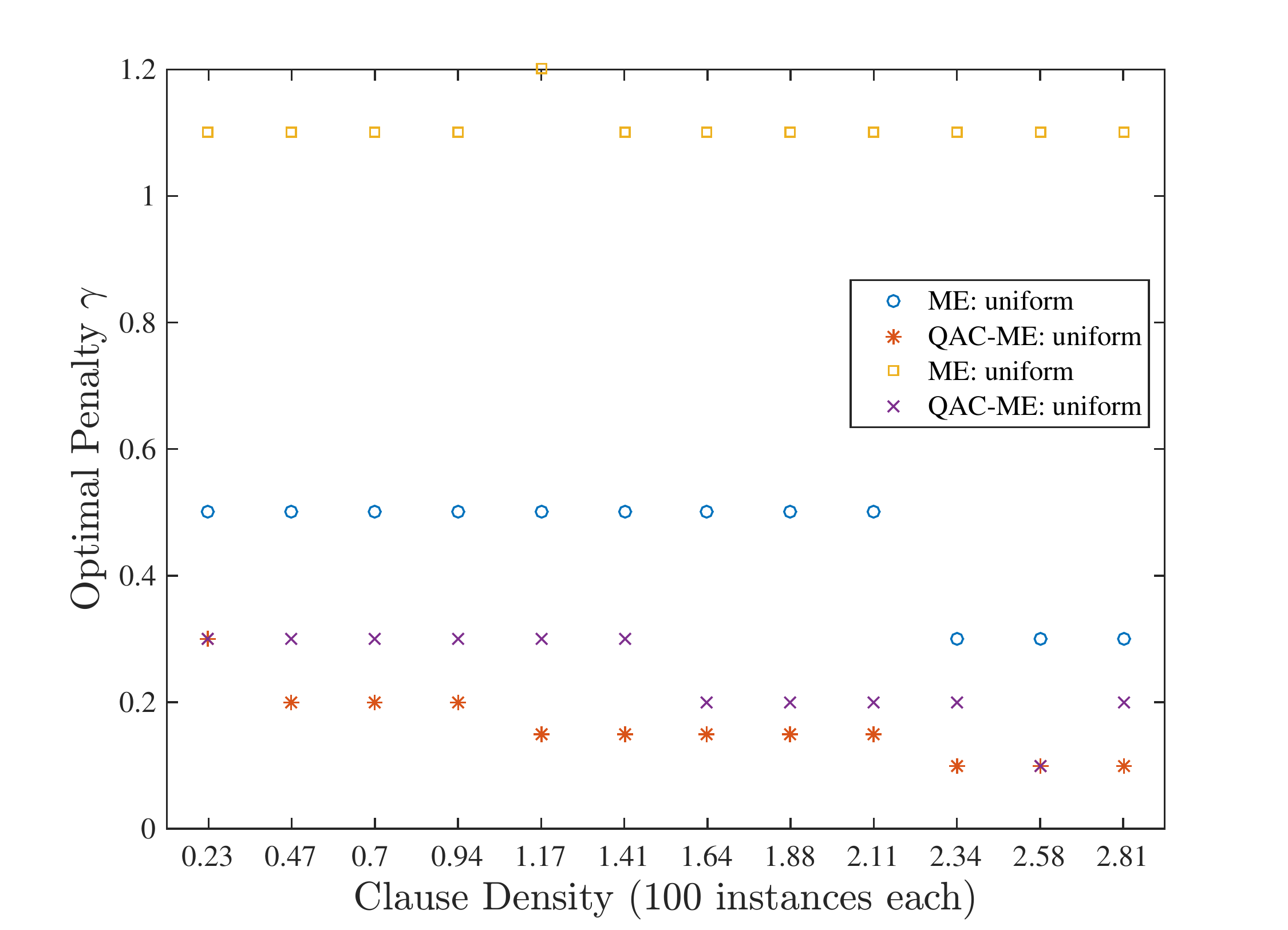}}
\subfigure[\ Deformed embeddable]{\includegraphics[width=0.48\textwidth]{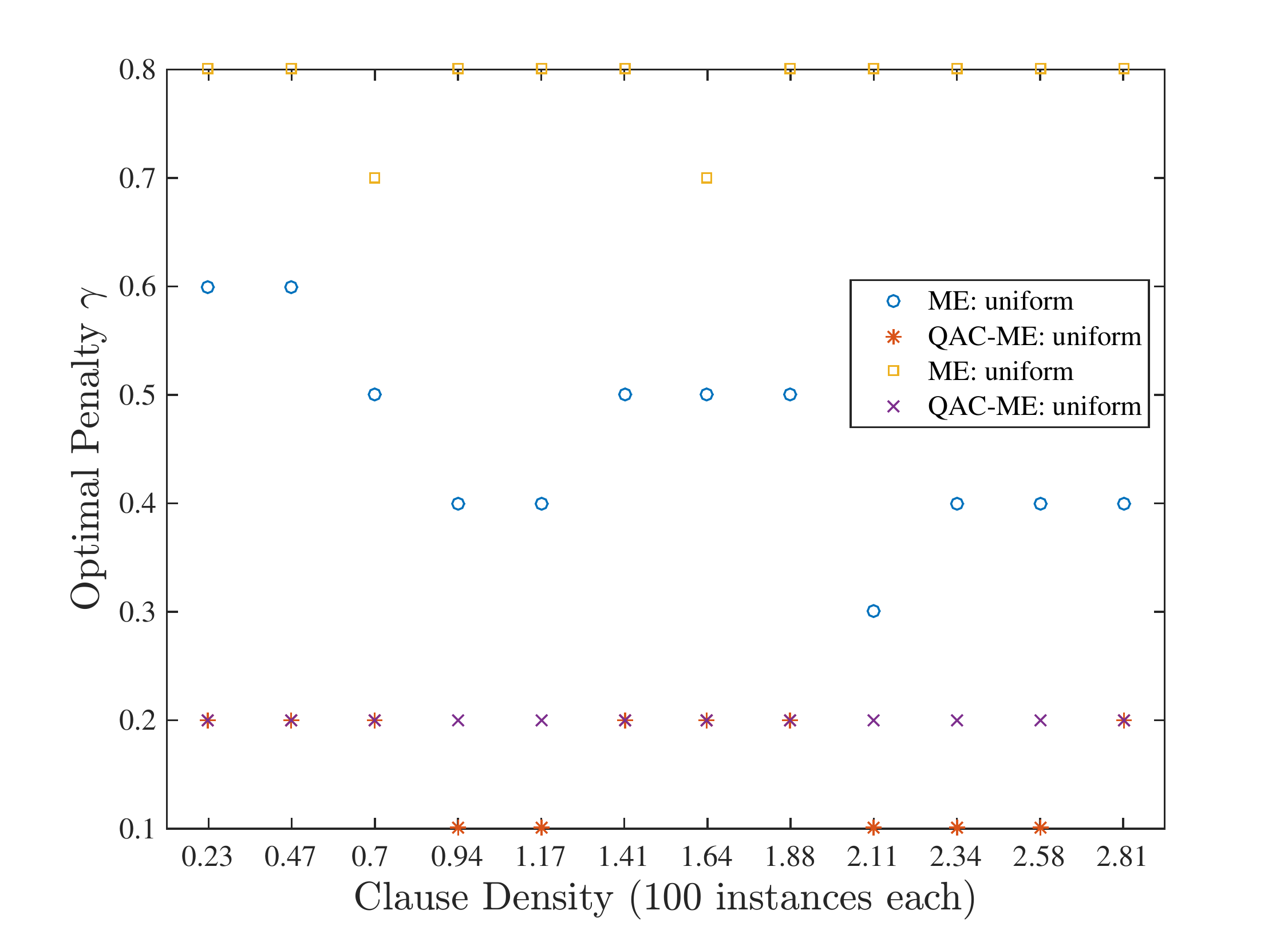}}
\caption{The four panels show the optimal penalty strength $\gamma$ for each set of instances and encoding choice (uniform/nonuniform). The optimal $\gamma$ is fairly constant over the different values of the clause density $\alpha$. } 
\label{fig:opt-gamma}
\end{center}
\end{figure*}

\begin{figure*}[ht]
\begin{center}
\subfigure[\ Embeddable planted, uniform]{\includegraphics[width=0.48\textwidth]{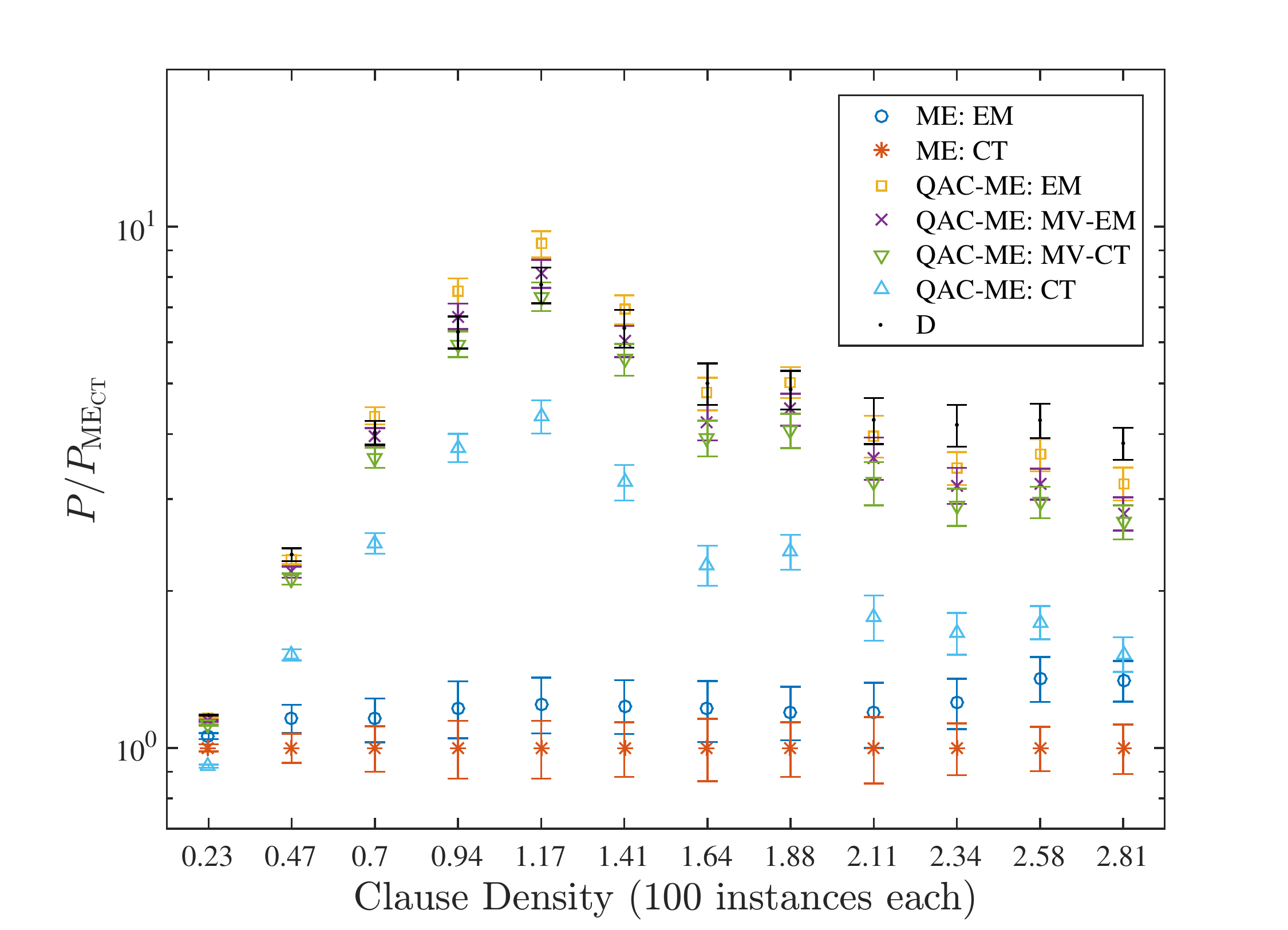}}
\subfigure[\ Embeddable planted, nonuniform]{\includegraphics[width=0.48\textwidth]{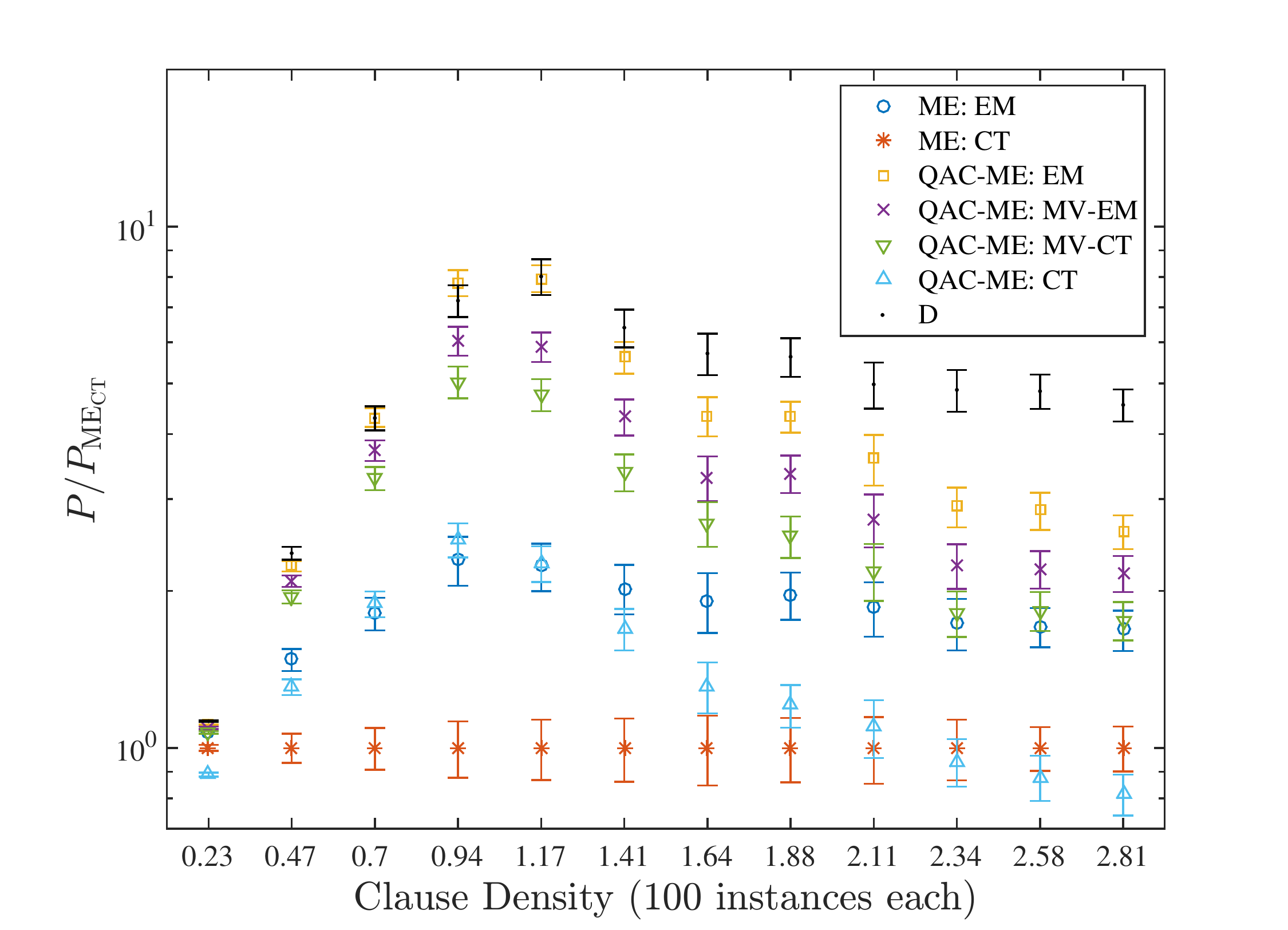}}
\subfigure[\ Embeddable planted, uniform.]{\includegraphics[width=0.48\textwidth]{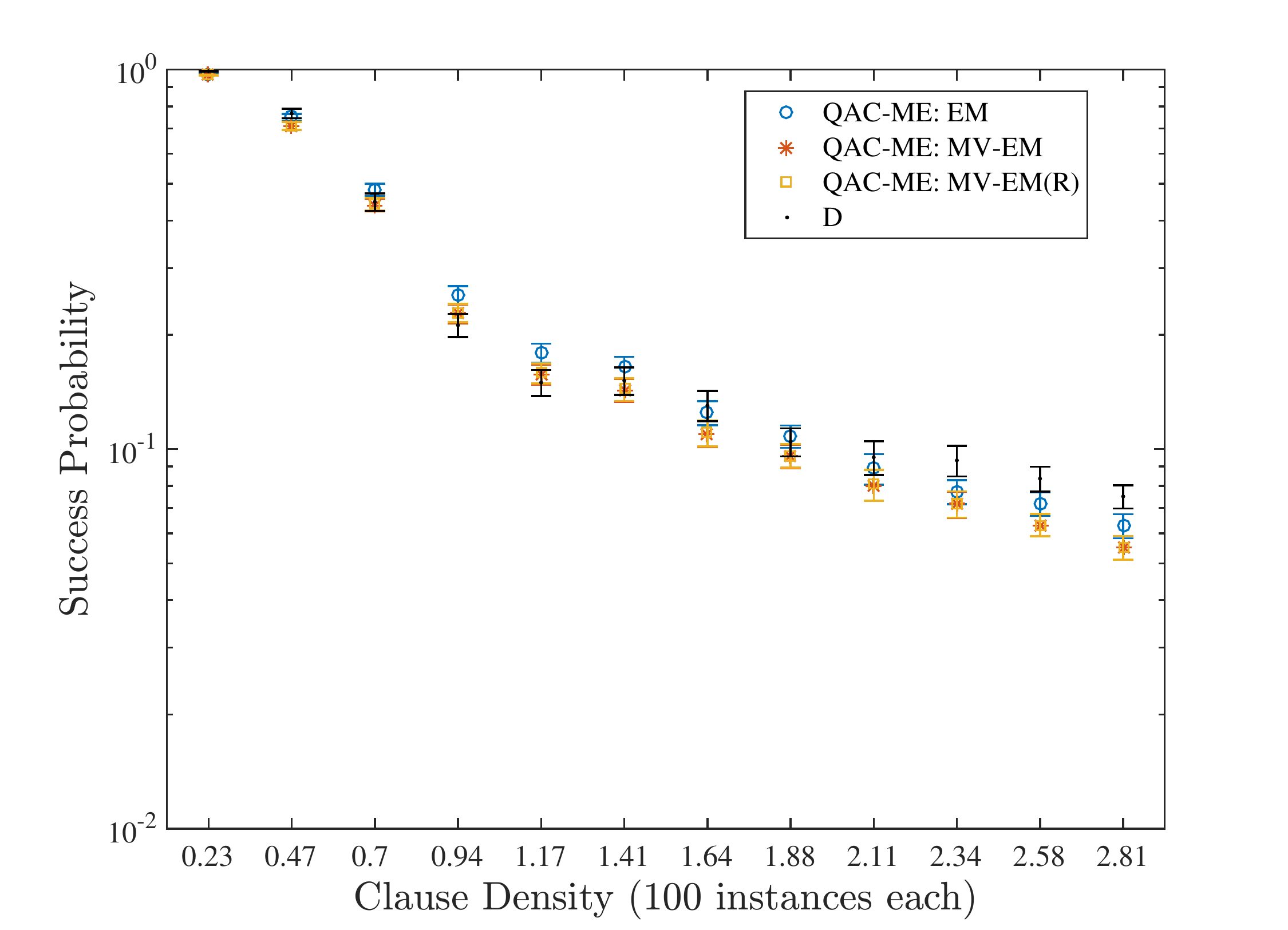} }
\subfigure[\ Embeddable planted, nonuniform.]{\includegraphics[width=0.48\textwidth]{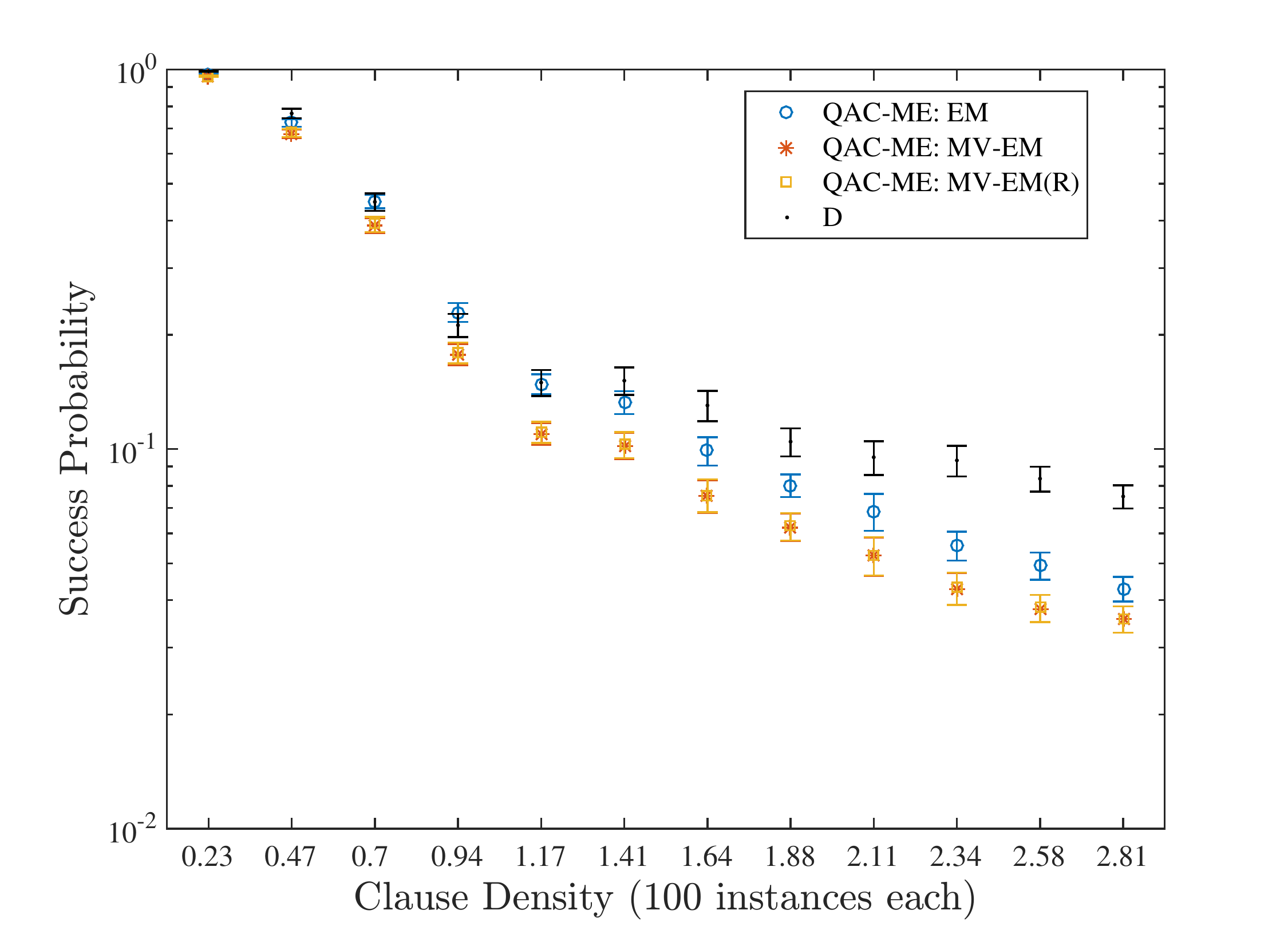} }
\caption{Comparison between the decoding strategies for the embeddable planted instances.} \label{fig:embeddableplantedfull2}
\end{center}
\end{figure*}

%\begin{figure*}[ht]
%\begin{center}
%\subfigure[\ weighted planted, uniform]{\includegraphics[width=0.45\textwidth]{Figures/Decoding_efficacy-WeightedPlanted-}}
%\subfigure[\ weighted planted, nonuniform]{\includegraphics[width=0.45\textwidth]{Figures/Decoding_efficacy-WeightedPlanted-mean}}
%\subfigure[\ weighted planted, uniform]{\includegraphics[width=0.45\textwidth]{Figures/EM_efficacy-WeightedPlanted-}}
%\subfigure[\ weighted planted, nonuniform]{\includegraphics[width=0.45\textwidth]{Figures/EM_efficacy-WeightedPlanted-mean}}
%\caption{Comparison between the decoding strategies for the weighted planted instances.} \label{fig:weightedplantedfull2}
%\end{center}
%\end{figure*}

\begin{figure*}[ht]
\begin{center}
\subfigure[\ Deformed embeddable, uniform]{\includegraphics[width=0.48\textwidth]{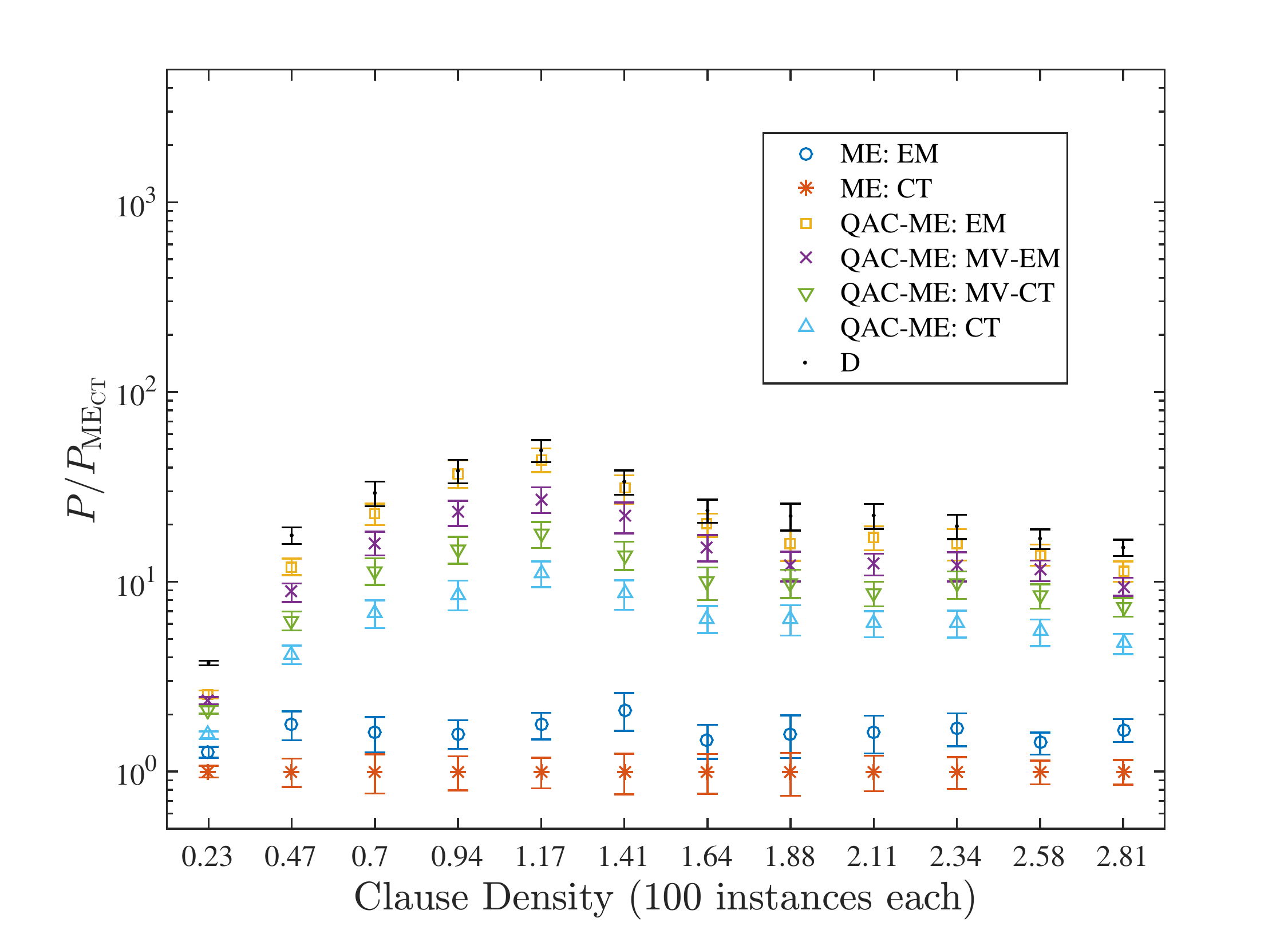}}
\subfigure[\ Deformed embeddable, nonuniform]{\includegraphics[width=0.48\textwidth]{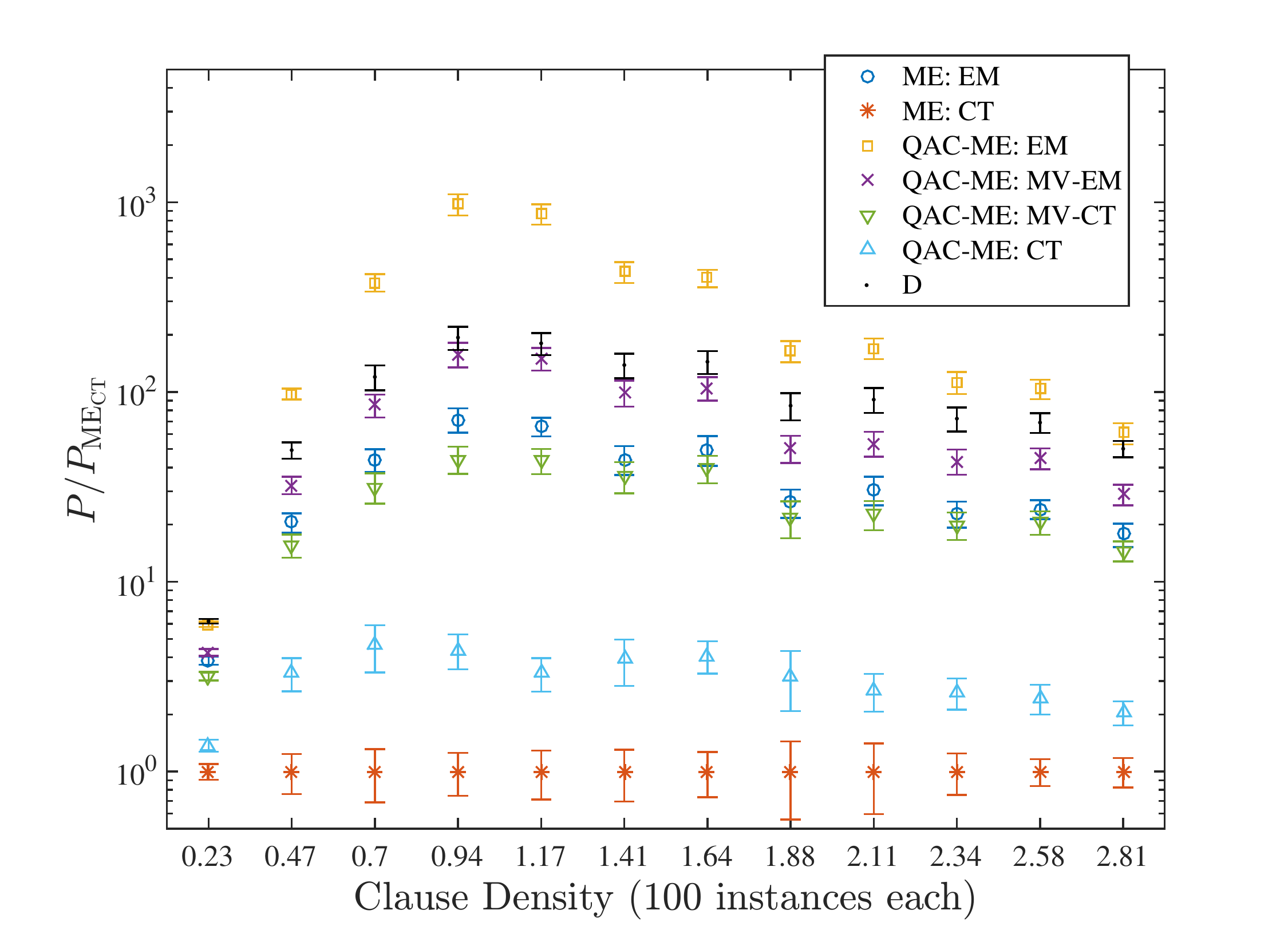}}
\subfigure[\ Deformed embeddable, uniform]{\includegraphics[width=0.48\textwidth]{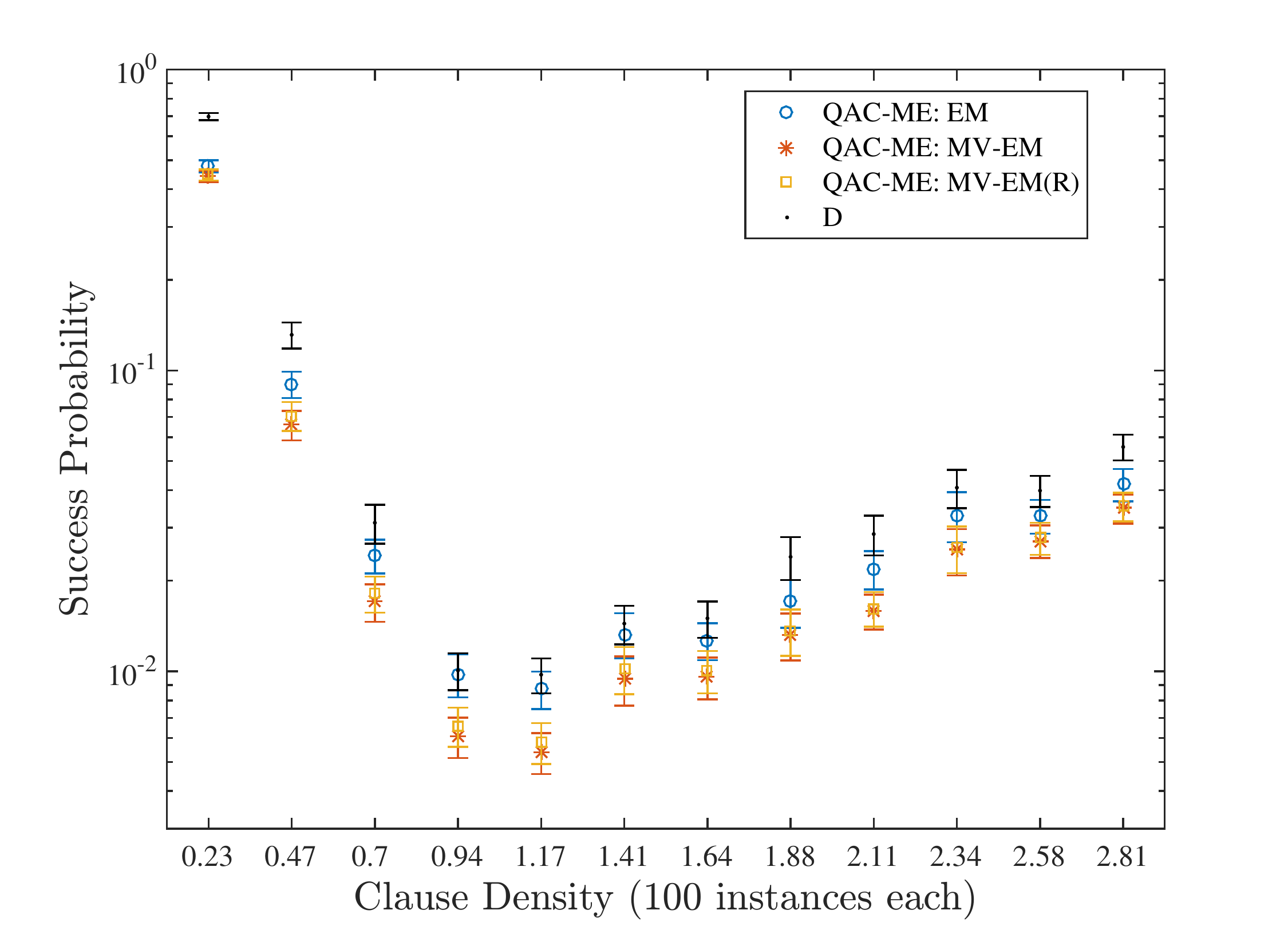}}
\subfigure[\ Deformed embeddable, nonuniform]{\includegraphics[width=0.48\textwidth]{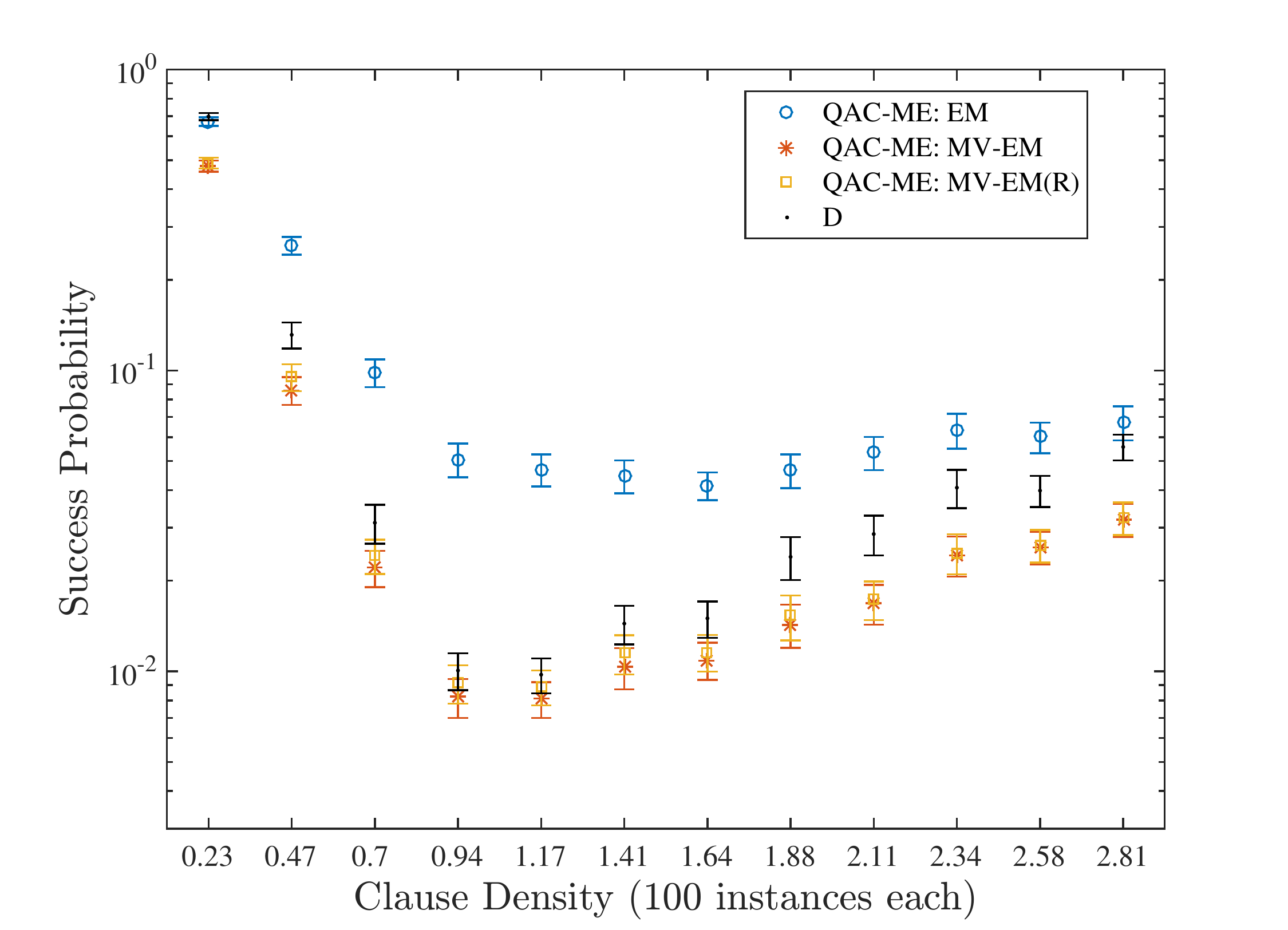}}
\caption{Comparison between the decoding strategies for the deformed embeddable instances.}  \label{fig:deformedembeddableplantedfull2}
\end{center}
\end{figure*}

\begin{figure*}[ht]
\begin{center}
\includegraphics[width=0.5\textwidth]{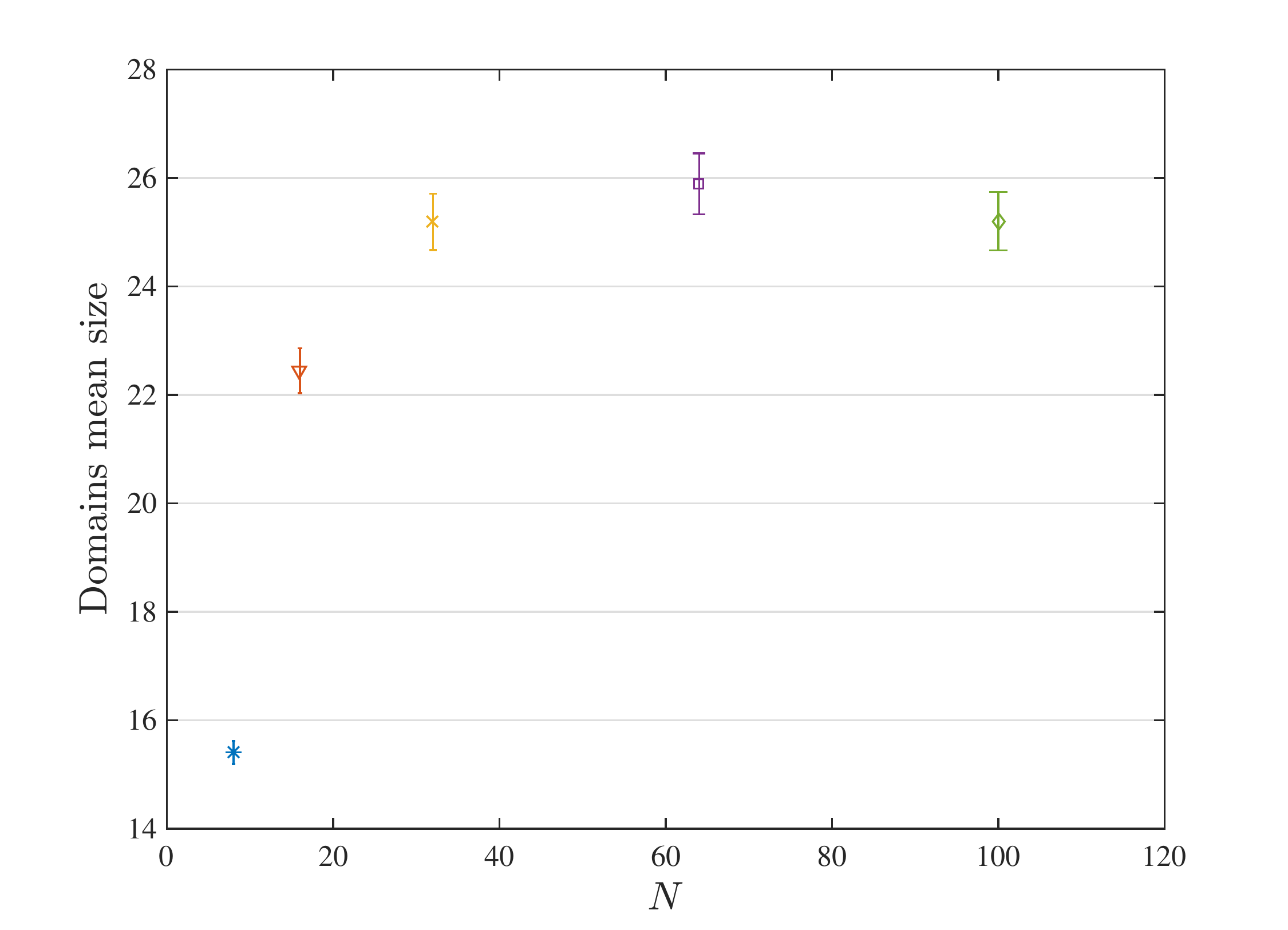}
\caption{Mean size of domains connected to a central spin for various $N\times N\times 2$ 2LG graph sizes for the worst-case percolation probability $p^{\mathrm{tie}}_{\mathrm{wc}} = 0.375$. The mean size of connected domains saturates at a size of about $25$ spins. We used $10^4$ random configurations per point.} 
\label{fig:treelength}
\end{center}
\end{figure*}
%%%%%%%%%%%%%%%%%%%%

%%%%%%%%%%%%%%%%%%%%
\section{Energy Minimization Decoding} 
\label{app:EMD}

We performed decoding through energy minimization by implementing simulated annealing on the ``decoding'' Hamiltonian defined in Eq.~\eqref{eq:DecHam}. We used a linear schedule for the temperature, with the initial and final temperatures equal to $T_{\mathrm{in}} = 4\, {\rm max}_{i,j} (J_{ij})$ and $T_{\mathrm{fin}} = 0.1\, {\rm min}_{i,j} (J_{ij})$ respectively. We chose the computational effort in the decoding step to be constant for all the decoding attempts. For each readout after an annealing cycle, decoding consisted of repeating $10$ simulated annealing runs with $10$ sweeps. For each decoding attempt, we defined a decoding success probability $P_{\mathrm{dec}}$ as the fraction of times the lowest decoded configuration was found. We consider this quantity as a measure of our decoding quality, with a $P_{\mathrm{dec}}$ close to $1$ meaning that the best decoded configuration is likely to be the optimally decoded state. A small value of $P_{\mathrm{dec}}$, on the other hand, means that a better decoded configuration could likely be found with increased computational effort (a larger number of sweeps and/or repetitions in the simulated annealing procedure).

$P_{\mathrm{dec}}$ is shown in Fig.~\ref{fig:deceff} for a few representative examples. Recall that the number of broken qubits is typically larger for QAC-ME (Fig.~\ref{fig:exptie}), which therefore requires a larger decoding effort in order to obtain a similar decoding quality. This explains why we observe lower decoding success probabilities for the QAC-ME case with energy minimization. Nonuniform penalties are also responsible for a larger number of broken qubits. Improving $P_{\mathrm{dec}}$ for QAC-ME by expending more effort on the simulated annealing step would thus further increase the advantage of QAC-ME with respect to ME, as well as the advantages of nonuniform penalties with respect to the uniform case.

\section{Concatenating the square code}
\label{app:concat}

In Sec.~\ref{sec:square-code} of the main text we mentioned that the square code can be concatenated, and described the first and second encoding levels. To see how this can be continued, consider, e.g., combining $n\times n$ encoded level-$2$ qubits to form an encoded level-$3$ qubit on each plane of the 2LG, to obtain a $[[4n^4,1,4n^4]]_0$ code. In general we have $[n,r]\equiv  [[4n^{2(r-1)},1,4n^{2(r-1)}]]_0$ for a level-$r$ code with integer $r\geq 1$.
The realizable stabilizer terms included in the penalty Hamiltonian suffice to penalize every possible bit-flip error, at every level of concatenation. 

It is interesting to consider the growth in the encoded fields and couplings. As illustrated in Fig.~\ref{SampleEncodedL1}, for levels $r\geq 2$ the encoded couplings grow in proportion to the side length of the square of encoded qubits, and the encoded fields must grow in the same proportion in order to maintain their relative strengths. E.g., the level-$2$ local fields correspond to $n$ level-$1$ realizations, each with two level-0 realizations. Analogously, every encoded level-$2$ coupling comprises $n$ level-$1$ couplings, which are in turn equivalent to $2$ level-0 couplings. Thus 
$\bar{h}^{(2)}_\ell =  n \bar{h}^{(1)}_\ell = 2 n h_\ell$ and $\bar{J}^{(2)}_{\ell\ell'} = n \bar{J}^{(1)}_{\ell\ell'} = 2 n J_{\ell\ell'}$. 
At level $r$ each square has side length $n^{r-1}$, so the encoded level-$r$ fields and couplings are
\beq
\bar{h}^{(r)}_\ell =  2 n^{r-1} h_\ell \ , \qquad \bar{J}^{(r)}_{\ell\ell'} = 2n^{r-1}  J_{\ell\ell'} \ ,
\eeq
i.e., the energy scale of the level-$r$ final Hamiltonian has been boosted by a factor of $2n^{r-1}$. Note that the encoded penalty strength grows in proportion to the area of each square if all the penalties are utilized. Since this would result in overly rigid qubits a simple remedy is to use only a linear number of the total available penalties.

Since the number of physical qubits required grows quadratically faster than the strength of the encoded local fields and couplings, it is not a priori clear that a higher concatenation level $r$ is beneficial. However, in a Markovian model where the probability of a thermal excitation at inverse temperature $\beta$ is proportional to the Gibbs factor $e^{-\beta \Delta}$ (e.g., Ref.~\cite{Albash:2015nx}), with $\Delta$ the gap from the ground state, it is possible to realize a gain by increasing $r$, provided the gap grows along with the energy scale of the level-$r$ final Hamiltonian. Determining the scaling of the gap with $r$ is a non-trivial problem since the energy scale of the initial transverse field Hamiltonian does not also grow under the present encoding, as remarked earlier. Our proposal for concatenating the square code for QAC can be viewed as a physically realizable (and hence severely compromised) version of Bacon's proposal for quantum concatenated-code Hamiltonians \cite{Bacon:2008dp}.

%%%%%%%%%%%%%%%%%%%%
\section{Recursive decoding of the square code}
\label{app:recursive}

In this section we propose an alternative (untested) recursive decoding scheme of the square code, that is similar to the energy minimization strategy. 

Suppose we already decoded all the untied encoded qubits with only the tied encoded qubits remaining to be decoded.  Consider a problem Hamiltonian $H^{(0)} = \sum_{(i,j)\in G} J_{ij} \sigma_i^z \sigma_j^z$, where the sum is over all connected pairs of logical qubits $i$ and $j$ in the encoded graph $\bar{G}$ (the graph obtained after, say, the first level of encoding using the square code), and for notational simplicity here we use undecorated Pauli operators to denote the encoded operators. At the end of a (quantum) annealing run some encoded qubits will be tied. The untied qubits now have fixed values $\{s_i\}$, so they act as local fields on the tied qubits, weighted by the couplings.  We can rewrite the problem Hamiltonian after this run as
\begin{equation}
H^{(1)} = \sum_{j'\in \mathrm{ties}} h_{j'} \sigma^z_{j'} +\sum_{(i',j')\in \bar{G}'} J_{i'j'} \sigma^z_{i'} \sigma^z_{j'}\ ,
\end{equation}
where the sum runs only over the tied qubits, which occupy the vertices of a subgraph $\bar{G}'$ of $\bar{G}$, and where the local fields are
\begin{equation}
h_{j'} = \sum_i J_{ij'} s_i \ ,
\end{equation}
with the sum including all the untied qubits coupled to the given tied qubit $j'$.

The renormalized Hamiltonian $H^{(1)}$ represents a new Ising problem. To find the values of the tied qubits we can run $H^{(1)}$ on the (quantum) annealer, i.e., implement the EM strategy by bootstrapping the problem. If after such a run there are still tied qubits we can repeat the same process, defining a new Hamiltonian $H^{(2)} = \sum_{j''\in \mathrm{ties}} h_{j''} \sigma^z_{j''} +\sum_{(i'',j'')\in \bar{G}''} J_{i''j''} \sigma^z_{i''} \sigma^z_{j''}$, with $h_{j''} = \sum_{i'} J_{i'j''} s_{i'}$, etc. Note that successive runs of the quantum annealer need not necessarily resolve the tied qubits, but since $\bar{G}''$ is by construction a subgraph of $\bar{G}'$, this process will converge. Any ties that remain after the process has converged can be decoded using the previously described EM strategy. Accounting for wall-clock times, it may be preferable to start running this sequence of recursive optimizations on a classical solver once the subgraph size has become small enough.
%%%%%%%%%%%%%%%%%%%%

\section{Broken Qubit Probability, Range and Frustration}
\label{sec:rang&frust}

In this section we present the complete data set in Figs.~\ref{fig:plantedfull3} (planted), \ref{fig:embeddableplantedfull3} (embeddable planted), \ref{fig:weightedplantedfull3} (weighted planted), and \ref{fig:deformedembeddableplantedfull3} (deformed embeddable) 
for the broken qubit (and ties) probabilities for the instances studied  for all values of the penalties, for both the ME and QAC-ME cases. Note how, as expected, the broken qubit probability ($p_{\mathrm{BQ}}$) strongly depends on the strength of the energy penalties. One more feature that is evident from the same figures is the different behavior of $p_{\mathrm{BQ}}$ in the uniform and nonuniform cases, at a given value of the penalty strength $\gamma$. In the uniform case, $p_{\mathrm{BQ}}$ is mostly a monotonically decreasing function of the clause density $\alpha$, while in the nonuniform case there is a clear peak that is located close to the critical clause density.
%, i.e., the clause density that gives the set with the hardest instances. 

There is a simple and intuitive explanation of this behavior in terms of range and frustration, that is consistent with the general discussion of Sec.~\ref{sec:MEQA-frustr}. Let us first give a quantitative definition of both frustration and range. We define frustration as the percentage of $J_{ij}$ couplings that are not satisfied on the planted ground state solution. The inverse range is the smallest (non-zero) absolute value of the $J_{ij}$ couplings of a given instance. Each time we implement a given instance, we always set the value of the largest coupling equal to $1$. This exploits the largest energy scale achievable with the DW2 processor. The range thus gives us an estimate of the required precision needed to correctly implement the given instance.

Figure~\ref{fig:frust-range} shows the computed values of the required (inverse) range and the frustration of the various classes of instances. We can see that the qualitative behavior of $p_{\mathrm{BQ}}$ in the uniform case follows that of the inverse range. This is related to the fact that a smaller inverse range implies the presence of a  larger number of weak couplings that are dominated by the uniform and relatively larger energy penalty, making the appearance of a broken qubit less likely. On the other hand, the value of the energy penalties in the nonuniform case is adjusted to the typical value of the logical couplings. Typically, a smaller inverse range results in weaker energy penalties, balancing the range effect. Thus, in the nonuniform case  $p_{\mathrm{BQ}}$ is more positively correlated with frustration. This is consistent with the general expectation that  $p_{\mathrm{BQ}}$ is large for highly frustrated instances.

%%%%%%%%%%%%%%%%%%%%
\section{Data Collection and Analysis}
\label{app:data}

As already mentioned in Section \ref{sec:ER-method}, the presence of control errors (ICE) due to low frequency noise and systematic biases prevents the physical couplings to be set with a precision better than $\sim 5\%$. To average out the effects of such errors, we run 10 programming cycles for all the encoding schemes (D, ME, QAC-ME)  and penalty values for every instance considered. Each programming cycle implements a gauge transformation and consists of $10^3$ annealing runs. For each instance, encoding strategy and penalty value, (altogether labeled here by the index $i$) we extract an ``intrinsic" success probability $P^i$ as the overall probability to find the ground state over the $10$ programming cycles. In other words, this is the mean of the success probabilities extracted from each programming cycle, and it serves as our best guess for the intrinsic (without ICE) success probability of each encoding of a given instance
\beq
P^i = \sum_{g} P^i_g, \quad g = 1,\dots ,10\,.
\eeq

We then use these gauge-averaged probabilities to compute the renormalized success probabilities for the D and and ME cases:
\bes
\begin{align}
P^i_{\mathrm{D}}    &\mapsto 1 - (1-P^i_{\mathrm{D}})^4\, ,\\
P^i_{\mathrm{ME}} &\mapsto 1 - (1-P^i_{\mathrm{ME}})^2\,,\\ 
P^i_{\mathrm{QAC-ME}}       & \mapsto  P^i_{\mathrm{QAC-ME}}\,. 
\end{align}
\ees
As explained in the main text, this renormalization is necessary to correctly take into account the amount of hardware resources that each implementation requires. We then use these renormalized probabilities to compute $5000$ means, obtained from $5000$ bootstrapped samples of $100$ instances (picked with replacement among the $100$ total number of instances) for each clause density $\alpha$ and penalty value $\gamma$. The quantities plotted in Fig.~\ref{fig:planted} (as well as in all other figures plotting experimental data) are the mean and the standard error of such $5000$ samples of bootstrapped means.

The same approach was followed to compute means and error bars of other similar plots. See, e.g., the plot showing broken qubits probabilities (Fig.~\ref{fig:exptie}). 
%%%%%%%%%%%%%%%%%%%%

%%%%%%%%%%%%%%%%%%%%
\section{Energy Penalties}
\label{app:penalties}

We show explicitly in Fig.~\ref{fig:opt-gamma} the value of the optimal $\gamma$ for all the cases considered in the main text. As can be seen, this value is fairly constant as a function of the clause density $\alpha$. There is a slight tendency for the optimal $\gamma$ to decrease with $\alpha$ in the case of uniform penalties. This is most pronounced in the ME implementation of the deformed embeddable instances. When choosing nonuniform penalties, however, the optimal $\gamma$ again becomes almost constant. This shows that the nonuniform choice of penalties has the potential of being closer to optimality than the uniform case with a single choice of the optimal penalty strength. Since optimization of the energy penalties can be time consuming, this property is important for any practical encoding scheme. Ideally, we would like an encoding scheme (for the choice of penalties) that is close to optimality and does not depend on any quantity that has to be optimized on an instance-by-instance case.

\section{Decoding Strategies}
\label{app:dec}

We show in Figs.~\ref{fig:embeddableplantedfull2} and \ref{fig:deformedembeddableplantedfull2} the comparison between various decoding strategies for the embeddable planted and the deformed embeddable cases. This complements the results shown in Figs.~\ref{fig:majvsmin}~and~\ref{fig:majvsmin3} of the main text, which showed the results for the planted and weighted planted instances. Results for the embeddable sets are very similar to the sets defined on the full 2LG. We notice here too that the use of the EM strategy is crucial for the QAC-ME encoding to give comparable (for the planted embeddable set) or better (for the deformed embeddable set) results than the direct implementation (D).

%\begin{figure*}[ht]
%\begin{center}
%\subfigure[\ planted, uniform]{\includegraphics[width=0.45\textwidth]{Figures/Decoding_efficacy-Planted-}}
%\subfigure[\ planted, nonuniform]{\includegraphics[width=0.45\textwidth]{Figures/Decoding_efficacy-Planted-mean}}
%\subfigure[\ planted, uniform.]{\includegraphics[width=0.45\textwidth]{Figures/EM_efficacy-Planted-} }
%\subfigure[\ planted, nonuniform.]{\includegraphics[width=0.45\textwidth]{Figures/EM_efficacy-Planted-mean} }
%\caption{Comparison between the decoding strategies for the planted instances. \mg{Isn't this exactly the same as Fig.~\ref{fig:majvsmin}?}}
%\label{fig:plantedfull2}
%\end{center}
%\end{figure*}

%%%%%%%%%%%%%%%%%%%%

%%%%%%%%%%%%%%%%%%%%
\section{Worst-case Scenario Decodability of the Square Code}
\label{app:worstcase}

We can set an upper limit (worst-case scenario) for the probability to obtain ties and broken qubits in general by assuming that all four physical qubits comprising an encoded qubit are uncorrelated. This corresponds to assuming that the output of the annealer corresponds to a thermal distribution with infinite temperature. Within a single encoded qubit there are $6$ physical qubit states that correspond to a tie (out of a total of $16$ states), so we obtain $p^{\mathrm{tie}}_{\mathrm{wc}} = 6/16 =0.375$ for the worst case probability to obtain a tie. The worst case probability to obtain a broken qubit is $p^{\mathrm{BQ}}_{\mathrm{wc}} = 14/16 =0.875$. While $p^{\mathrm{BQ}}_{\mathrm{wc}}$ is definitely above the percolation threshold of the square code, $p^{\mathrm{tie}}_{\mathrm{wc}}$ is almost certainly below it since the device output is better approximated by a finite temperature thermal distribution (as opposed to an infinite temperature distribution). Moreover, the non-vanishing penalty term and (if present) longitudinal field disfavor broken qubits and ties. This strongly suggests that in the more realistic scenario the number of broken qubits is well below the worst-case value. Fig.~\ref{fig:treelength} shows the mean size of the connected domains over $10^4$ randomly generated configurations for increasingly large $N\times N$ 2LG graphs. The fact that this mean size saturates instead of growing with $N$ is evidence that $p^{\mathrm{tie}}_{\mathrm{wc}} < p_{\textrm{2LG}}$. This means that the square code is efficiently decodable (at least using using the MV-EM strategy) even in the worst-case scenario.

%%%%%%%%%%%%%%%%%%%%
\section{Simulated Quantum Annealing}
\label{app:SQA}
We reported SQA results in Sec.~\ref{sec:NS} of the main text. Here we briefly review this technique. SQA is a quantum Monte Carlo based algorithm whereby Monte Carlo dynamics are used to sample from the instantaneous Gibbs state associated with the Hamiltonian in Eq.~\eqref{eq:adiabatic}.  The state at the end of the quantum Monte Carlo simulation of the quantum Hamiltonian $H(t)$ is introduced as the initial state of the Monte Carlo simulation with Hamiltonian $H(t+\Delta t)$.  This proceeds until $H(t_f)$ is reached.  Originally proposed as an optimization algorithm \cite{sqa1,Santoro}, it has gained some traction as a classically efficient model for finite temperature quantum annealers \cite{q108,speedup,Albash:2014if,Hen:2015rt}.  SQA does not capture the unitary dynamics of the quantum system, but it is hoped that the sampling of the instantaneous Gibbs state captures thermal processes in the quantum annealer, which may be the dominant dynamics if the evolution is sufficiently slow.  Although there is strong evidence that SQA does not completely capture the final-time statistics of the D-Wave processors \cite{Albash:2014if,Boixo:2014yu}, at present it is the only viable means to simulate large open quantum annealing systems.

We used discrete-time quantum Monte Carlo in our simulations, which we briefly review.  For Hamiltonians of the form of the transverse Ising model, such as in Eqs.~\eqref{eq:adiabatic} and \eqref{eq:HP}, the sampling from the instantaneous Gibbs state is done by sampling from the dual classical system with Hamiltonian $\mathcal{H}_C$ and inverse temperature $\beta$ :
\begin{align}
\beta \mathcal{H}_C(t) &=J_{\perp}(t)  \sum_{i,\tau} s_{i,\tau} s_{i,\tau+1} \\
&\quad + \frac{\beta}{N_{\tau}} B(t) \sum_\tau \left[ \sum_{i} h_i s_{i,\tau} + \sum_{i<j} J_{ij} s_{i,\tau} s_{j,\tau} \right] \notag \ ,
\end{align}
where $N_\tau$ is the number of Trotter slices used along the Trotter direction, $s_{i,\tau}$ denotes the $i$th classical spin on the $\tau$th Trotter slice, and 
\begin{equation}
J_{\perp}(t) \equiv  \frac{1}{2}  \ln\left[\tanh( \beta A(t) /N_{\tau})\right] < 0
\end{equation}
is the nearest-neighbor (ferromagnetic) coupling strength along the Trotter direction.  

In the simulations presented in the main text, the number of Trotter slices was fixed to $64$.  Spin updates were performed via Wolff-cluster updates \cite{PhysRevLett.62.361} along the Trotter direction only.
%%%%%%%%%%%%%%%%%%%%

\end{document}